\documentclass[useAMS,usenatbib]{mn2e}
\usepackage{graphicx}
\usepackage{amsmath}
\usepackage{amsmath,amssymb}
 \usepackage{color}
\newcommand{\msun}{{\rm M}_{\sun}}
\newcommand{\rsun}{{\rm R}_{\sun}}

\title[Possible Indirect Confirmation of the Existence of Pop III Massive Stars by Gravitational Wave]{
Possible Indirect Confirmation of the Existence of Pop III Massive Stars by Gravitational Wave
}
\author[T. Kinugawa et al.]{Tomoya Kinugawa$^{1}$ \thanks{E-mail:
kinugawa@tap.scphys.kyoto-u.ac.jp}, Kohei Inayoshi$^{1}$, Kenta Hotokezaka $^{1}$, Daisuke Nakauchi$^{1}$,\and  Takashi Nakamura$^{1}$\\
\\
$^{1}$Department of Physics, Graduate School of Science, Kyoto University,
Kyoto 606-8502, Japan}
\begin{document}

\date{\today}
\maketitle
\begin{abstract}
We perform population synthesis simulations for Population III (Pop III) coalescing compact binary which merge within the age of the universe. We found that the typical mass of Pop III binary black holes (BH-BHs) is $\sim30~\msun$ so that the inspiral chirp signal of gravitational waves can be detected up to z=0.28 by KAGRA, Adv. LIGO, Adv. Virgo and GEO network. Our simulations suggest that the detection rate of the coalescing Pop III BH-BHs is $140(68)\ {\rm events/yr ~(SFR_p/(10^{-2.5}\msun/yr/Mpc^3))\cdot Err_{sys}}$ for the flat (Salpeter) initial mass function (IMF), respectively, where $\rm SFR_{p}$ and $\rm Err_{sys}$ are the peak value of the Pop III star formation rate and the possible systematic errors due to the assumptions in Pop III population synthesis, respectively. $\rm Err_{sys}=1$ correspond to conventional parameters for Pop I stars. From the observation of the chirp signal of the coalescing Pop III BH-BHs, we can determine both the mass and the redshift of the binary for the cosmological parameters determined by Planck satellite. Our simulations suggest that the cumulative redshift distribution of the coalescing Pop III BH-BHs depends almost only on the cosmological parameters. We might be able to confirm the existence of Pop III massive stars of mass $\sim 30~\msun$ by the detections of gravitational waves if the merger rate of the Pop III massive BH-BHs dominates that of Pop I BH-BHs. 
\end{abstract}


\section{Introduction}
Gravitational-wave astronomy with KAGRA\footnote{http://gwcenter.icrr.u-tokyo.ac.jp/en/}, 
Adv. LIGO\footnote{http://www.ligo.caltech.edu/},  Adv. Virgo\footnote{http://www.ego-gw.it/index.aspx/}, 
and GEO\footnote{http://www.geo600.org/}  will reveal the formation and evolution of
binaries through the observed merger rates of compact binaries, such as binary neutron stars (NS-NSs), neutron
star -- black hole binaries (NS-BHs), and binary black holes (BH-BHs).  For this gravitational wave astronomy, 
estimates of the merger rate of compact binaries play key roles to develop observational
strategy and to translate the observed merger rates into the binary formation and evolution processes. 

There are two 
methods to estimate the merger rate of compact binaries.   One is to use observational facts such as the observed NS-NSs whose 
coalescence time due to the emission of gravitational waves is less than the age of the universe. Taking into account
the observation time,  the sensitivity of the radio telescope, the luminosity function of pulsars and the 
beaming factor so on, the probability distribution function of the merger rate can be found. For example, 
\citet{Kalogera2004b} found that the event rate of the coalescing NS-NSs is in the range from $10^{-5}\ {\rm events}\ 
{\rm yr}^{-1}\ {\rm galaxy}^{-1}$ to $4\times 10^{-4}\ {\rm events}\ {\rm yr}^{-1}\ {\rm galaxy}^{-1}$ at the 
$99\ \%$ confidence level (see their Fig.~2)\footnote{Note here that there are errors in \citet{Kalogera2004a} 
so that the rates in \citet{Kalogera2004b}  are the correct ones.}.

The merger rate of NS-NSs can be restricted by  the rate of the observed Type Ib and Ic supernovae,
supposing that the formation of NS-NSs really starts from the massive binary zero age main sequence~(ZAMS) stars. 
This is because the formation of the second neutron star should occur in association with 
Type Ib and Ic supernovae in which the H-rich envelope and the He-layer are lost, respectively,
otherwise the binary disrupts due to the sudden large mass loss at the supernova explosion\footnote{If more than
half of the total mass is suddenly lost at the supernova explosion, the binary disrupts.}. Under the assumption
of the equality of the formation rate to the merger rate, the merger rate of the NS-NSs is limited by the Type Ib and Ic
supernova rate of $\sim 10^{-3}\ {\rm events}\ {\rm yr}^{-1}\ {\rm galaxy}^{-1}$ \citep{Cappellaro1997,
Cappellaro1999,Li2011}.  Therefore  the maximum rate of $4\times 10^{-4}\ {\rm events}\ {\rm yr}^{-1}\
{\rm galaxy}^{-1}$ by \citet{Kalogera2004b} implies that $\sim 40\ \%$ of the Type Ib and Ic supernovae is associated with the formation of NS-NSs with the coalescence time less than the age of the universe. This percentage seems to be too large.  We also note here that under the assumption that central engine of short gamma ray bursts are coalescing binary neutron stars, one  can use the observations of short gamma ray bursts to estimate the coalescing rate (\citet{coward12} ~and references cited there).

The dynamical interaction in a globular cluster is another route to the formation of NS-NSs
since there exists PSR2127+11C, which is contained in a NS-NS system, in the globular cluster M15 \citep{Prince1991}. 
The age of the globular cluster is $\sim 10^{10}\ {\rm yr}$ so that all the
massive stars ended their life and the young pulsars do not exist.  The coalescence time of PSR2127+11C is
$\sim 2 \times 10^{8}\ {\rm yr}$ which is much smaller than the age of the globular cluster so that it was 
formed most likely by three body interactions such as the collision of neutron star -- white dwarf binary or
neutron star -- dwarf star binary
with a single neutron star. Since there exists only one NS-NS observed in the globular cluster, it is difficult to estimate
the merger rate. Theoretical simulation is the only method at present~\citep{Grindlay2006, Ivanova2008}.

The another method to estimate the merger rate of compact binaries is to use theoretical computation based on
the hypothetical assumptions of binary formation and evolution. For NS-BHs and BH-BHs, in particular, 
there exists no observations so that we can only use theoretical estimates. The merger
rates of compact binaries of Population I~(Pop I) stars were estimated  by many authors~\citep{Belczynski2002, 
Belczynski2007, Belczynski2012}.  \citet{Dominik2012} computed the merger rates for the progenitor stars  of  
metallicity $Z=0.1\ Z_\odot$ and found that the number of the coalescing BH-BHs increases compared with that for
$Z=Z_\odot$. \citet{Dominik2013} adopted a certain model of the star formation rate and the chemical evolution
of the metallicity $Z$ to compute the cumulative redshift distribution of the coalescing compact binaries.

In this paper, we focus on the compact binary merger originated from 
Population III stars (Pop III stars) as gravitational wave sources. 
Pop III stars are the first stars after the Big Bang which are formed from
metal-free gas~\citep{Omukai1998,  Bromm2002,  Abel2002, Yoshida2008, Greif2012}. 
The simulations of a rotating primordial gas cloud suggest that the formation of Pop III 
star binaries and multiple star systems are frequent \citep{Machida2008a, Stacy2010}. 
The main differences of our work from \cite{Dominik2012,Dominik2013}
are the following two:
(1) we focus on metal-free Pop III stars and 
(2) consider the star formation history including 
the transition to metal-enriched stars (see \S4). 
The observed merger rate will be the sum of our work and \citet{Dominik2013}.

There are at least three differences between Pop III and Pop I compact binaries. 
First of all, Pop III stars are more massive than Pop I stars 
\citep{McKee2008, Hosokawa2011, Hosokawa2012, Stacy2012} 
with mass $10-100\ {\rm M}_\odot$ so that  Pop III star binaries 
probably evolve into BH-BHs.  Secondly, since the typical formation time of Pop III stars is at $z\sim 10$, 
even if the coalescence time is comparable to the age of the universe, 
they merge at present and contribute to the sources of gravitational waves 
for KAGRA, Adv. LIGO, Adv. Virgo, and GEO network. 
Therefore, if Pop III NS-NSs were formed, they might merge at present 
so that their rate is free from the constraint of the observed NS-NSs as well as 
Type Ib and Ic supernova rate discussed in the previous paragraphs. 
Thirdly, Pop III black holes are expected to be more massive than Pop I
black holes due to less mass loss so that the resulting gravitational
waves are easier to detect, since the detectable distance is
proportional to 5/6 power of the chirp mass ($M_{\rm chirp}$) of a binary defined
by $ M_{\rm chirp} = (M_1M_2)^{3/5}/(M_1 + M_2)^{1/5}$ (Peters 1964 and Peters \& Mathews 1963,
see also Sathyaprakash \& Schutz(2009)), where $M_1$ and $M_2$ are the mass of
each compact object.

The idea of Pop III compact binaries as gravitational-wave sources has 
been considered by \cite{Belczynski2004}, \cite{Kulczycki2006} and \cite{Kowalska2012}. 
However they considered very massive Pop III stars with mass over hundred solar masses. 
Recent study shows that the typical mass of Pop III stars is set to 10--100$\rm{M_{\odot}}$ by the stellar
radiation feedback on the accretion flow~\citep{Hosokawa2011, Hosokawa2012}. Therefore, in this paper, we
calculate $10^6$ Pop III binary evolutions with the mass range of 10--100$\rm{M_{\odot}}$ to estimate merger rates
and mass distribution of Pop III compact binaries. In order to calculate Pop III binary evolutions, we upgrade
Hurley's binary population synthesis code \citep{Hurley2002} for the Pop I star to Pop III star case.  
    
This paper is organized as follows. We describe Pop III single star evolution in \S \ref{sec:single}, the
method to calculate Pop III binary star evolutions in \S \ref{sec:binary}, numerical calculation methods in
\S \ref{sec:initial condition}. In \S 3, we present the results of simulations and argue properties of Pop III
compact binaries. We compare Pop III compact binary mergers with Pop I compact binary mergers in 
\S \ref{sec:comparison models}. In \S 4, we describe the Pop III compact binary merger rates. 
\S 5 is devoted to the discussions. In Appendix, we show the details of our numerical methods, the
comparison of our results with Hurley's ones and the convergence check of our simulations. We adopt the
cosmological parameters of $(\Omega_{\Lambda}, \Omega_{\rm m}) = (0.6825, 0.3175)$ and the Hubble parameter of 
$H_0 = 67.11\ {\rm km}\ {\rm s}^{-1}\ {\rm Mpc}^{-1}$~\citep{Planck2013}. Those who are not interested in the details of
the methods in  numerical simulations  can skip \S 2.


\section{method of binary population synthesis simulations}

\subsection{Single star evolution}\label{sec:single}

\subsubsection{Population III stars}

Pop III stars are  formed  in the early universe from primordial gas, 
i.e., without heavy elements.
The star formation process of Pop III stars has been investigated 
by many authors \citep{Tegmark1997, Omukai1998, Bromm2002, Yoshida2008, Greif2012}. 
According to their studies, the differences of the chemical compositions
lead to the following three features of Pop III stars compared with Pop I stars:(1) more massive $> 10\ 
{\rm M}_\odot$ (2) smaller stellar radius for the same mass  (3) less mass loss by stellar wind. Since these
features play key roles in a single stellar evolution and binary interactions~(see also Sec. 
\ref{sec:binary}), we briefly summarize these features of Pop III stars in what follows.

In primordial gas, the H$_2$-line emission is 
the main cooling process, which is less efficient than 
the dust cooling as in Pop I star formation. 
Since the gas temperature is kept hotter, typically massive cloud collapses and 
forms protostars at the center.
Recent numerical simulations 
\citep[e.g.,][]{Hosokawa2011,Hosokawa2012, Stacy2012} suggest
that the Pop III protostar can grow to $\sim$ several $10~\msun$
until the radiation feedback halts the gas accretion onto the central protostar. 
Therefore, Pop III stars at the ZAMS stage are typically more massive than Pop I stars of mass $\sim 1~\msun$.

\begin{figure}
\centering
\includegraphics[width=8cm]{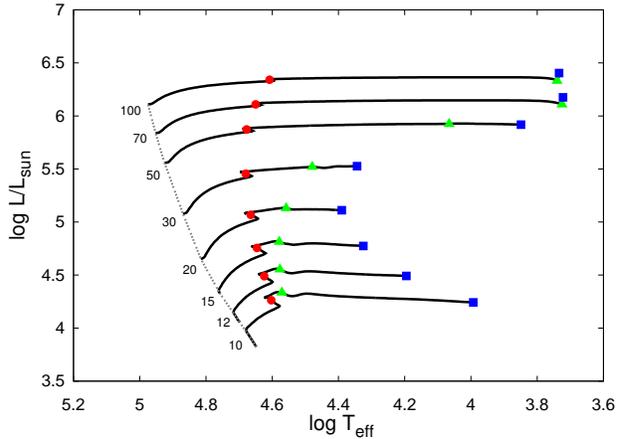}
\caption{The Hertzsprung-Russell (HR) diagram for the Pop III stars of  mass 
$10~\msun \leq M \leq 100~\msun$ using the data taken from \citet{Marigo2001}. 
The number attached to each solid curve is the mass of each star in unit of $\msun$. 
The dashed line shows  the ZAMS (Zero Age Main Sequence) stars.
Red circles, green triangles and blue squares correspond to  the beginning of He-burning, 
the end of the He-burning and the beginning of the C-burning, respectively.
}
\label{popIIIHR}
\end{figure}

When the protostar reaches the Zero Age Main Sequence (ZAMS)  stage,  
the star contracts until the central temperature rises above $10^8$~K
to generate C via triple-alpha reaction so that CNO-cycle starts~\citep{Marigo2001}. 
Thus, stable structure of Pop III ZAMS star has the smaller  radius than that of Pop I stars.
As a result, the binary interaction for Pop III stars becomes more weak than those for Pop I stars. 
Figure \ref{popIIIHR} shows the Hertzsprung-Russell (HR) diagram for Pop III
stars over the mass range of $10~\leq M\leq 100~\msun$ from the ZAMS stage 
to the beginning of the C-burning stage. 
In Pop III star case, the central temperature is so high that the He-burning soon begins 
after the end of the H-burning. 
Therefore, the resultant stellar evolution at the post main sequence stage is  
different from the usual Pop I star case \citep{Kippenhahn1990}.

The mass loss due to the stellar wind and pulsation has 
impacts on the stellar evolution and the mass of the remnant compact objects. 
For Pop III star case, such mass-loss processes do not operate 
because of no heavy elements at the stellar surface \citep[e.g.][]{Baraffe2001, Inayoshi2013}. 
Therefore, we neglect the effect of the mass loss on the stellar evolution.

\subsubsection{Fitting formulae of Pop III steller evolution}\label{sec:fitting formula}

In order to include the single PopIII star evolution  to the binary population synthesis simulation
code given by \cite{Hurley2002}, we need to construct the fitting formula to the  stellar radius and the core
mass as a function of time since it consumes too long cpu time to numerically evolve Pop III stars up to the
C-burning phase in each population synthesis. Using the results of stellar evolution for Pop III
stars calculated by \cite{Marigo2001},  we here present fitting formulae of the stellar radius and core mass as functions of the
stellar mass $M$ and the time ($t$) from the birth of a star.

We basically fit the stellar radii of Pop III stars in the same way as \cite{Hurley2000} did for Pop I stars.
As shown in Fig.~\ref{popIIIHR}, we divide the life of Pop III stars into the four characteristic phases: 
(1) H-burning phase~(from the ZAMS to red circle), (2) the He-burning phase~(from red circle to green triangle), (3)
the He-shell burning phase~(blue square), and (4) after the C-ignition.
In the followings, we show the fitting formulae in each phase.
We use the subscripts H, He, HeS and C to each physical variables such as the radius and the mass to show the H-burning
phase, the He-burning phase,  the He-shell burning phase and the C-burning phase, respectively. The superscripts
b and e denote the beginning and the end of each phase, respectively. 
Basically, the fitting formulae are expressed as the forms of polynomials of  the mass and age.
In other cases, we will mention how to obtain each formulae.

\subsubsection*{(1) H-burning phase}

\begin{figure*}
\centering
\begin{minipage}{.35\linewidth}
\includegraphics[width=1.1 \linewidth]{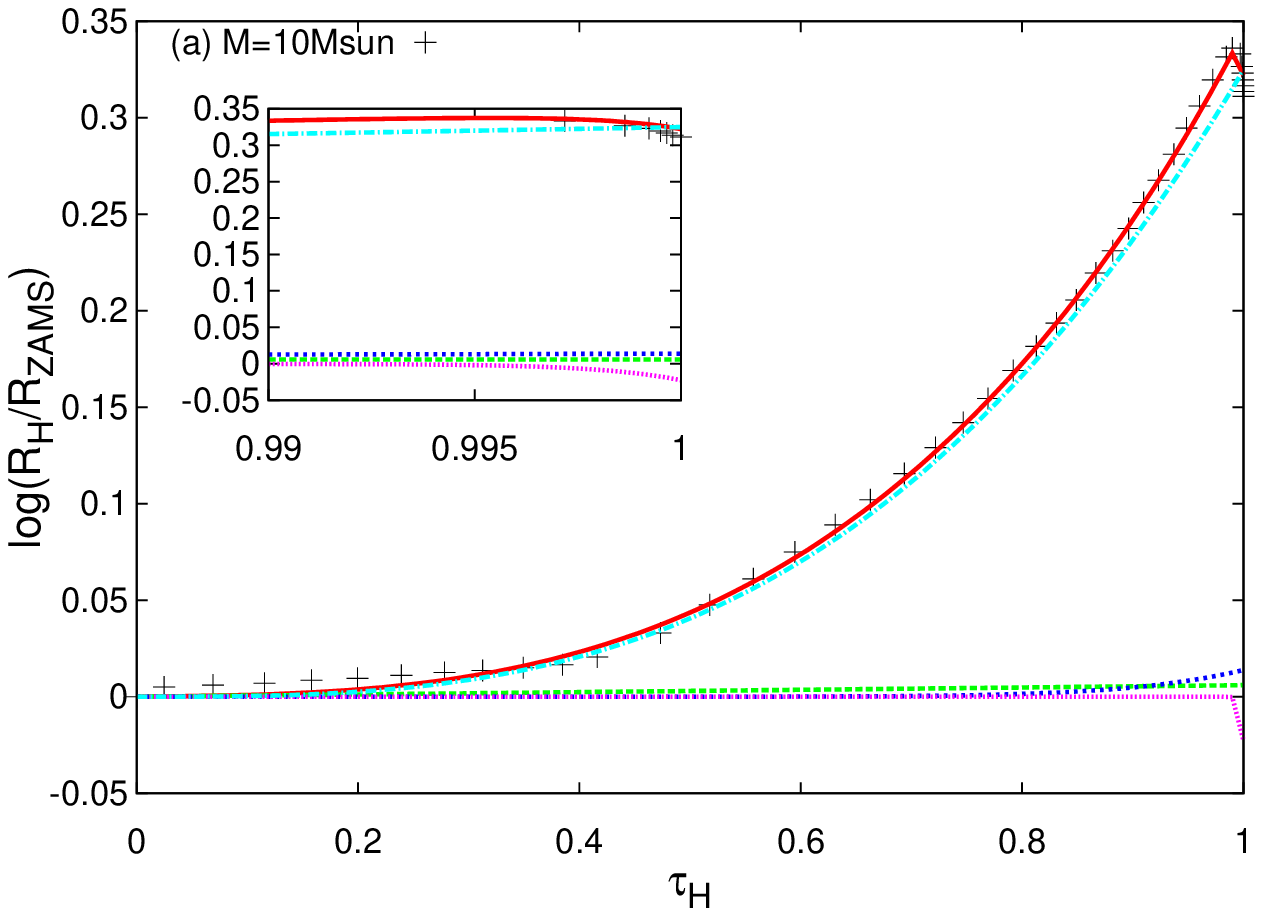}
\smallskip
\end{minipage}
\hspace{2.0pc}
\begin{minipage}{.35\linewidth}
\includegraphics[width=1.1 \linewidth]{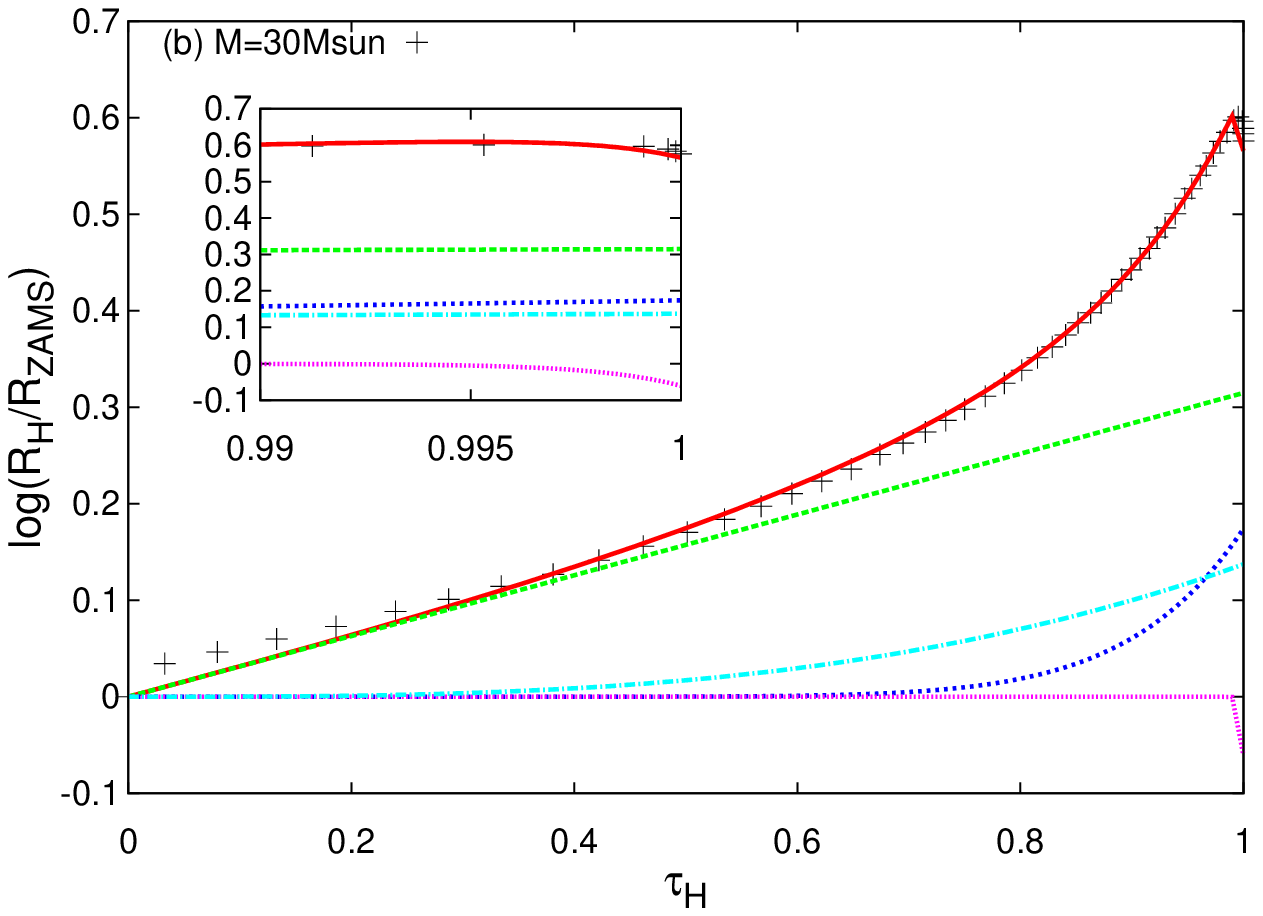}
\smallskip
\end{minipage}
\begin{minipage}{.35\linewidth}
\includegraphics[width=1.1 \linewidth]{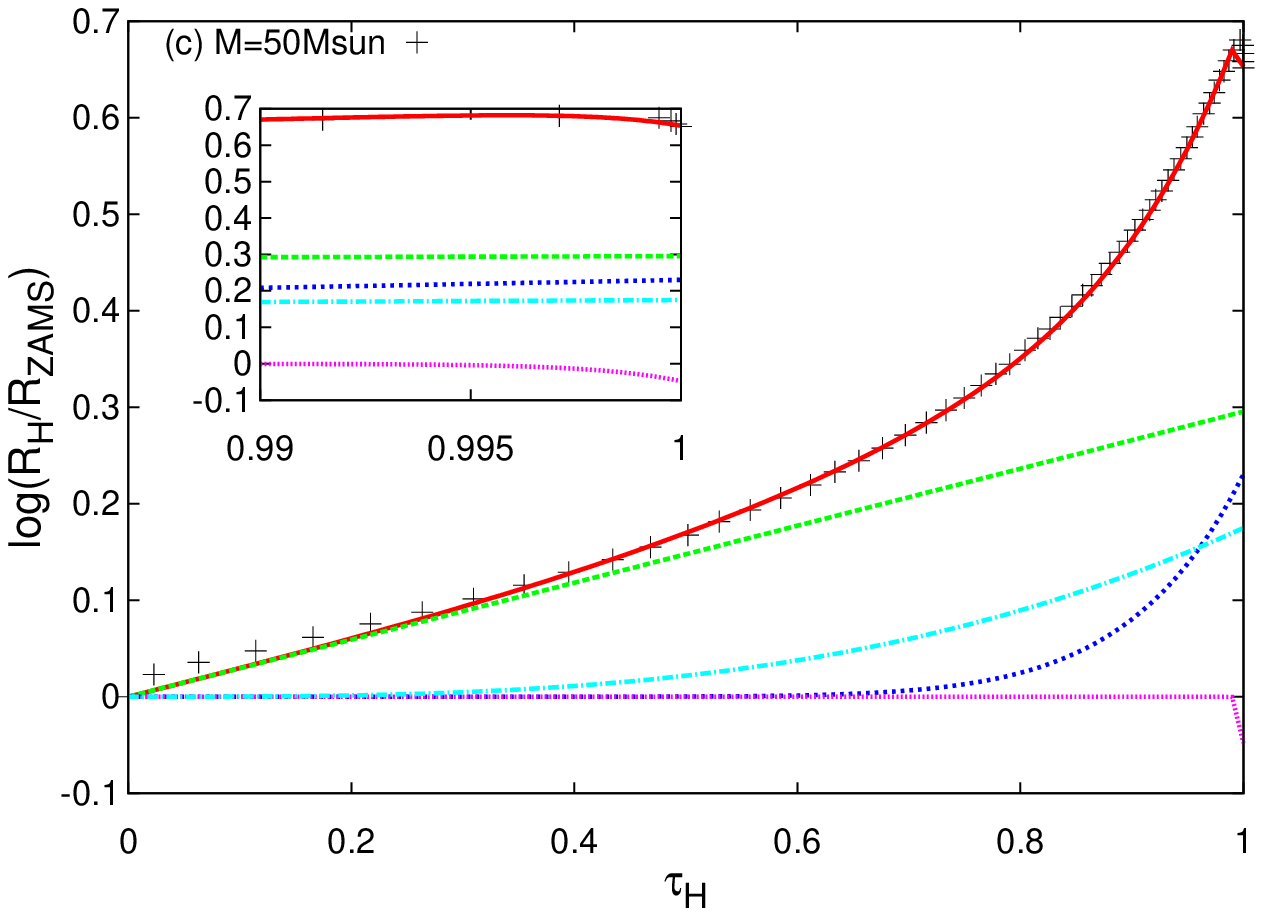}
\end{minipage}
\hspace{2.0pc}
\begin{minipage}{.35\linewidth}
\includegraphics[width=1.1 \linewidth]{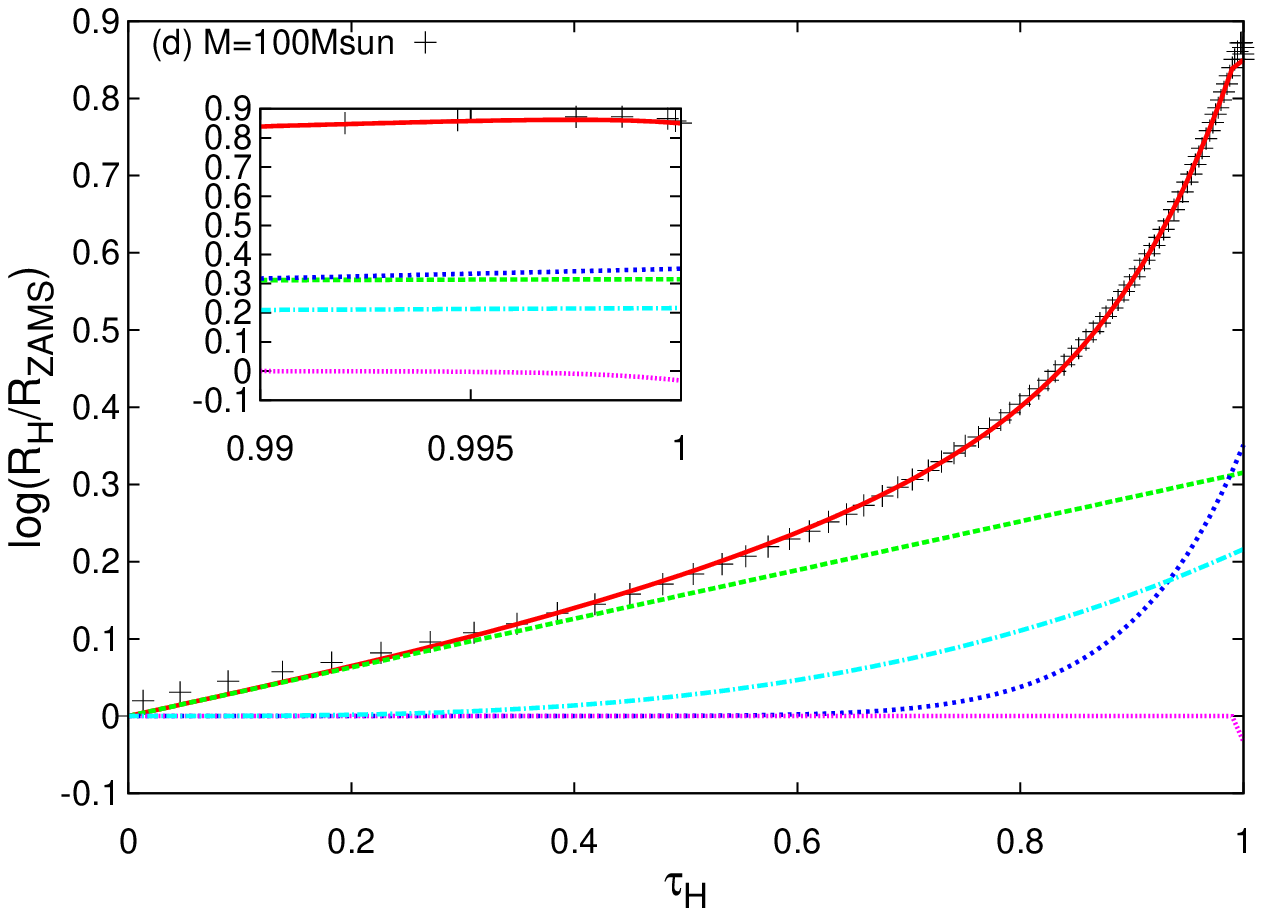}
\end{minipage}
\caption[]{The comparison of the  fitting formula with the numerical data as a function of time.  The vertical and
horizontal axises are ${\rm{log}}(R_{\rm{H}}/R_{\rm{ZAMS}})$ and $\tau_{\rm{H}}\equiv {t}/{t_{\rm{H}}}$ with $t_{\rm{H}}$ being the H-burning time,
respectively. The red line is the fitting formula of stellar radius~(Eq. \ref{eq:fitMS}) and the crosses are
computed data given by \cite{Marigo2001}.  The  green, blue, pink and light blue lines, represent the
contributions from the second, third, fourth and fifth term of the fitting formula~(Eq.~\ref{eq:fitMS}), respectively. Each panel refers to the stellar mass (a) $10~\rm{M}_{\odot}$, (b) $30~\rm{M}_{\odot}$, (c) 
$50~\rm{M}_{\odot}$ and (d) $100~\rm{M}_{\odot}$, respectively.  For the low mass case, the stellar radius can be
expressed mainly by the fifth term of $d_{\rm{H}}\tau_{\rm{H}}^3$~(light blue line), whereas for high mass case 
they are mainly expressed by the terms of $a_{\rm{H}}\tau_{\rm{H}}$~(green line) and $b_{\rm{H}}\tau_{\rm{H}}
^{10}$~($\tau_{\rm H}\ga 0.5$, blue line). Around the end of the main sequence lifetime~($\tau_{\rm H}\ga 0.99$),
the stellar radii dramatically shrink, because H has been exhausted in the central core. This
prominent feature is called as the {\it main sequence hook}~(see also Fig.~\ref{popIIIHR}) and is well
described by the term of $c_{\rm{H}} \tau_{\rm{H}}^{500}$~(pink line) in Eq. \eqref{eq:fitMS}. The inset in each
figure is the magnification of the contribution from each term for $0.99 \le \tau_{\rm H} \le 1$ to show the effect of this term.}
\label{msfit}
\end{figure*}

In order to characterize the stellar radius of the H-burning phase, we first need to obtain the stellar radius
of the ZAMS ($R_{\rm{ZAMS}}$), the stellar radius at the end of the main sequence, and the H-burning time
$t_{\rm H}$, which can be expressed as 
\begin{align}
& (R_{\rm{ZAMS}}/{\rm R}_\odot)=1.22095+2.70041\times 10^{-2}(M/10~\msun)\notag\\
&+0.135427(M/10~\msun)^2-1.95541\times10^{-2}(M/10~\msun)^3\notag\\
&+8.7585\times10^{-4}(M/10~\msun)^4, \;\; \label{eq:rzams10}
\end{align}

\begin{align}
& (R_{\rm{H}}^{\rm{e}}/{\rm R}_\odot)=0.581309+2.27745(M/10~\msun)\notag\\
&+6.63321\times10^{-3}(M/10~\msun)^3,\label{eq:reh}
\end{align}
and
\begin{align}
&(t_{\rm{H}}/{\rm Myr})=1.78652+10.4323(M/10~\msun)^{-1}\notag\\
                     &+3.70946(M/10~\msun)^{-2}+2.04264(M/10~\msun)^{-3},  \label{eq:th}
\end{align}
respectively.

For simplicity, we introduce the time $\tau_{\rm H}$ by $\tau_{\rm H}=t/t_{\rm{H}}$ and express the stellar
radius  $R_{\rm{H}}$ during H-burning phase  as a function of time  as
\begin{eqnarray}
&&\log (R_{\rm{H}}/{\rm R}_\odot)=\log (R_{\rm{ZAMS}}/{\rm R}_\odot)
+a_{\rm{H}}\tau_{\rm{H}}+b_{\rm{H}}\tau_{\rm{H}}^{10}\nonumber\\
&&+c_{\rm{H}}\tau_{\rm{H}}^{500}+d_{\rm{H}}\tau_{\rm{H}}^3,\label{eq:fitMS}
\end{eqnarray}
where
\begin{equation}
a_{\rm{H}} = \begin{cases}
                   -0.430873+0.520408 (M/10~\msun)\\
                 -7.99762\times10^{-2}(M/10~\msun)^2\\
                 -3.55095\times10^{-3}(M/10~\msun)^3\\
 \;\;\;\;\;\;\;\;\;\;\;\; \text{($10 ~{\rm{M}_{\odot}} \le M < 30 ~{\rm{M}_{\odot}}$)},\\
                   0.476498-9.07537\times10^{-2}(M/10~\msun)\\
                 +1.43538\times10^{-2}(M/10~\msun)^2\\
                 -6.89108\times10^{-4}(M/10~\msun)^3 \\ \;\;\;\;\;\;\;\;\;\;\;\;\text{($30 ~{\rm{M}_{\odot}} \le M \le 100 ~{\rm{M}_{\odot}}$)},\\
                   \end{cases}\\
\end{equation}
\begin{equation}
b_{\rm{H}} = \begin{cases}
                   0.669345-1.5518(M/10~\msun) +1.15116(M/10~\msun)^2\\
                  -0.254811(M/10~\msun)^3\\
                 \;\;\;\;\;\;\;\;\;\;\;\; \text{($10 ~{\rm{M}_{\odot}} \le M < 20 ~{\rm{M}_{\odot}}$)},\\
                   3.02801\times10^{-2}+6.48197\times10^{-2}(M/10~\msun)\\
                   -6.64582\times10^{-3}(M/10~\msun)^2\\
                   +3.37205\times10^{-4}(M/10~\msun)^3\\
\;\;\;\;\;\;\;\;\;\;\;\; \text{($20 ~{\rm{M}_{\odot}} \le M \le 100 ~{\rm{M}_{\odot}}$)},\\
                  \end{cases}
\end{equation}
\begin{equation}
c_{\rm{H}} = \begin{cases}
                  5.63328\times10^{-2}-9.88927\times10^{-2}(M/10~\msun)\\
                  +2.00071\times10^{-2}(M/10~\msun)^2  \\
\;\;\;\;\;\;\;\;\;\;\;\;  \text{($10 ~{\rm{M}_{\odot}} \le M < 30 ~{\rm{M}_{\odot}}$)},\\
                 -0.128025+3.63928\times10^{-2}(M/10~\msun)\\
                -5.43719\times10^{-3}(M/10~\msun)^2\\
+2.75137\times10^{-4}(M/10~\msun)^3\\
\;\;\;\;\;\;\;\;\;\;\;\;  \text{($30 ~{\rm{M}_{\odot}} \le M \le 100 ~{\rm{M}}_{\odot}$)},
                  \end{cases}
\end{equation}
and
\begin{equation}
d_{\rm{H}} =\log (R_{\rm{H}}^{\rm{e}}/R_{\rm{ZAMS}})-a_{\rm{H}}-b_{\rm{H}}-c_{\rm{H}}.
\end{equation}

In Fig.~\ref{msfit}, we compare the fitting formula with the numerical data as a function of time.  The vertical 
and horizontal axes are  ${\rm{log}}(R_{\rm{H}}/R_{\rm{ZAMS}})$ and  $\tau_{\rm{H}}\equiv {t}/{t_{\rm{H}}}$,
respectively. The red line and the crosses denote the fitting formula of stellar radius~Eq.~(\ref{eq:fitMS}) and 
the computed data given by \cite{Marigo2001}, respectively.  The green, blue, pink and light blue lines represent
the contributions from the second, third, fourth and fifth term of the fitting formula ~(Eq.
(\ref{eq:fitMS})), respectively. Each panel refers to the stellar mass (a) $10~\rm{M}_{\odot}$, (b) 
$30~\rm{M}_{\odot}$, (c) $50~\rm{M}_{\odot}$ and (d) $100~\rm{M}_{\odot}$, respectively.  For low mass cases,
stellar radii can be expressed mainly by the fifth term of $d_{\rm{H}}\tau_{\rm{H}}^3$~(light blue line),
whereas for high mass cases they are mainly approximated by the terms of $a_{\rm{H}}\tau_{\rm{H}}$~(green line)
and $b_{\rm{H}}\tau_{\rm{H}}^{10}$~($\tau_{\rm H}\ga 0.5$, blue line).
Just before the end of the main sequence~($\tau_{\rm H}\ga 0.99$), the stellar radius dramatically shrinks,
because H has been exhausted in the central core. This prominent feature is called as the {\it main
sequence hook}~(see also Fig.~\ref{popIIIHR}) and is well described by the term of 
$c_{\rm{H}}\tau_{\rm{H}}^{500}$~(pink line) in Eq.~\eqref{eq:fitMS}. The inset in each figure is the
magnification of the contribution from each term for $0.99 < \tau_{\rm H} <  1$  to show the effect of this term.

\begin{figure}
\centering
\includegraphics[width=7.5cm]{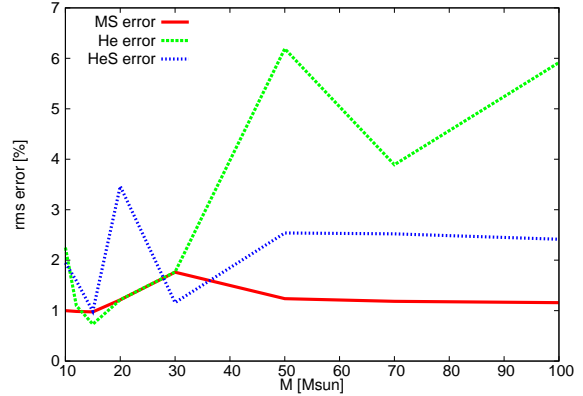}
\caption[]{The time averaged root mean square~(rms) errors of our fitting formulae relative to the numerical
results given in \cite{Marigo2001}, as a function of stellar mass. The red, green and blue lines correspond to
those fitting formulae during the H-burning phase (Eq. \ref{eq:fitMS}), He-burning phase (Eq. \ref{eq:fitHeB})
and He-shell burning phase (Eq. \ref{eq:fitAHeB}), respectively.  We can see that our fitting formulae have
relative accuracy within 2\ \%, 6\ \% and 3.5\ \% of numerical calculations by  \cite{Marigo2001} for the
H-burning, He-burning and He-shell burning phase, respectively.}
\label{error}
\end{figure}

Fig.~\ref{error} shows the time averaged root mean square (rms)  errors of our fitting formula as a function of the stellar mass. The red line is rms
during the H-burning phase~(Eq.~\ref{eq:fitMS}), which shows that
our fitting formula has the relative accuracy within 2~\% of the models of \cite{Marigo2001}.

\subsubsection*{(2) He-burning phase}
At the end of the main sequence phase, the He-burning smoothly begins in the central core for massive Pop III
stars ($\ga 10~\msun$) without the Hertzsprung gap, because the central temperature during the H-burning
phase is already high enough to ignite He~($\ga 10^8$ K). Therefore, in this paper, the beginning of the
He-burning phase is assumed to be the same time as the end of the H-burning phase~,i.e., $R_{\rm He}^{\rm b}=
R_{\rm H}^{\rm e}$. 

The stellar radii at the end of the He-burning phase~$R_{\rm{He}}^{\rm e}$ and the lifetime of the He-burning
phase~($ t_{\rm{He}}$) is approximated by
\begin{equation}
\log (R_{\rm{He}}^{\rm e}/{\rm R}_\odot)=\begin{cases}
                            -7.23005\times10^{-2}+0.814329(M/10~\msun)\\
                            -0.252995(M/10~\msun)^2\\
                            +5.88465\times10^{-2}(M/10~\msun)^3\\
                            -4.28501\times10^{-3}(M/10~\msun)^4\\
\;\;\;\;\;\;\;\;\;\;\;\; \text{($10 ~{\rm{M}}_{\odot} \le M < 50 ~{\rm{M}}_{\odot}$)},\\
                          -2.40224+1.32865\times(M/10~\msun)\\
                        -7.65293\times10^(-2)(M/10~\msun)^2\\
\;\;\;\;\;\;\;\;\;\;\;\; \text{($50 ~{\rm{M}}_{\odot} \le M \le 100 ~{\rm{M}}_{\odot}$)},\\
                            \end{cases}                  
\end{equation}  
and 
\begin{eqnarray}
\log (t_{\rm{He}}/{\rm Myr})=\begin{cases}
                      6.95516-1.17529(M/10~\msun)\\
                       +0.264783(M/10~\msun)^2\\
 \;\;\;\;\;\;\;\;\;\;\;\; \text{($10 ~{\rm{M}}_{\odot} \le M < 20 ~{\rm{M}}_{\odot}$)},\\
                      6.13-0.331059(M/10~\msun)\\
                     +5.16053\times10^{-2}(M/10~\msun)^2\\
                     -2.8\times10^{-3}(M/10~\msun)^3\\
 \;\;\;\;\;\;\;\;\;\;\;\; \text{($20 ~{\rm{M}}_{\odot} \le M \le 100 ~{\rm{M}}_{\odot}$)},\\
                      \end{cases}
\end{eqnarray}
where the normalized time $\tau_{\rm{He}}$ in He-burning phase is defined by
\begin{equation}
\tau_{\rm{He}}\equiv \frac{t-t_{\rm{H}}}{t_{\rm{He}}}.
\end{equation}
Using $\tau_{\rm{He}}$, the fitting formula of the stellar radius during the He-burning phase is given by
\begin{eqnarray}
&&\log (R_{\rm{He}}/{\rm R}_\odot)=\log (R_{\rm{H}}^{\rm{e}}/{\rm R}_\odot)+a_{\rm{He}}\tau_{\rm{He}}
+b_{\rm{He}}\tau_{\rm{He}}^2+c_{\rm{He}}\tau_{\rm{He}}^3
+d_{\rm{He}}\tau_{\rm{He}}^4\nonumber\\
&&+\left(\log (R_{\rm{He}}^{e}/R_{\rm{H}}^{\rm{e}})
-a_{\rm{He}}-b_{\rm{He}}-c_{\rm{He}}-d_{\rm{He}}\right)\tau_{\rm{He}}^5,
\label{eq:fitHeB}
\end{eqnarray}
where
\begin{equation}
a_{\rm{He}}= \begin{cases}
                     -0.891114+0.992291(M/10~\msun)\\
                    -0.500532(M/10~\msun)^2+7.46275\times10^{-2}(M/10~\msun)^3 \\
\;\;\;\;\;\;\;\;\;\;\;\; \text{($10 ~{\rm{M}}_{\odot} \le M < 20 ~{\rm{M}}_{\odot}$)},\\
           3.08883-3.85847(M/10~\msun)+1.40618(M/10~\msun)^2\\
           -0.178175(M/10~\msun)^3+7.32187\times10^{-3}(M/10~\msun)^4\\
                    \;\;\;\;\;\;\;\;\;\;\;\;\text{($20 ~{\rm{M}}_{\odot} \le M \le 100 ~{\rm{M}}_{\odot}$),}
                    \end{cases}
\end{equation}
\begin{equation}
b_{\rm{He}}= \begin{cases}
                    -0.433454+0.768418 (M/10~\msun)\\
                    \;\;\;\;\;\;\;\;\;\;\;\;\text{($10 ~{\rm{M}}_{\odot} \le M < 15 ~{\rm{M}}_{\odot}$)},\\
                   -2.10737+1.88553(M/10~\msun)\\
                    \;\;\;\;\;\;\;\;\;\;\;\;\text{($15 ~{\rm{M}}_{\odot}\le M < 20 ~{\rm{M}}_{\odot}$)},\\
                    -28.3697+33.7648(M/10~\msun)-12.2469(M/10~\msun)^2\\
                    +1.56514(M/10~\msun)^3-6.4361\times10^{-2}(M/10~\msun)^4\\
                    \;\;\;\;\;\;\;\;\;\;\;\;\text{($20 ~{\rm{M}}_{\odot}\le M \le 100 ~{\rm{M}}_{\odot}$),}\\
                    \end{cases}
\end{equation}
\begin{equation}
c_{\rm{He}}=\begin{cases}
                    45.8092-114.873(M/10~\msun)+110.156(M/10~\msun)^2\\
                   -46.1519(M/10~\msun)^3 +6.88478(M/10~\msun)^4 \\
\;\;\;\;\;\;\;\;\;\;\;\;\text{($10 ~{\rm{M}}_{\odot} \le M < 20 ~{\rm{M}}_{\odot}$)},\\
                    85.996-100.37(M/10~\msun)+36.7017(M/10~\msun)^2\\
                   -4.68789(M/10~\msun)^3 +0.191704(M/10~\msun)^4\\
                     \;\;\;\;\;\;\;\;\;\;\;\; \text{($20 ~{\rm{M}}_{\odot} \le M \le 100 ~{\rm{M}}_{\odot}$),}\\
                   \end{cases}
\end{equation}
and
\begin{equation}
d_{\rm{He}}=\begin{cases}
                    -51.6917+125.87(M/10~\msun)-121.373(M/10~\msun)^2\\
                   +51.3681(M/10~\msun)^3-7.74452(M/10~\msun)^4\\
                    \;\;\;\;\;\;\;\;\;\;\;\;\text{($10 ~{\rm{M}}_{\odot} \le M < 20 ~{\rm{M}}_{\odot}$)},\\
                   -103.871+120.228(M/10~\msun)-44.0198(M/10~\msun)^2\\
                   +5.58876(M/10~\msun)^3-0.226361(M/10~\msun)^4\\
                      \;\;\;\;\;\;\;\;\;\;\;\; \text{($20 ~{\rm{M}}_{\odot} \le M \le 100 ~{\rm{M}}_{\odot}$)}.\\
                   \end{cases}
\end{equation}
The green line in Fig.~\ref{error} shows the rms error of our fitting formula in the He-burning phase. 
We find that our fitting formula for each mass has accuracies within 6\% of the stellar models of
\cite{Marigo2001} during this phase.

In the He-burning phase, a star evolves into a giant star, which has the core-envelope structure. 
The structure can be characterized by the He-core mass at the beginning and the end of the He-burning.
These core masses are approximated by
\begin{equation}
(M_{\rm{He}}^{\rm{b}}/\msun)= \begin{cases}
                     -0.47466+2.49981(M/10~\msun)^{1.13274} \\
\;\;\;\;\;\;\;\;\;\;\;\; \text{($10 ~{\rm{M}}_{\odot} \le M < 15 ~{\rm{M}}_{\odot}$)},\\
                    -2.3546+3.61261(M/10~\msun)^{1.12392}\\ 
\;\;\;\;\;\;\;\;\;\;\;\; \text{($15 ~{\rm{M}}_{\odot} \le M \le 100 ~{\rm{M}}_{\odot}$)},\\
                    \end{cases}\label{eq:mbhe}
\end{equation}
and
\begin{align}
&(M_{\rm{He}}^{\rm{e}}/\msun)=1.31569(M/10~\msun)+0.993475(M/10~\msun)^2\notag\\
&-0.112405(M/10~\msun)^3+4.60669\times10^{-3}(M/10~\msun)^4.\label{eq:mehe}
\end{align}
Then, the He-core mass as a function of the total stellar mass and time can be given by
\begin{align}
(M_{\rm{He}}/\msun)=&(M_{\rm{He}}^{\rm{b}}/\msun)+A_{\rm{He}}\tau_{\rm{He}}+B_{\rm{He}}\tau_{\rm{He}}^2\notag\\
&+((M_{\rm{He}}^{\rm{e}}/\msun)-(M_{\rm{He}}^{\rm{b}}/\msun)-A_{\rm{He}}-B_{\rm{He}})\tau_{\rm{He}}^3,
\end{align}
where
\begin{equation}
A_{\rm{He}}= \begin{cases}
                    -301.285+1210.26(M/10~\msun)-1808.76(M/10~\msun)^2\\
                   +1191.99(M/10~\msun)^3-292.114(M/10~\msun)^4 \\\;\;\;\;\;\;\;\;\;\;\;\;\text{($10 ~{\rm{M}}_{\odot} \le M < 12 ~{\rm{M}}_{\odot}$)},\\
                    -1.27007+2.97787(M/10~\msun)-1.66077(M/10~\msun)^2\\
                    +0.307506(M/10~\msun)^3\\\;\;\;\;\;\;\;\;\;\;\;\;\text{($12 ~{\rm{M}}_{\odot} \le M < 30 ~{\rm{M}}_{\odot}$)},\\
                    5.55735\times10^{-2}-4.91742\times10^{-2}(M/10~\msun)\\
                    +9.62294\times10^{-2}(M/10~\msun)^2\\
                    -9.4471\times10^{-3}(M/10~\msun)^3\\\;\;\;\;\;\;\;\;\;\;\;\;\text{($30 ~{\rm{M}}_{\odot} \le M \le 100 ~{\rm{M}}_{\odot}$)},\\
                    \end{cases}
\end{equation}
and
\begin{equation}
B_{\rm{He}}=\begin{cases}
                    20.771-47.8361(M/10~\msun)+38.9548(M/10~\msun)^2\\
                   -13.6227(M/10~\msun)^3+1.70524(M/10~\msun)^4 \\
\;\;\;\;\;\;\;\;\;\;\;\;\text{($10 ~{\rm{M}}_{\odot} \le M < 30 ~{\rm{M}}_{\odot}$)},\\
                    -9.30219+4.79562(M/10~\msun)\\
                    -0.937401(M/10~\msun)^2\\
                   +5.62695\times10^{-2}(M/10~\msun)^3\\
\;\;\;\;\;\;\;\;\;\;\;\;\text{($30 ~{\rm{M}}_{\odot} \le M \le 100 ~{\rm{M}}_{\odot}$).}\\
                   \end{cases}
\end{equation}

\subsubsection*{(3) He-shell burning phase}
After the He-burning ends in the core, the He-shell burning starts until the onset of the C-burning.
The He-shell burning phase is characterized by the stellar radius at the end of the He-burning 
$R_{\rm{He}}^{\rm e}$, the stellar radius at the beginning of the C-burning $R_{\rm{C}}^{\rm{b}}$, 
and the ignition time of the C-burning $t_{\rm{C}}^{\rm{b}}$.
$R_{\rm C}^{\rm b}$ and $t_{\rm C}^{\rm b}$ are approximated by
\begin{equation}
\log (R_{\rm{C}}^{\rm{b}}/{\rm R}_\odot)=\begin{cases}
                           5.4491-5.78767(M/10~\msun)\\
                           + 1.99667(M/10~\msun)^2\\
\;\;\;\;\;\;\;\;\;\;\;\;\text{($10 ~{\rm{M}}_{\odot} \le M < 15 ~{\rm{M}}_{\odot}$)},\\
                          1.39753-0.254317(M/10~\msun)\\
                          +0.106221(M/10~\msun)^2 \\
\;\;\;\;\;\;\;\;\;\;\;\;\text{($15 ~{\rm{M}}_{\odot} \le M \le 50 ~{\rm{M}}_{\odot}$)}\\
                          0.51943+0.621622(M/10~\msun)\\
                          -3.48026\times10^{-2}(M/10~\msun)^2 \\
\;\;\;\;\;\;\;\;\;\;\;\;\text{($50 ~{\rm{M}}_{\odot} \le M \le 100 ~{\rm{M}}_{\odot}$)},\\
                           \end{cases}
\end{equation}
and 
\begin{equation}
(t_{\rm{C}}^{\rm{b}}/{\rm Myr})=2.09464+\frac{106.25}{10(M/10~\msun) -3.90499},
\end{equation}
respectively.
Then, using the normalized time which is defined by
\begin{equation}
\tau_{\rm{HeS}}\equiv \frac{t-t_{\rm{H}}-t_{\rm{He}}}{t_{\rm{C}}^{\rm{b}}-t_{\rm{H}}-t_{\rm{He}}},
\end{equation}
the fitting formula of the stellar radius at the He-shell burning phase is obtained as
\begin{equation}
\log (R_{\rm{HeS}}/{\rm R}_\odot) =\begin{cases}
\log (R_{\rm{He}}^{e}/{\rm R}_\odot)+a_{\rm{HeS}}\tau_{\rm{HeS}}+b_{\rm{HeS}}\tau_{\rm{HeS}}^2\\
+c_{\rm{HeS}}\tau_{\rm{HeS}}^3+(\log (R^{\rm b}_{\rm{C}}/R_{\rm{He}}^{e})\\
-a_{\rm{HeS}}-b_{\rm{HeS}}-c_{\rm{HeS}})\tau_{\rm{HeS}}^{15}\\
\;\;\;\;\;\;\;\;\;\;\;\;\text{($10~\rm{M}_{\odot} \le M \le 50~\rm{M}_{\odot}$)},\\
\log (R_{\rm{He}}^{e}/{\rm R}_\odot)+\log(R_{\rm{C}}^{\rm{b}}/R_{\rm{He}}^{e})\tau_{\rm{HeS}} \\
\;\;\;\;\;\;\;\;\;\;\;\;\text{($50~\rm{M}_{\odot} < M \le 100~\rm{M}_{\odot}$)},\\
\end{cases}
\label{eq:fitAHeB}
\end{equation}
where
\begin{equation}
a_{\rm{HeS}}=\begin{cases}
                    0.198773-8.62031\times10^{-2}(M/10~\msun)\\
                    -6.9987\times10^{-2}(M/10~\msun)^2\\
\;\;\;\;\;\;\;\;\;\;\;\; \text{($10 ~{\rm{M}}_{\odot} \le M < 15 ~{\rm{M}}_{\odot}$)},\\
                   -2.17094+2.46127(M/10~\msun)\\
                   -0.866681(M/10~\msun)^2\\
                   +9.41554\times10^{-2}(M/10~\msun)^3  \\
                   \;\;\;\;\;\;\;\;\;\;\;\;\text{($15 ~{\rm{M}}_{\odot} \le M \le 50 ~{\rm{M}}_{\odot}$),}\\
                   \end{cases}
\end{equation}
\begin{equation}
b_{\rm{HeS}}=\begin{cases}
                  0.45\;\;\;\;\;\;\text{($10 ~{\rm{M}}_{\odot} \le M < 15 ~{\rm{M}}_{\odot}$)},\\
                  5.85223-5.9911(M/10~\msun)+2.05449(M/10~\msun)^2\\
                  -0.217241(M/10~\msun)^3 \\
\;\;\;\;\;\;\;\;\;\;\;\;\text{($15 ~{\rm{M}}_{\odot} \le M \le 50 ~{\rm{M}}_{\odot}$)},\\
                    \end{cases}
\end{equation}
and
\begin{equation}
c_{\rm{HeS}}=\begin{cases}
                    0.15\;\;\;\;\;\text{($10 ~{\rm{M}}_{\odot} \le M < 15 ~{\rm{M}}_{\odot}$)},\\
                   -2.34416+2.5736(M/10~\msun)-0.920019(M/10~\msun)^2\\
                   +0.100612(M/10~\msun)^3 \\
\;\;\;\;\;\;\;\;\;\;\;\;\text{($15 ~{\rm{M}}_{\odot} \le M \le 50 ~{\rm{M}}_{\odot}$)},\\
                   \end{cases}
\end{equation}
The rms error in this phase is shown with the blue line in Fig.~\ref{error}.
We find that our fitting formula has an accuracy within 3.5~\% of the stellar models of \cite{Marigo2001}.

During the He-shell burning phase, we suppose that the CO-core mass, which is formed in the He-burning phase, remains constant. This is because the duration of the He-shell burning is so short that the CO-core mass does not change so much
by the end of the He-shell burning.

For later use, we fit the stellar luminosity at the beginning of He-shell burning as
\begin{align}
 \log \left(\frac{L}{\rm{L_{\odot}}}\right)=&6.74298-4.72995/(M/10~\msun)\notag\\
&+3.59526/(M/10~\msun)^2-1.27068/(M/10~\msun)^3.
\end{align}
For simplicity, we assume that the luminosity does not depend on time after the He-shell burning phase,
because the luminosity is almost constant at this phase~(see Figure~\ref{popIIIHR}).

\subsubsection*{(4) Compact remnants}

After the C-ignition, the nuclear fusion  further proceeds in the core
and finally the Fe-Ni core is formed.
The final fate of a star depends on the Fe-Ni core mass.
However, at the C-ignition, we stop to trace the stellar evolution 
and regard the star to be a compact object,
since the evolution time of the final stage is so short that the whole stellar structure hardly changes~\citep{Kippenhahn1990}.
From the numerical results of Pop III single stellar evolution, the CO-core mass is described
as a function of the stellar mass as
\begin{align}
(M_{\rm{CO}}/\msun)&=0.618397-0.57395(M/10~\msun)\notag\\
&+1.73053(M/10~\msun)^2 -0.312008(M/10~\msun)^3\notag\\
&+2.99858\times10^{-2}(M/10~\msun)^4 \notag\\
&-1.12942\times10^{-3}(M/10~\msun)^5.
\label{eq:mco}
\end{align} 
From the CO core mass, we can estimate the Fe-Ni core mass using the fitting
formula given by \cite{Belczynski2002} as,
\begin{equation}
M_{\rm{FeNi}}=\begin{cases}
                0.161767M_{\rm{CO}} + 1.067055~\msun\;\;\;\;\;\text{($ M_{\rm{CO}} \le 2.5 ~{\rm{M}}_{\odot}$)},\\
                0.314154M_{\rm{CO}}+ 0.686088~\msun\;\;\;\;\;\text{($ 2.5 ~{\rm{M}}_{\odot} \le M_{\rm{CO}}$)}.\\
                 \end{cases}
\end{equation}

As for the criterion of whether a supernova explosion occurs or not after the stellar death,
we here adopt the model adopted in \cite{Belczynski2002}.
The assumptions of the model is as follows: (1) supernovae can occur for stars with
$M_{\rm{CO}}\le5 ~{\rm{M}_{\odot}}$,
(2) some fractions of envelope fall back onto the compact remnant after a supernova explosion for stars with the
intermediate mass range of $5 ~{\rm{M}_{\odot}}<M_{\rm{CO}}\le7.6 ~{\rm{M}_{\odot}}$, 
(3) a star directly collapses so that a supernova explosion does not occur for a star with mass of 
$M_{\rm{CO}} > 7.6 ~{\rm{M}_{\odot}}$. The remnant mass of the compact object in their model is given by
\begin{equation}
M_{\rm{rem}}=\begin{cases}
                M_{\rm{FeNi}}\;\;\;\;\;\text{($ M_{\rm{CO}} \le 5 ~{\rm{M}_{\odot}}$)},\\
                M_{\rm{FeNi}}+\frac{M_{\rm{CO}}-5\msun}{2.6\msun}(M-M_{\rm{FeNi}}),\\
                \;\;\;\;\;\;\;\;\;\;\;\;\;\;\text{($ 5 ~{{\rm{M}}_{\odot}} < M_{\rm{CO}}< 7.6 ~{\rm{M}}_{\odot}$)},\\
                M \;\;\;\;\;\;\;\;\;\;\text{($ 7.6 ~{{\rm{M}}_{\odot}} \le M_{\rm{CO}}$)}.\\
                 \end{cases}
\end{equation}

From the value of a remnant mass, we determine the type of a compact object,
i.e., a neutron star or a black hole.
We assume that the maximum mass of the neutron star is $3~\msun$, which is higher than the mass of
the observed massive pulsars $\sim 2M_{\odot}$. 
Thus a remnant is regarded as a black hole if its mass is higher than $3~{\rm{M}}_{\odot}$.
Although the stellar evolution after the CO burning has been well studied \citep{Woosley1986, Timmes1996,  Fryer1999, Fryer2012}, there are uncertainties for the formation of a compact object. 
In particular, the mechanism of supernova explosions has not been theoretically established.
Thus, our results might change depending on the models for supernova explosions. In this paper,
we employ the same condition of the formation of a compact object as the previous
studies~\citep{Belczynski2002,Belczynski2004}.
\subsection{Binary evolution}
\label{sec:binary}
For the calculation of binary stellar evolution, we need to consider binary interactions such as tidal
evolution, mass transfer, effect of supernova explosions, and the radiation reaction by the gravitational wave.
Here magnetic braking is not taken into account, because Pop III stars are expected to have magnetic
fields much weaker than those of Pop I stars \citep[e.g.,][]{Pudritz1989, Kulsrud1997,Langer2003,
Widrow2002,Doi2011}.

\subsubsection{Tidal evolution}
The orbital angular momentum of a binary system $J_{\rm orb}$ is effectively
transferred to the spin angular momentum $J_{\rm spin, i}$ through tidal interaction
between the two stars. Here $i=1$ and $i=2$ correspond to the primary star with the mass of $M_1$ 
and the secondary star with the mass of $M_2$, respectively.
The time variation of the parameters of a binary orbit such as the semi-major axis $a$, eccentricity $e$, 
$J_{\rm orb}$, and $J_{\rm spin, i}$  are given by \cite{Hurley2002} as follows.
The time variation of $a$ is given by
\begin{equation}
\frac{\dot a}{a}=\frac{2e\dot e}{1-e^2}+2\frac{\dot J_{\rm orb}}{J_{\rm orb}}.
\label{eq:adot}
\end{equation}
Since the total angular momentum (= $J_{\rm orb}+J_{\rm spin, 1}+J_{\rm spin, 2}$) is conserved,
$\dot J_{\rm orb}$ is given by $\dot J_{\rm orb}=-(\dot J_{\rm spin, 1}+\dot J_{\rm spin, 2})$.
Denoting $I_i$ and $ \Omega_{\rm{spin,i}}$ as the moment of the inertia and the spin angular velocity of
each star, $\dot J_{\rm{spin,i}}$ can be written as 
\begin{equation}
\dot{J}_{\rm spin,i}=\dot{ I}_i\Omega_{\rm  spin,i}+I_i\dot{\Omega}_{\rm spin,i} , \label{eq:tidal_spin}
\end{equation}
where the first term of r.h.s expresses the contribution of 
the change of the internal structure of the star $i$ and the second term is due to the tidal force from the
other star. 
\citet{Hut1981} showed that the time evolution of the spin angular velocity can be calculated by 
\begin{eqnarray}
&\dot \Omega_{{\rm spin},1}=3\frac{k}{T}\frac{q_2^2}{r_{\rm{g}}^2}\left(\frac{R_1}{a}\right)^6\frac{\Omega_{\rm{orb}}}{(1-e^2)^6}\nonumber\\
&~~~\times \left[f_1(e^2)-(1-e^2)^{\frac{3}{2}}f_2(e^2)\frac{\Omega_{\rm{spin,1}}}{\Omega_{\rm{orb}}}\right],\\
&f_1(e^2)=1+\frac{15}{2}e^2+\frac{45}{8}e^4+\frac{5}{16}e^6\label{eq:f2},\\
&f_2(e^2)=1+3e^2+\frac{3}{8}e^4,\label{eq:f5}\\
&q_2\equiv M_2/M_1,\label{eq:q2}
\end{eqnarray}
where  $T$, $k$, $r_{\rm{g}}$ and $\Omega_{\rm{orb}}$ are the tidal timescale, the apsidal motion constant
of the primary star, the gyration radius which is defined by $\sqrt{I_1/M_1/R_1^2}$
and the orbital angular velocity, respectively.
$T$, $k$, and $r_{\rm g}$ depend on the properties of the internal structure of the primary star and
their specific forms are given later.
The time evolution of $ \dot \Omega_{\rm{spin,2}}$ is given by
changing 1 to 2 and 2 to 1 in the above equations. Once $T$, $k$, $r_{\rm g}$, and the binary parameters
are given, one can determine $ \dot J_{\rm orb}$ from $ \dot J_{\rm{spin,1}} + \dot J_{\rm{spin,2}}$.

\citet{Hut1981} also gave the equations for $\dot e$ as
\begin{align}
\dot e=&-27\frac{k}{T}q_2(1+q_2)\left(\frac{R_1}{a}\right)^8\frac{e}{(1-e^2)^{\frac{13}{2}}}\nonumber\\
&\times\left[f_3(e^2)-\frac{11}{18}(1-e^2)^{\frac{3}{2}}f_4(e^2)\frac{\Omega_{\rm{spin,1}}}{\Omega_{\rm{orb}}}\right],\label{eq:dedt}
\end{align}
\begin{align}
f_3(e^2)&=1+\frac{15}{4}e^2+\frac{15}{8}e^4+\frac{5}{64}e^6,\label{eq:f3}\\
f_4(e^2)&=1+\frac{3}{2}e^2+\frac{1}{8}e^4.\label{eq:f4}
\end{align}
Substituting Eqs.~(34)--(41) into r.h.s of Eq.~(33), we   can determine the time evolution of the semi major axis  $ a$ .

The  moment of inertia $I_i$ depends on the stellar evolutionary phase.
$I_i$ at H-burning phase can be written as $I_i=k_{\rm H}(M_i,\tau_{\rm H}) M_iR_i^2$.
\citet{Hurley2002} constructed a fitting formula of $k_{\rm H}(M_i,\tau_{\rm H})$ in their open code
so that we adopt the same formula. On the other hand, when a star has core-envelope structure, 
$I_i=k_{\rm env}(M_i-M_{{\rm c},i})R_{i}^2 + k_{\rm core}M_{{\rm c},i}R_{{\rm c},i}^2$
\citep{Hurley2000, Hurley2002}, where $M_{{\rm c},i}$ and $R_{{\rm c},i}$ are the stellar core
mass and radius, and $k_{\rm env}$ which is the  same as $k_{\rm H}(M_i,\tau_{\rm H})$ in Hurley's open code and
$k_{\rm core}=0.21$, respectively. In this paper, we approximate the core radius using the core mass 
following \citet{Tout1997} as
\begin{equation}
\frac{R_{{\rm c},i}}{{\rm R}_\odot}=0.9334\left(\frac{M_{{\rm c},i}}{10~\msun}\right)^{0.62},
\end{equation}
where the core mass corresponds to the He-core mass for a star with H-envelope and a CO-core mass 
for a star without H-envelope due to the binary interaction  so-called naked-He star, respectively.

As for the initial stellar spin at the ZAMS phase, we follow \cite{Hurley2000} as
\begin{align}
&\Omega_{{\rm spin},i}=45.35\left(\frac{v_{\rm{rot}}}{1~\rm{km~s^{-1}}}\right)\left(\frac{R_{\rm{ZAMS}}}{{\rm R}_{\odot}}\right)^{-1}~\rm{yr^{-1}},
\\
&v_{\rm{rot}}(M_i)=\frac{658437{(M_i/10~\msun)^{3.3}}}{15+2818({M_i}/10~\msun)^{3.45}}~\rm{km ~s^{-1}}.
\end{align}

Next, we argue the apsidal motion constant $k$ and tidal time scale $T$.
In the case that the primary envelope is convective, the energy dissipation due to the convective motions
causes the lag of the tidal deformation, which yields the misalignment of the direction of the maximum
tidal deformation and the direction to the secondary star. This misalignment generates the torque to the
primary star so that the angular momentum is transferred between the spin one and the orbital one.
According to \citet{Verbunt1995, Rasio1996}, the apsidal motion constant decided by the tidal time scale for
the convective envelope is  given by
\begin{equation}
\frac{k}{T}=\frac{2}{21}\frac{
f_{\rm{con}}}{\tau_{\rm{con}}}\frac{M_{\rm{env,1}}}{M_{\rm{1}}},
\end{equation}
where $M_{\rm{env,1}}\equiv M_1-M_{\rm{c,1}}$ is
the primary envelope mass and the factor $f_{\rm{con}}$ is the correction of the tidal torque.
The eddy turnover timescale $\tau_{\rm{con}}$,
which describes the contribution of the turbulent viscosity due to the
convective motions, is given by \cite{Hurley2002},
\begin{equation}
\tau_{\rm{con}}=\left[\frac{M_{\rm{env,1}}R_{\rm{env,1}}
\left(R_1-\frac{1}{2}R_{\rm{env,1}}\right)}{3L_1}\right]^{1/3},
\end{equation}
where $L_1$ and $R_{\rm{env,1}}\equiv R_1-R_{\rm{c,1}}$ are
the stellar luminosity  and the envelope radius of the primary star, respectively.
If $\tau_{\rm{con}}\ll P_{\rm{tid}}/2~(=\pi|\Omega_{\rm{orb}}-\Omega_{\rm{spin1}}|^{-1}$),
the turbulent viscosity due to the convective motions can be affected. If $\tau_{\rm{con}}> P_{\rm{tid}}/2$,
on the other hand, the contribution of the convective motions to the viscosity is negligible.
The factor $f_{\rm con}$ is obtained by Rasio et al. (1996) as
\begin{equation}
f_{\rm{con}}={\rm{min}}\left[1,\left(\frac{P_{\rm{tid}}}{2\tau_{\rm{con}}}\right)^2\right].
\end{equation}

If the envelope is radiative, a tide is a dynamical tide with radiative damping \citep{Zahn1975}.
For a star which has the radiative envelope, the energy dissipation due to radiation
is so small that the equilibrium tide cannot be effective.
However, the non-radial oscillations at the surface are driven by gravity waves due
to the tide and the resonances of those oscillations are damped by radiation.
The value of $k$ devided by $T$ is given by \cite{Zahn1977, Hurley2002} as
\begin{align}
\frac{k}{T}=&4.3118\times10^{-8}\left(\frac{M_1}{\msun}\right)\left(\frac{R_1}{{\rm R}_{\odot}}\right)^2\notag\\
&\times\left(\frac{a}{1~{\rm{AU}}}\right)^{-5}(1+q_2)^{5/6}E~\rm{yr^{-1}},
\end{align}
where the tidal coefficient $E$ is described by \cite{Zahn1975} as
\begin{equation}
E=1.101\times10^{-6}\left(\frac{M_1}{10~\msun}\right)^{2.84}.
\end{equation}

\subsubsection{Roche lobe overflow}
\label{sec:RLOF}

When the primary star in a binary system fills its Roche lobe, 
its stellar envelope is transferred to the secondary star,
which is called the Roche lobe overflow (RLOF).
The radius of the Roche lobe of the primary star ($R_{\rm{L,1}}$) is approximately given  
by \cite{Eggleton1983} as
\begin{equation}
\frac{R_{\rm{L,1}}}{a} \approx \frac{0.49q_1^{2/3}}{0.6q_1^{2/3}+\ln(1+q_1^{1/3})},
\label{eq:RL}
\end{equation}
where $q_1\equiv{M_1}/{M_2}$ is the mass ratio. 
This equation is within 1\% accuracy over the whole range.
When the RLOF occurs and the primary star loses its envelope,
the stellar radius changes depending on the properties of the stellar envelope
\citep{Paczynski1972}.
Since the dynamical time of the star  given by
\begin{align}
\tau_{\rm{dyn,1}}=
\frac{\pi}{2}\left(\frac{R_1^3}{2GM_1}\right)^{1/2}
\end{align}
is much shorter than the thermal timescale (Kelvin-Helmholtz timescale) given by 
\begin{align}
\tau_{\rm{KH,1}}
=\frac{GM_1(M_1-M_{\rm{c},1})}{L_1R_1},
\end{align}
the radius of the primary star after the mass transfer is adjusted to {\it the adiabatic radius} $R_{\rm ad,1}$ first, 
that is, the star reaches the hydrostatic equilibrium state but not the thermal equilibrium.
After the thermal time scale, the primary radius approaches {\it the thermal equilibrium radius}
$R_{\rm th,1}$.
The mass transfer via the RLOF actually depends on the responses
of the Roche lobe radius, $R_{\rm ad,1}$ and $R_{\rm th,1}$ after the mass transfer.

We here introduce the following two quantities for convenience 
to understand the fate of the binary after the RLOF;
\begin{equation}
\zeta_{\rm{L}}=\frac{d{\rm{log}}R_{\rm L,1}}{d{\rm{log}}M_1},
\end{equation}
and
\begin{equation}
\zeta_{\rm{ad}}=\frac{d{\rm{log}}R_{{\rm ad},1}}{d{\rm{log}}M_1}.
\end{equation}
Since it is difficult to obtain the exact forms $\zeta_{\rm{L}}$ and $\zeta_{\rm{ad}}$, 
we here use the approximated expressions.
Assuming that the mass transfer is conservative 
i.e., the total mass is conserved during the mass transfer, 
we have $\zeta_{\rm{L}}$ as~\citep{Tout1997}
\begin{equation}
\zeta_{\rm{L}} \approx 2.13q_1-1.67\;\;\;\; (0<q_1<50),\label{eq:zetal}
\end{equation} 
where we use  Eq. (\ref{eq:RL}). The  value of $\zeta_{\rm{ad}}$ depends on the property of the stellar envelope.
When the primary star is in the giant phase,  it  has a deep convective envelope
with the polytropic index of  1.5 so that $\zeta_{\rm{ad}}$ is given by 
\begin{equation}
\zeta_{\rm{ad}}\approx-1+\frac{2}{3}\frac{M_1}{M_1-M_{c1}},\label{eq:zetaad}
\end{equation}
under the assumption that the envelope mass is neglected compared to the total mass
\citep{Hjellming1987}.
When the primary star is in the  other stages, 
$\zeta_{\rm ad}=2.59, 6.85, 1.95$ and 5.79 for the main sequence,
  the giant phase with the radiative envelope \citep{Hjellming1989}, 
 the naked-He main sequence and  the naked-He giant star \citep{Ivanova2002, Belczynski2008}, respectively.


Now let us compare $\zeta_{\rm ad}$ and $\zeta_{\rm L}$. We first consider the case of 
$\zeta_{\rm ad}<\zeta_{\rm L}$, which means $d\log R_{{\rm ad},1} > d\log R_{\rm L,1}$ since $d\log M_1  < 0$.
In this case, the radius of the primary star continues to exceed the Roche lobe radius 
at the dynamical timescale as the primary star loses its envelope.
Thus, the mass transfer violently occurs and the stars rapidly approach each other
\footnote{For simplicity, we here assume the conservation of the angular momentum and 
neglect the mass ejection from the binary system during the mass transfer on the dynamical timescale.}.
When the primary star is a giant with the outer envelope, which is either radiative or convective, 
the primary envelope rapidly swallows the secondary star.
After that, the binary stars will be in the common envelope (CE) phase \citep{Paczynski1976}.
We describe the subsequent evolution of the binary in Sec.~\ref{sec:CE}. 
On the other hand, when the primary star does not have the core-envelope structure
like in the H-burning main sequence and naked-He main sequence,
the binary will merge via the rapid mass transfer.


Next, we consider  the case for $\zeta_{\rm ad}>\zeta_{\rm L}$.
In this case,  the primary star shrinks within the Roche lobe radius 
($R_{\rm ad,1}<R_{\rm L,1}$)
on the dynamical timescale by the mass transfer of the envelope so that
RLOF stops for a while. However in  the thermal timescale,
the  radius of the primary star approaches $R_{\rm th,1}$.
If the thermal equilibrium radius is larger than the Roche lobe radius, 
the mass transfer begins and the transfer rate is expressed by \citet{Paczynski1972, Tout1997,Hurley2002}  as
\begin{equation}
\dot{M_1}=F(M_1)\left[{\rm{ln}} \left(\frac{R_{\rm th ,1}}{R_{\rm L,1}}\right)\right]^3~\rm{M_{\odot}~yr^{-1}}
\label{eq:m1dotf}
\end{equation}
and 
\begin{equation}
F(M_1)=3\times10^{-6}\left\{{\rm{min}}\left[\left(10\frac{M_1}{10~\msun}\right),5.0\right]\right\}^2,
\end{equation}
where the expression of $R_{\rm th,1}$ is shown   in Appendix~A.1. 
Since the stellar radius changes on the thermal timescale (or more slowly),
the maximum value of the mass transfer rate is  
\begin{equation}
\dot{M}_{1,\rm{max}}=\frac{M_1}{\tau_{\rm{KH,1}}}.
\label{eq:mkhmax}
\end{equation}
We assume that the binary stars merge
if $R_{\rm th,1}>10R_{\rm L,1}$ for the star without the core-envelope structure
since the mass transfer rate is comparable to the above upper limit.

\begin{figure}
\centering
\includegraphics[width=8cm]{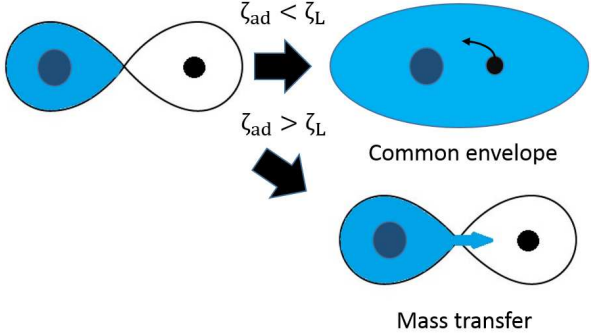}
\caption{Schematic diagram of the mass transfer when the primary becomes a giant star. Let us define 
$\zeta_{\rm{L}}=d{\rm{log}}R_{\rm L,1}/d{\rm{log}}M_1$ and $\zeta_{\rm{ad}}=d{\rm{log}}R_{\rm ad,1}/d{\rm{log}}
M_1$. When the primary star fulfills the Roche lobe as in the upper left of the figure, there are two destinies.
1) If $\zeta_{\rm{ad}}<\zeta_{\rm{L}}$, then  $d\log R_{{\rm ad},1} > d\log R_{\rm L,1}$ since $d\log M_1  < 0$
so that the mass transfer is dynamically unstable. The secondary star is swallowed by the primary envelope to be
the common envelope phase as the upper right of the figure. 2) If $\zeta_{\rm{ad}}>\zeta_{\rm{L}}$, the mass 
transfer is dynamically stable so that the radius of the primary star becomes smaller than the Roche lobe radius
 on the dynamical timescale after losing the small fraction of the envelope mass. However in the thermal time 
scale~(Kelvin-Helmholz time), the radius increases again and fulfills the Roche lobe so that the stable mass 
transfer from the primary star to the secondary star occurs like in the lower right of the figure.}
\label{MT}
\end{figure}

In the case of the stable mass transfer~($\zeta_{\rm ad}>\zeta_{\rm L}$), the accretion rate onto the secondary 
varies with its evolutionary stage~\citep{Hurley2002}.
If the secondary is in the main sequence or in the He-burning phase, the
accretion timescale
\begin{equation}
\tau_{\dot{M}}\equiv\frac{M_2}{\dot{M}_1}
\end{equation}
is much shorter than the thermal timescale of the secondary $\tau_{\rm KH,2}$.
Therefore the secondary does not always receive all the accreting matter, and the accretion rate onto the 
secondary is calculated by ~\citep{Hurley2002} as
\begin{equation}
\dot{M}_{2}={\rm{min}}\left(10\frac{\tau_{\dot{M}}}{\tau_{{\rm{KH,2}}}},1\right)\dot{M}_{1}.
\end{equation}
If the secondary is in the He-shell burning phase or
the naked-He star, the secondary can receive all the accreting mass~\citep{Hurley2002}.
For the secondary of the compact object, the accretion rate is limited by 
the Eddington limit \citep{Cameron1967, Hurley2002} as
\begin{align}
\dot{M}_{\rm{Edd}}&=\frac{4\pi c R_2}{\kappa_{\rm T}}\\\nonumber
&=2.08\times10^{-3}(1+X)^{-1}
\left(\frac{R_2}{\rsun}\right)~\rm{M_{\odot}~yr^{-1}},
\end{align}
where $\kappa_{\rm T} =0.2(1+X)\ \rm cm^2\ g^{-1}$ is the opacity of the Thomson scattering
and $X(=0.76)$ is the H-mass fraction.

\subsubsection{Common envelope phase}
\label{sec:CE}

If $\zeta_{\rm ad}<\zeta_{\rm L}$,  the mass transfer occurs violently 
from the primary star with the core-envelope structure to the secondary so that
the binary system becomes the CE phase as discussed in the previous Sec.~\ref{sec:RLOF}.
Once the secondary star is engulfed into the primary envelope, 
it  spirals into the core of the primary star due to the gas friction.
Then, the orbital energy is converted into the thermal energy of the primary envelope, 
which results in the mass ejection from the binary system.  As we showed in Sec.~2.2.2, stars 
with convective envelope typically take smaller values of $\zeta_{\rm ad}$ than stars with radiative envelope. 
Thus the former stars are easier to satisfy the condition of the onset of the CE phase of
$\zeta_{\rm ad}<\zeta_{\rm L}$ than the latter ones.
For Pop III stars, we determine whether they have the radiative or convective envelope from the HR diagram
in Fig.~\ref{popIIIHR}.
We found that Pop III stars with mass $\ge70~\msun$ reach the Hayashi track 
at the beginning of the He-shell burning.
While  Pop III stars less massive than $50~\msun$ do not reach the Hayashi track.
Thus,  Pop III stars with mass  $>50~\msun$ develop the deep convective envelope
during the He-shell burning phase and are easier to be the CE phase.

In order to take into account the CE phase,
we here adopt the prescription  given by \cite{Webbink1984}.
For simplicity, we assume that all gas in the primary envelope is ejected.
Then, the change of the orbital energy when the secondary star spirals in is expressed as 
\begin{equation}
\Delta E_{\rm orb}=\frac{GM_{\rm{c,1}}M_2}{2a_{\rm{f}}} -\frac{GM_1M_2}{2a_{\rm{i}}}, \label{eq:eorbi}
\end{equation}
where $a_{\rm i}$ and $a_{\rm f}$ are the separation before and after 
the secondary star spirals in, respectively. A fraction of the orbital energy is actually converted to the 
kinetic energy of the ejected matter.
The binding energy of the primary envelope is somewhat smaller than the gravitational energy,
and thus, it is expressed by
\begin{equation}
E_{\rm{bind}}=\frac{GM_{\rm{1}}M_{\rm{env,1}}}{\lambda R_1},\label{eq:ebind}
\end{equation}
where $\lambda$ is a parameter depending on the properties of the envelope
and $M_{\rm env,1}$ is the mass of the primary envelope.
Therefore, we can estimate the final separation $a_{\rm f}$ by
\begin{equation}
\alpha\left(\frac{GM_{\rm{c,1}}M_2}{2a_{\rm{f}}}-\frac{GM_1M_2}{2a_{\rm{i}}}\right)=\frac{GM_{\rm{1}}M_{\rm{env,1}}}{\lambda R_1}, \label{eq:ce2}
\end{equation}
where $\alpha$ is the efficiency factor of the energy conversion which depends on
the interaction between the primary envelope and the secondary star.  When the secondary star is also a giant star, we need to modify the above prescription. 
For simplicity, we assume that the remnants after the mass ejection are two cores and the values of the $\lambda$ parameter are the
same for both envelopes. Then, we can obtain the final separation $a_{\rm f}$ from
\begin{equation}
\alpha\left(\frac{GM_{\rm{c,1}}M_{\rm{c,2}}}{2a_{\rm{f}}}-\frac{GM_1M_2}{2a_{\rm{i}}}\right)=\frac{GM_{\rm{1}}M_{\rm{env,1}}}{\lambda R_1}+\frac{GM_{\rm{2}}M_{\rm{env,2}}}{\lambda R_2}.
\label{eq:ce3}
\end{equation}
To determine the values of $\alpha$ and $\lambda$, the sophisticated numerical simulations 
including the hydrodynamics in the CE phase and the stellar evolution are required
\citep{Xu2010, Loveridge2011}.
However, to do these simulations for each star is actually time consuming.
So, in this paper, we assume $\alpha \lambda=1$ as in the previous studies
\citep{Hurley2002, Belczynski2002, Belczynski2008}.

Next, we describe the treatment of the fate of the binary after the mass ejection at the CE phase.
When the final separation estimated from  Eqs. (\ref{eq:ce2}) or (\ref{eq:ce3}) 
is small, the binary will merge under following two conditions. The first condition is  a simple one by  
\cite{Belczynski2002, Belczynski2004, Belczynski2008, Dominik2012}, 
where the core of the binary merges if $R'_{\rm{1}} +R'_{\rm{2}}>a_{\rm{f}}$.
Since the binary merges in this case only if the binary stars contact each other,
we call this condition as the {\it conservative core merger criterion}.
The second condition is the same as that used in \cite{Hurley2002},
that is,  if $R'_{\rm{1}} > R'_{\rm{L,1}}$ or $R'_{\rm{2}} > R'_{\rm{L,2}}$
where the quantities with prime mean the value after the mass ejection.
In other words, this means that the cores of the binary stars merge if the mass transfer occurs 
during the CE phase. 
However, there are arguments against this by  
\cite{Ivanova2002, Podsiadlowski2010} so that we call
 this condition  as  {\it optimistic core merger criterion} .

\subsubsection{Effect of supernova explosion}
\label{sec:SN}
When a supernova explosion occurs, binary parameters
($M_{\rm{total}}\equiv M_1+M_2,a,e$) change  due to the instant mass ejection and the kick velocity. 
In our present simulations, we assume zero kick velocity for
simplicity. The neutron star formation with the kick velocity is easier to disrupt NS-NSs than the spherically
symmetric supernova explosion so that the merger rate of NS-NSs and NS-BHs from our simulations should be 
considered as the upper limit.  While in the formation of a black hole, zero kick velocity is reasonable so that
the formation rate of BH-BHs is reliable.

Before the supernova explosion, the relative velocity ${\bf v}$ is described by the orbital speed $v$ and the 
angle ${\bf \beta}$ between the relative velocity and the  separation vector ${\bf r}\equiv{\bf r}_1-{\bf r}_2$ as
\begin{equation}
{\bf{v}}=(-v\sin\beta,-v\cos\beta,0),
\end{equation}
where the orbital speed is expressed by the specific orbital energy $GM_{\rm{total}}/2a$ as
\begin{equation}
v=\sqrt{GM_{\rm{total}}\left(\frac{2}{r}-\frac{1}{a}\right)} \label{eq:v1},
\end{equation}
where $a$ is the semi major axis before the supernova explosion.  After the supernova of the primary star, 
it ejects mass instantly is a good approximation since the expansion velocity of the supernova ejecta 
($\sim {\rm 10^{9}cm/s}$) is much larger than the orbital velocity. Thus, the total mass immediately after 
the supernova explosion is
\begin{equation}
 M_{\rm{total}} \to M_{\rm{total}}'=M_{\rm{total}}-\Delta M_1 
\end{equation}
where the subscript ' means the value immediately after the supernova explosion and $\Delta M_1$ is the primary
ejected mass. The relative velocity immediately after the supernova explosion is described by
\begin{equation}
v'=\sqrt{GM_{\rm{total}}'\left(\frac{2}{r}-\frac{1}{a'}\right)}.\label{eq:v2}
\end{equation}
The relative velocity does not change immediately after the supernova explosion, because the spherically 
symmetric supernova explosion does not change the specific angular momentum. Thus, $v=v'$. Therefore, using the 
equation 
(\ref{eq:v1}) and (\ref{eq:v2}), we have the separation after supernova explosion as
\begin{equation}
a'=\left(\frac{v^2}{GM_{\rm{total}}}-\frac{v^2}{GM_{\rm{total}}'}+\frac{1}{a}\right)^{-1}\label{eq:an}.
\end{equation}
The eccentricity after the supernova explosion is calculated by the conservation of the specific angular 
momentum as  
\begin{equation}
e'=\sqrt{1-\frac{|\mathbf{r}\times\mathbf{v}|^2}{GM'_{\rm{total}}a'}}\label{eq:en}.
\end{equation}

For example, let us  consider the case of  the initial $e=0$. In this case, $r=a$ and $v=(GM_{\rm{total}}/a)^{1/2}$. Thus, the
separation and the eccentricity after the supernova explosion  are
\begin{align}
a'&=\left(\frac{2}{a}-\frac{M_{\rm{total}}}{M_{\rm{total}}'a}\right)^{-1}, \\
e'&=\frac{M_{\rm{total}}}{M'_{\rm{total}}}-1.
\end{align}
The mass ejection of the supernova explosion decelerates the escape velocity of the binary. On the other hand,
the velocity of compact object in the supernova remnant does not change. Thus, if the mass ejection is lager than
a half of the total mass of the binary, the velocity of the compact object can be larger than the escape velocity. 
Therefore, if $M'_{\rm{total}}<\frac{1}{2}M_{\rm{total}}$, the binary is disrupted.

Note that in Hurley's original code, when the supernova explosion occurs as soon as after the CE phase, 
the primary mass before the CE phase is treated as the primary mass before the supernova explosion. 
On the other hand, we assume that the primary mass after the CE phase is treated as the primary 
mass before the supernova explosion.

\subsubsection{Coalescence time due to the emission of gravitational waves }

When the stars of a binary system explode or collapse at the end of their lifetime, the compact star binary is
formed. The compact binary loses the angular momentum and the orbital energy by the emission of gravitational
waves. We use the slow-motion and weak-field approximation formalism described by \cite{Peters1963} and
\cite{Peters1964}. The equations of  the change of the angular momentum, the semi major axis and the eccentricity are given by
\begin{equation}
\frac{\dot{J}}{J}=-\frac{32G^3M_1M_2M_{\rm{total}}}{5c^5a^4}\frac{1+\frac{7}{8}e^2}{(1-e^2)^{5/2}},\label{eq:angular}
\end{equation}
\begin{equation}
\frac{\dot{a}}{a}=-\frac{64G^3M_1M_2M_{\rm{total}}}{5c^5a^4}\frac{1+\frac{73}{24}e^2+\frac{37}{96}e^4}{(1-e^2)^{7/2}},\label{eq:semimajor}
\end{equation}
and
\begin{equation}
\frac{\dot{e}}{e}=-\frac{304G^3M_1M_2M_{\rm{total}}}{15c^5a^4}\frac{1+\frac{121}{304}e^2}{(1-e^2)^{5/2}}\label{eq:edot}.
\end{equation} 
From Eqs. (\ref{eq:semimajor}) and (\ref{eq:edot}), we can express $a$ by $e$ as
\begin{equation}
\frac{a}{a_0}=\frac{1-e_0^2}{1-e^2}\left(\frac{e}{e_0}\right)^{12/19}\left(\frac{1+\frac{121}{304}e^2}{1+\frac{121}{304}e_0^2}\right)^{870/2299}\label{eq:aerelation},
\end{equation}
where $a_0$ and $e_0$ are the initial values of $a$ and $e$, respectively.
For $a/a_0\ll 1$, Eq. (\ref{eq:aerelation}) is approximated by
 \begin{equation}
e\sim \left(\frac{a}{a_0(1-e_0^2)}\right)^{19/12}e_0.
\end{equation}
For $e_0=0$, Eq. (\ref{eq:semimajor}) is integrated as
\begin{eqnarray}
&t_{\rm coal}(e_0=0)=\frac{5}{256}\frac{a_0^4}{c}\left(\frac{GM_1}{c^2}\right)^{-1}
\left(\frac{GM_2}{c^2}\right)^{-1}\left(\frac{GM_{\rm total}}{c^2}\right)^{-1}\\
&=10^{10}(\frac{a_0}{16~{\rm R}_\odot})^4\left(\frac{M_1}{10~\msun}\right)^{-1}
\left(\frac{M_2}{10~\msun}\right)^{-1}\left(\frac{M_{\rm total}}{10~\msun}\right)^{-1}\  {\rm yr}. \notag
\end{eqnarray}
\cite{Peters1963} and \cite{Peters1964} found numerically that for $e_0 > 0$, $t_{\rm merge}(e_0)$ is
approximately given by
\begin{equation}
t_{\rm coal}(e_0)\sim (1-e_0^2)^{7/2}t_{\rm merge}(e_0=0).
\end{equation}
 However in our simulations, we solve Eqs. (75) and (77).

\subsection{Initial condition}
\label{sec:initial condition}
In this paper, we calculate the evolution of $10^6$ binaries using the fitting formulae (see Sec.
\ref{sec:fitting formula}) and prescriptions for the binary interactions (see Sec.~\ref{sec:binary}). As
initial conditions, we should set the binary parameters, that is, the primary mass $M_1$, the mass ratio of
secondary to primary $q_2=M_2/M_1 <  1$ with $M_2$ being the secondary mass, the eccentricity $e$ and the
semi major axis $a$. In this section, we describe how to generate the initial conditions of $10^6$ binaries.

\subsubsection{Distribution function of the binary parameters}
\label{initial}
Since  the  distribution functions of the binary parameters are not known for Pop III stars, as a first step we
use those of  the observed Pop I stars except for the initial mass function~(IMF).

\subsubsection*{(1) Initial mass}
We consider two kinds of the IMF to study the dependence of the results on IMF.  The first one is the Salpeter
IMF \citep{Salpeter1955} given by
\begin{equation}
\Psi(M_1)\propto M_1^{-2.35},\label{eq:IMF}
\end{equation}
where $\Psi(M_1)$ is the number of stars per unit  mass. The second one is the flat IMS given by
\begin{equation}
\Psi(M_1)\propto \rm{const}.\label{eq:IMFflat}
\end{equation}
This mass function is suggested by some numerical simulations of the Pop III star formation\citep{Clark2011}. 
We set the mass range of $10~\msun \leq M_1\leq 100~\msun$ to the mass of the primary star, as suggested by the
recent numerical simulations \citep{Hirano2013}.

\subsubsection*{(2) Initial mass ratio}
The distribution function of the initial mass ratio $q_2  <  1$  is given by 
\begin{equation}
\Phi(q_2)\propto \rm{const}.\label{eq:mass ratio}
\end{equation}
This distribution is suggested by the recent observations of binary systems \citep{Kobulnicky2007,
Kobulnicky2012}. We set the minimum mass ratio to be $q_{\rm{2, min}} \equiv 10\ \msun/M_1$, because we assume
that the secondary mass range is the same as that of the primary.

\subsubsection*{(3) Initial eccentricity}
Following \cite{Heggie1975} and \cite{Duquennoy1991}, we use the distribution function of the initial
eccentricity in the form of
\begin{equation}
\Xi(e)\propto e,\label{eq:initial ecc}
\end{equation}
for the range of $0\leq e\leq 1$.

\subsubsection*{(4) Initial separation}
We adopt the logarithmically flat distribution for the initial semi major axis following \cite{Abt1983} as
\begin{equation}
\Gamma(a)\propto \frac{1}{a},
\end{equation}
for the range of $A_{\rm min}\leq a \leq 10^6~\rsun$. $A_{\rm min}$ is determined from Eq. (\ref{eq:RL}) as
\begin{align}
A_{\rm{min}}&=\frac{A_{\rm L}}{1-e}\\\nonumber
&=\frac{0.6q_1^{2/3}+\ln(1+q_1^{1/3})}{0.49q_1^{2/3}}\frac{R_1}{1-e}. 
\end{align}
$A_{\rm{L}}$ corresponds to the separation when the primary star fills its Roche lobe  at the peri-astron at the initial time.
A binary should not fill its Roche lobe from the beginning so that we adopt this minimum separation.

\subsubsection{Monte Calro method}
We generate the initial conditions of binaries using the above distribution functions
and the Monte Carlo method. In this paper, we mainly set the total number of binaries $N_{\rm total}$ to be 
$10^6$. For example, we describe how to generate the initial condition of the primary mass.
We prepare the homogeneous random variable $X$ in the interval of $0\le X\le1$.
For random choice of $X$,  we define $M_1$  by 
\begin{equation}
X\equiv\frac{\int_{M_{\rm{min}}}^{M_1}\Psi(M)dM}
{\int_{M_{\rm{min}}}^{M_{\rm{max}}}\Psi(M)dM}.
\label{eq:MCmass}
\end{equation}
For example in the case of the Salpeter IMF 
with the mass range of $10~\msun \leq M_1 \leq 100~\msun $, $M_1$ is given by
\begin{equation}
M_1=[10^{-1.35}-X(10^{-1.35}-100^{-1.35})]^{-1/1.35}\ \msun.
\end{equation}
We assign a random number generated by the Mersenne twister method \citep{Matsumoto1998}
to the number $X$ and set the primary mass $M_1$.
For the other parameters ($q_2$, $e$, and $a$), 
we also generate the initial parameters in the same way.
 To check the reliability of our Monte Carlo method, in Appendix A.2, we perform the same simulations as \cite{Hurley2002} and obtain the similar  results.
 In Appendix A.3, we show the convergence check, that is, for one model, we performed the simulations with $10^5$, $10^6$ and $10^7$ binaries. The results agree with each other within the statistical errors.

\section{Results}
We compute the evolution of $10^6$ binaries having random values of binary parameters. In this paper, we adopt the
four models as shown in Table 1. 
Each column represents the name of the model, population of stars, IMF, mass range of the
primary star, and that of the secondary star, respectively. Models III.s and III.f are simulations of Pop III 
binaries with the mass range of $10~\msun \le M \le100~\msun$. 
For Models III.s and III.f, the Salpeter and flat IMF are adopted, respectively.
Models I.h and I.l are simulations of Pop I binaries with Hurley's single stellar fitting formulae
\citep{Hurley2000} for comparison. In both models, the Salpeter IMF is adopted.
For Model I.h, the initial mass range is $10~\msun \le M \le100~\msun$.
For Model I.l, the initial mass range is $1~\msun \le M \le100~\msun$ to take into account the typical mass
of a Pop I star of  $\sim 1\msun$.

\begin{figure*}
  \begin{minipage}{.40\linewidth} 
 \includegraphics[width=1.18 \linewidth]{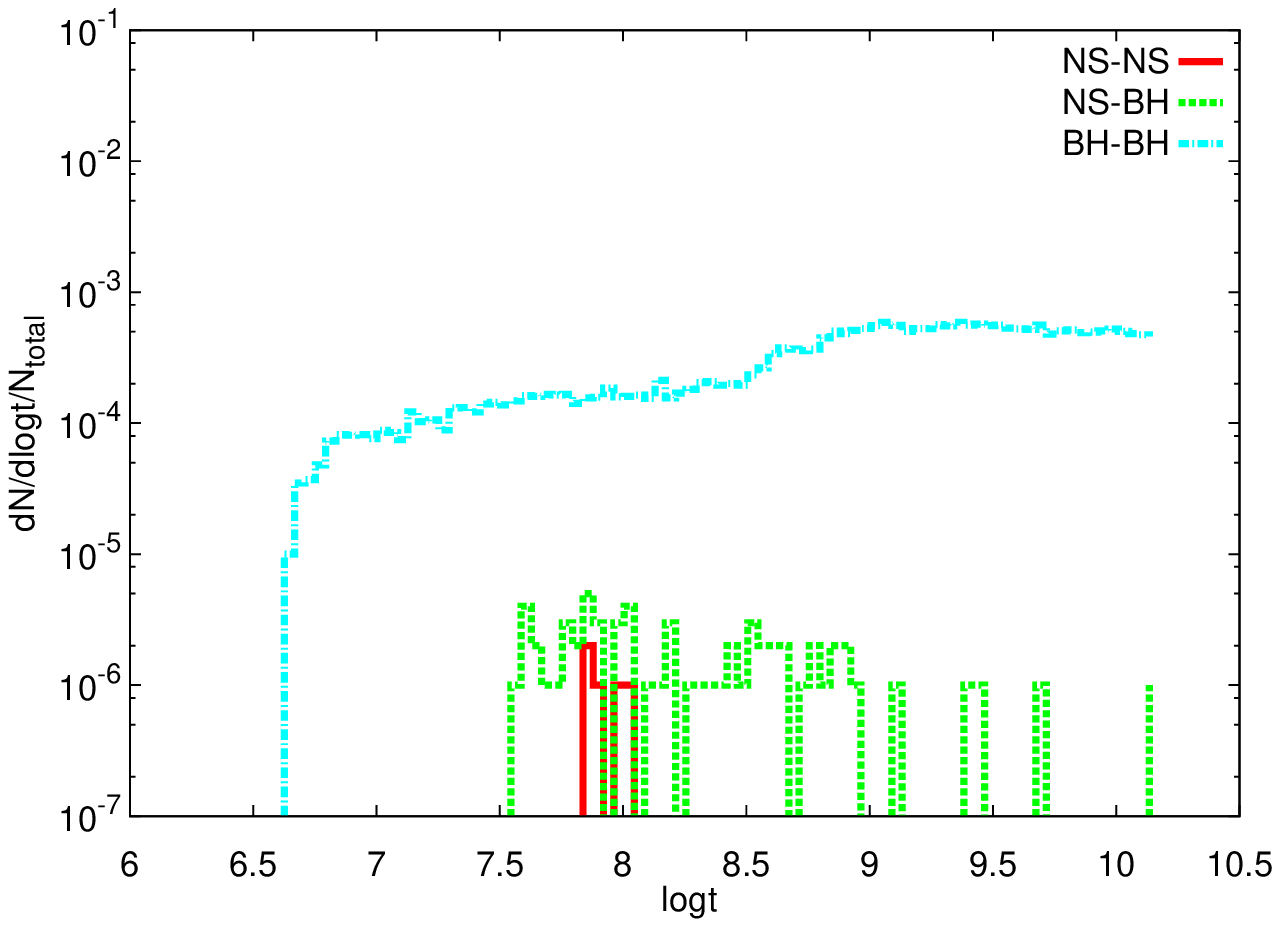}  
\smallskip

~~~~~~~~~~~~~~~~~~~~~~~~~~~~~~~~~~~~~(a) Model III.s
\medskip

  \end{minipage}
  \hspace{2.5pc} %
 \begin{minipage}{.40\linewidth} %
  \includegraphics[width=1.18 \linewidth]{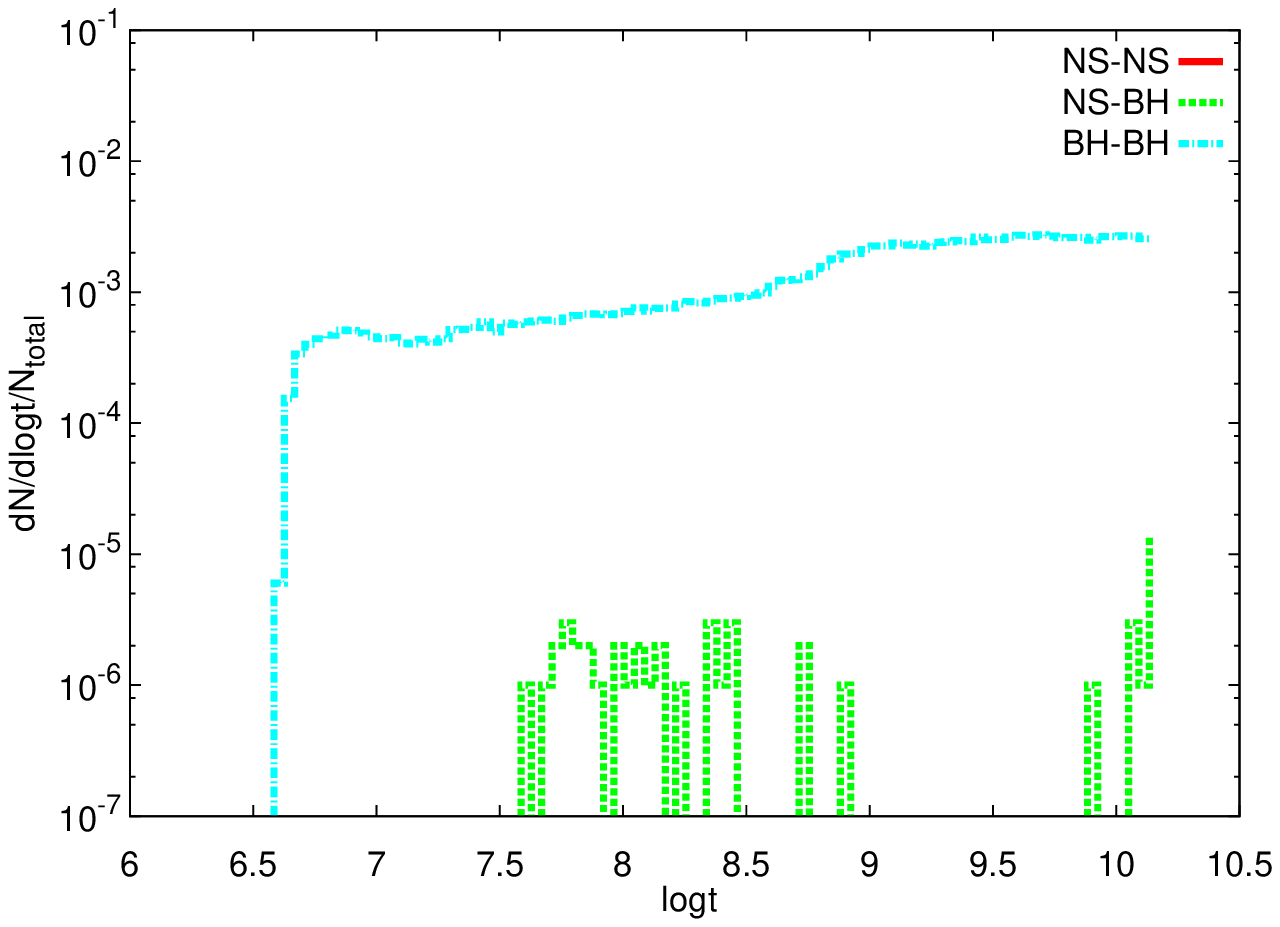} %
\smallskip

~~~~~~~~~~~~~~~~~~~~~~~~~~~~~~~~~~~~~(b) Model III.f
\medskip
  \end{minipage}

  \begin{minipage}{.40\linewidth} %
  \includegraphics[width=1.18 \linewidth]{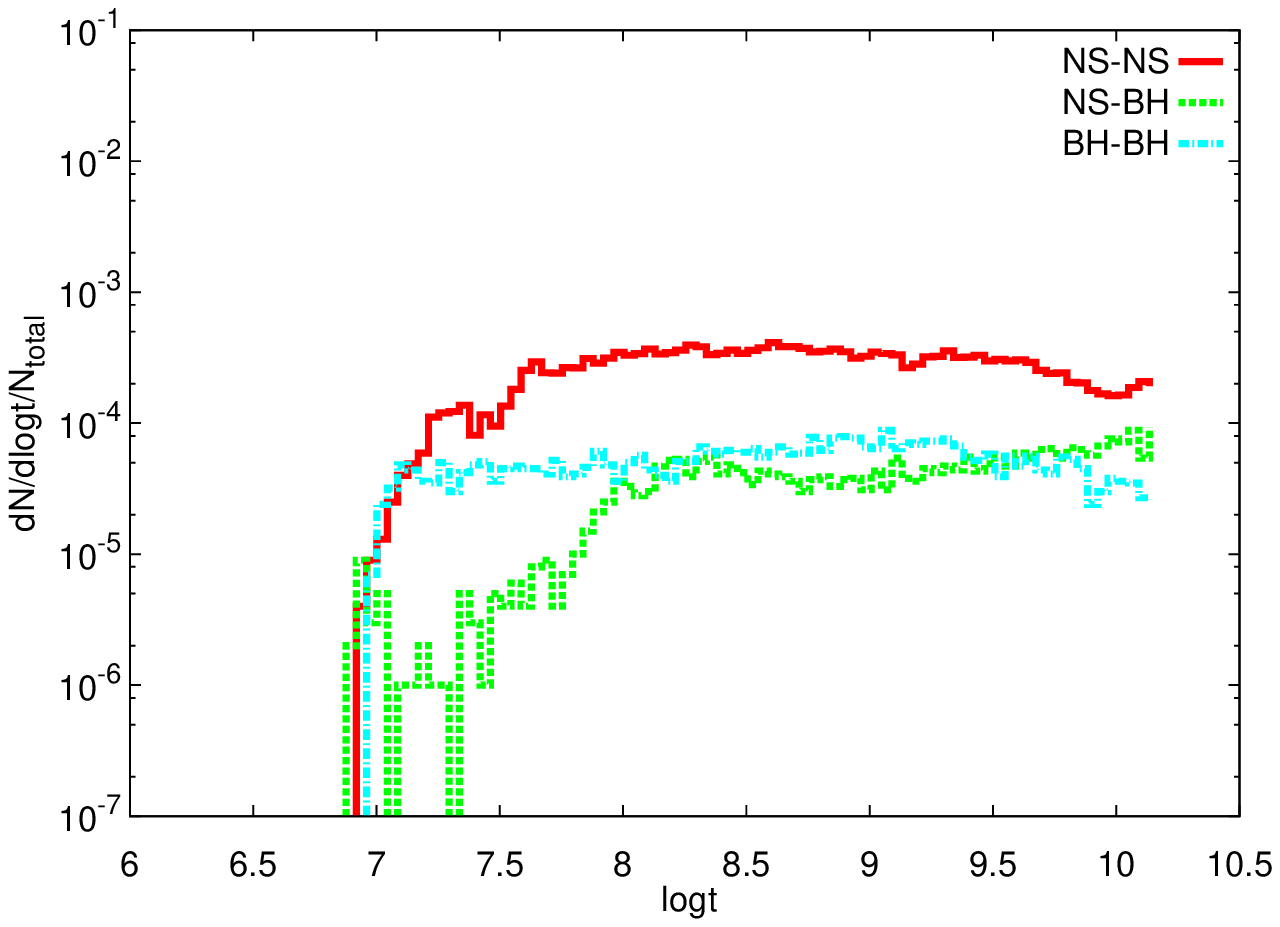} 
\smallskip

~~~~~~~~~~~~~~~~~~~~~~~~~~~~~~~~~~~~~(c) Model I.h

  \end{minipage}
 \hspace{2.5pc} 
  \begin{minipage}{.40\linewidth} %
  \includegraphics[width=1.18 \linewidth]{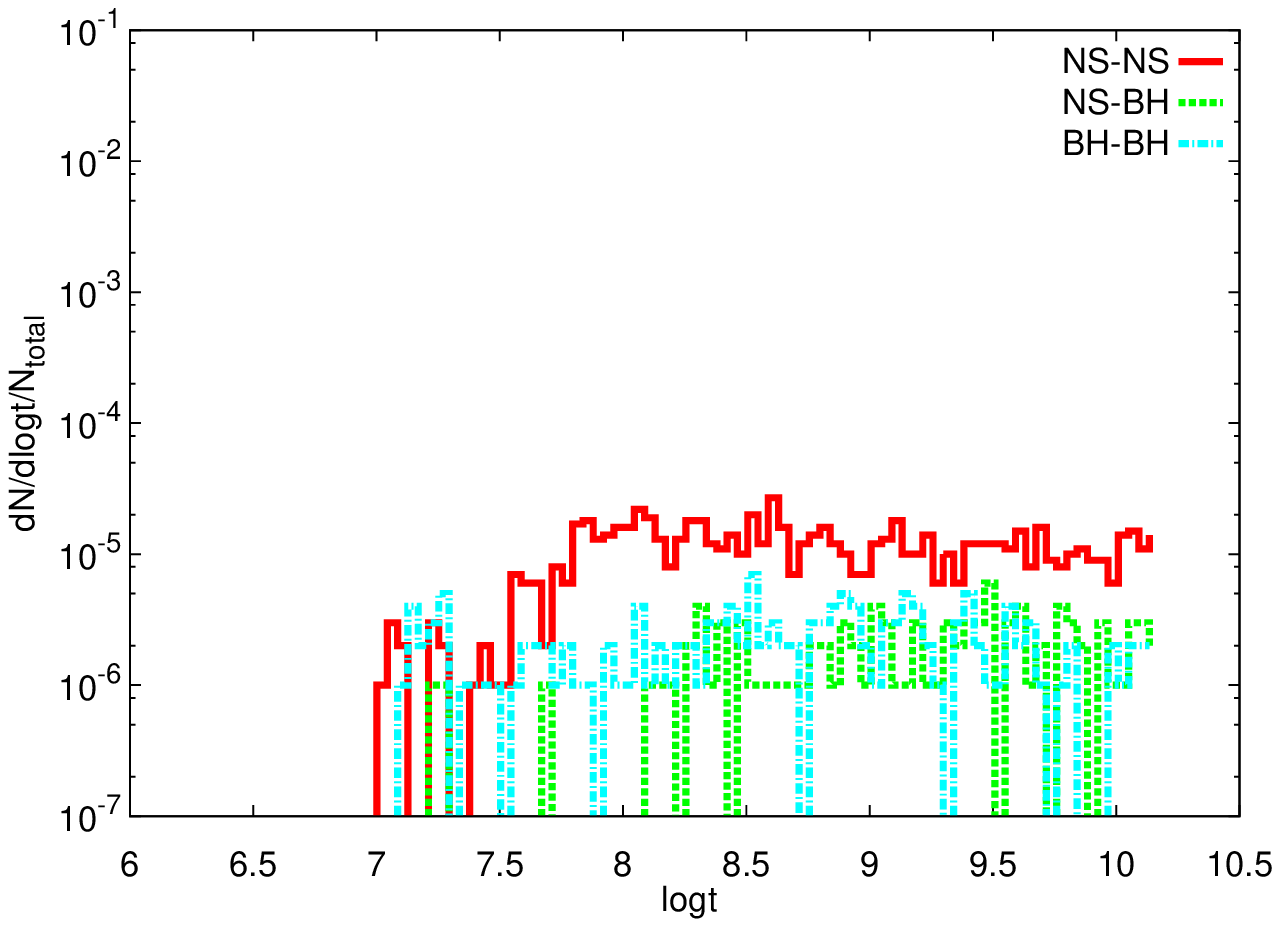} 
\smallskip

~~~~~~~~~~~~~~~~~~~~~~~~~~~~~~~~~~~~~(d) Model I.l

  \end{minipage}

\medskip
\caption{The normalized coalescence time distribution of compact binaries for each model. Each panel corresponds to 
Models III.s (a), III.f (b), I.h (c), and I.l (d), respectively. In each figure, the red, green, and blue lines 
correspond to the NS-NSs, NS-BHs, and BH-BHs, respectively.}
\label{mergertime}
\end{figure*}

The number of the resulting compact binaries formed in each model is listed in Table~2.
The numbers in the parenthesis are the numbers of the resulting compact binaries for the case of the
conservative core merger criterion, which is mentioned in Sec.~\ref{sec:CE}. 
In addition, the number of the compact binaries with coalescence time less than 15~Gyr is shown in Table~3.
For compact binaries which merge within 15~Gyr, we show four tables and three figures to see more details for
each model. In Tables~4,~5,~6 and~7, we describe the formation channels and the
evolution histories of the compact binaries for each model. The abbreviated terms should be referred to the 
caption of Table~4. Figure~\ref{mergertime} gives the distribution of the coalescence time, which is defined as the
time between the birth of the binary and the merger. The normalization of the vertical axis is
$d N/d\log t/N_{\rm total}$. Figures~\ref{mergermass} and \ref{mergercmass} are the distribution of the total
mass ($M_{\rm tot}=M_1+M_2$) and the chirp mass ($M_{\rm chirp}=(M_1M_2)^{3/5}/(M_1+M_2)^{1/5}$) of the compact
binaries, respectively.

\subsection{Pop III compact binaries with the Salpeter IMF}\label{sec:Pop III binary}

We now discuss the evolution of Pop III binaries and their final fate in more detail.
Here we focus on Model III.s as the reference model. 

\subsubsection{Binary black hole}
From Tables~2 and 3, we find that $\sim 58$ \% of Pop III compact binaries are BH-BHs. Remarkably, $\sim 20$ \% 
of the BH-BHs merge within the age of the universe ($\sim$ 15~Gyr). The coalescence time of BH-BHs distributes almost 
uniformly in $\log t$ and its value is $\sim 10\mbox{-}10^2$ times larger than that of NS-BHs and NS-NSs~(Fig.~5).
The total and chirp mass distributions of the coalescing BH-BHs has a peak at $\sim 50\ \msun$~(Fig.~6) and $\rm \sim 30\ 
\msun$~(Fig.~7), respectively. It is worth to note that more than $\sim 50\%$ 
of these BH-BHs does not experience the 
CE phase~(see Table~4) but RLOF.
Therefore, we expect that the uncertainties of the
parameters $\alpha$ and $\lambda$ in the CE phase do not affect the major part of the PopIII BH-BH mergers.

\subsubsection{Neutron star -- black hole binaries}

Although the total number of NS-BHs is comparable to that of BH-BH binaries, 
the number of the coalescing NS-BH within 15 Gyr is very small~(Tables~2 and 3). 
Specifically, only 0.2 \% of NS-BHs merges within 15 Gyr.
In our results, the typical mass of a NS-BH is $(M_{\rm NS}, M_{\rm BH}) = (1.4\ \msun, 30\ \msun$) 
and the typical separation is $10^{4}\mbox{-}10^6\ {\rm R}_\odot$ so that the merging time is much larger than 15Gyr from Eq. (79).
For NS-BH formation, the secondary star evolves into a neutron star via
a supernova explosion after the formation of the black hole of the primary star. 
In order to follow this evolutionary path, the secondary should have  mass less 
than $\sim 50\ \msun$, otherwise the binary disrupts by the sudden mass loss at the supernova explosion.  
However such a star  evolves via a blue supergiant~(BSG) with the radiative
envelope for the Pop III case~(Fig.~1). 
Since a star with radiative envelope typically takes a larger value of
$\zeta_{\rm ad}$ than those with convective envelope, 
such a star is more difficult to satisfy the condition of the onset 
of the CE phase ($\zeta_{\rm ad} < \zeta_{\rm L}$; Sec.~2.2.3). 
Therefore, Pop III binaries which form NS-BH tend to avoid the CE phase. 
 As we described in Sec.~2.2.4, 
if the binary system ejects the mass  comparable to half of the
total mass at the moment of the secondary supernova explosion, 
the separation of the NS-BH is extremely widened. Thus, the coalescence time due to 
the emission of  the gravitational wave tends to be so long that the 
NS-BHs seldom merge within 15~Gyr.

\subsubsection{Binary neutron star}

We find that the NS-NSs are rarely formed and merge within 15 Gyr~(Tables~2 and 3). 
Most Pop III stars with mass $\sim 10-20~\msun$
 evolve to NSs through the BSGs with radiative envelope.
Thus, the Pop III binary which evolves to the NS-NSs hardly  experiences 
the mass-losing processes in the  CE phase. 
As a result, the binary is easily disrupted by ejecting more than half of 
the mass of the binary system at the moment of supernova explosions. 
In Table~3,  there  is  a tiny number of NS-NSs. This comes from the rare binaries who experience common
envelope due to the initial small separation.

\begin{table*}
\caption{The model description for the Monte Carlo simulations. Each column represents the name of the model, 
population of stars, IMF, mass range of the primary star and that of the secondary star, 
respectively. Models III.s and III.f are simulations of Pop III binaries with the mass range of
$10~\msun \le M \le100~\msun$. 
For Models III.s and III.f, the Salpeter and flat IMF is adopetd, respectively.
Models I.h and I.l are simulations of Pop I binaries with Hurley's single stellar fitting formulae
\citep{Hurley2000} for comparison. In both models, the Salpeter IMF is adopted.
For Model I.h, the initial mass range is $10~\msun \le M \le100~\msun$.
For Model I.l, the initial mass range is $1~\msun \le M \le100~\msun$ to take into account the typical mass
of a Pop I star is $\sim 1\msun$.
}
\label{model}
\begin{center}
\begin{tabular}{c c c c c} 
\hline
model & population & IMF & primary mass range& secondary mass range\\ 
\hline
III.s& III & Salpeter &$10~\rm{M}_{\odot}\le M_1\le 100~\rm{M}_{\odot}$&$10~\rm{M}_{\odot}\le M_2\le M_1$\\
III.f &III & Flat  & $10~\rm{M}_{\odot}\le M_1\le 100~\rm{M}_{\odot}$ &$10~\rm{M}_{\odot}\le M_2\le M_1$\\
I.h & I & Salpeter & $10~\rm{M}_{\odot}\le M_1\le 100~\rm{M}_{\odot}$&$10~\rm{M}_{\odot}\le M_2\le M_1$\\
I.l & I & Salpeter & $~1~\rm{M}_{\odot}\le M_1\le 100~\rm{M}_{\odot}$&$0.5~\rm{M}_{\odot}\le M_2\le M_1$\\  \hline
\end{tabular}
\end{center}
\end{table*}
\begin{table*}
\caption{The number of the compact binaries formed in each model. Each column represents the model name, and the
number of NS-NSs, NS-BHs, and BH-BHs, respectively. The numbers in the parenthesis  are for the case of the 
conservative core-merger criterion while  those without the parenthesis are for the case of the optimistic 
core-merger criterion. The definition of optimistic and conservative core-merger criteria are shown in Sec.~\ref{sec:CE}.}
\label{binarynumber}
\begin{tabular}{@{} l c c c} 
\hline
 & NS-NS & NS-BH &BH-BH\\
\hline
Model III.s & $5 (1994)$ & 93085 (93793)  & 132534 (133485) \\
Model III.f & 0 (279) & 185335 (187638) & 517067 (522581) \\
Model I.h &  58724 (60715) & 73193 (76277) & 108184 (108734)  \\
Model I.l & 1847 (1865) & 2264 (2354) & 3559 (3578)\\
\hline
\end{tabular}
\end{table*}
\begin{table*}
\caption{The number of the compact binaries with coalescence time less than 15~Gyr among those in Table~2. Notations
are the same as Table~2.}
\label{mergernumber}
\begin{tabular}{@{} l c c c} 
\hline
 & NS-NS & NS-BH &BH-BH\\
\hline
Model III.s &  5 (1994) & 64  (164)  & 25536 (26468) \\
Model III.f &  0 (279) & 50 (149) & 115056 (120532) \\
Model I.h &  20149 (21155) & 2703 (3664) & 3928 (3976) \\
Model I.l &  776 (785) & 99 (134) & 150 (151) \\
\hline
\end{tabular}
\end{table*}

\begin{table*}
\caption{The formation channels of each compact binaries which merge within 15~Gyr for the case of Model III.s. 
Each column represents the formation channel, the fraction which each channel occupies, and the evolution 
history. Here, RLOF, CE, DCE, SN, CE+SN, and DCE+SN represents the Roche lobe over flow, CE phase, double CE phase, supernova explosion or direct collapse,  supernova explosion or direct collapse as soon as 
after the CE phase, and supernova explosion or the direct collapse as soon as after the double 
CE phase, respectively.}
\label{channel1}
\begin{tabular}{@{} l c  l} 
\hline
 Channel & Fraction & Evolution History\\
\hline
NSNS 1 & 80.0\% (0.4\%) &  SN:1, CE+SN:2 \\
NSNS 2 & 20\% (99.6\%) & RLOF:1$\rightarrow$2, SN:1, CE+SN:2 \\
NSNS others & 0\% (0\%) & The others\\
\hline
NSBH 1 & 86.8\% (90.5\%) & RLOF:1$\rightarrow$2, SN:1, CE+SN:2\\
NSBH 2 & 8.8\% (3.6\%) & RLOF:1$\rightarrow$2, SN:1, RLOF:2$\rightarrow$1, SN:2 \\
NSBH 3 & 2.9\% (1.2\%) &  SN:1, CE+SN:2 \\
NSBH 4 & 1.5\% (0.6\%) & CE:1, SN:1, RLOF:2$\rightarrow$1, SN:2 \\
NSBH 5 & 0\% (1.8\%) &  CE:1, RLOF:1$\rightarrow$2, SN:1, RLOF:2$\rightarrow$1, SN:2 \\
NSBH 6  & 0\% (1.8\%)   & CE+SN:1, RLOF:2$\rightarrow$1, SN:2\\
NSBH others & 0\% (0.5\%) & The others\\
\hline
BHBH 1 & 55.3\% (53.5\%) &  RLOF:1$\rightarrow$2, SN:1, RLOF:2$\rightarrow$1, SN:2 \\ 
BHBH 2 & 13.3\% (12.7\%) & RLOF:1$\rightarrow$2, SN:1, RLOF:2$\rightarrow$1, CE+SN:2 \\
BHBH 3 & 8.1\% (7.9\%) &  RLOF:1$\rightarrow$2, CE+SN:1, RLOF:2$\rightarrow$1, SN:2\\
BHBH 4 & 6.2\% (6.0\%) & CE:1, SN:1, RLOF:2$\rightarrow$1, SN:2\\
BHBH 5 & 5.5\% (5.3\%) & RLOF:1$\rightarrow$2, CE+SN:1, RLOF:2$\rightarrow$1, CE+SN:2\\
BHBH 6 & 2.9\% (2.8\%) & CE:1, SN:1,  RLOF:2$\rightarrow$1, CE+SN:2\\
BHBH 7 & 1.5\% (1.7\%) & CE+SN:1, RLOF:2$\rightarrow$1, SN:2\\
BHBH 8 & 1.3\% (1.3\%) & RLOF:1$\rightarrow$2, DCE+SN:1, SN:2\\
BHBH 9 & 1.1\% (1.1\%) & DCE+SN:1, SN:2\\
BHBH 10 & 1.1\% (1.1\%) & DCE, SN:1, SN:2\\
BHBH 11 & 1.1\% (1.1\%) & RLOF:1$\rightarrow$2, SN:1, CE+SN:2\\ 
BHBH others & 2.6\% (5.5\%) & The others \\
\hline
\end{tabular}
\end{table*}
\begin{table*}
\caption{The same as Table~4, but for Model III.f.}
\label{channel2}
\begin{tabular}{@{} l c  l} 
\hline
 Channel & Fraction & Evolutionary History\\
\hline
NSNS 2 & 0\% (100\%) & RLOF:1$\rightarrow$2, SN:1, CE+SN:2 \\
NSNS others & 0\% (0\%) & The others\\
\hline
NSBH 1 & 54.4\% (52.2\%) & RLOF:1$\rightarrow$2, SN:1, CE+SN:2\\
NSBH 2 & 12.3\% (4.5\%) & RLOF:1$\rightarrow$2, SN:1, RLOF:2$\rightarrow$1, SN:2 \\
NSBH 3 & 1.7\% (0.6\%) &  SN:1, CE+SN:2 \\
NSBH 4 & 28.1\% (10.2\%) & CE:1, SN:1, RLOF:2$\rightarrow$1, SN:2 \\
NSBH 5 & 3.5\% (29.9\%) &  CE:1, RLOF:1$\rightarrow$2, SN:1, RLOF:2$\rightarrow$1, SN:2 \\
NSBH 6  & 0\% (1.9\%)   & CE+SN:1, RLOF:2$\rightarrow$1, SN:2\\
NSBH others & 0\% (0.7\%) &The others\\
\hline
BHBH 1 & 36.9\% (35.4\%) &  RLOF:1$\rightarrow$2, SN:1, RLOF:2$\rightarrow$1, SN:2 \\ 
BHBH 2 & 16.3\% (15.7\%) & RLOF:1$\rightarrow$2, SN:1, RLOF:2$\rightarrow$1, CE+SN:2 \\
BHBH 3 & 8.6\% (8.3\%) &  RLOF:1$\rightarrow$2, CE+SN:1, RLOF:2$\rightarrow$1, SN:2\\
BHBH 4 & 8.5\% (8.2\%) & CE:1, SN:1, RLOF:2$\rightarrow$1, SN:2\\
BHBH 5 & 11.8\% (11.3\%) & RLOF:1$\rightarrow$2, CE+SN:1, RLOF:2$\rightarrow$1, CE+SN:2\\
BHBH 6 & 6.3\% (6.1\%) & CE:1, SN:1,  RLOF:2$\rightarrow$1, CE+SN:2\\
BHBH 7 & 0.8\% (0.9\%) & CE+SN:1, RLOF:2$\rightarrow$1, SN:2\\
BHBH 8 & 2.2\% (2.1\%) & RLOF:1$\rightarrow$2, DCE+SN:1, SN:2\\
BHBH 9 & 1.9\% (1.8\%) & DCE+SN:1, SN:2\\
BHBH 10 & 2.3\% (2.2\%) & DCE, SN:1, SN:2\\
BHBH 11 & 0.8\% (0.8\%) & RLOF:1$\rightarrow$2, SN:1, CE+SN:2\\ 
BHBH others & 3.6\% (7.2\%) & The others \\
\hline
\end{tabular}
\end{table*}
\begin{table*}
\caption{The same as Table 4, but for Model I.h.}
\label{channel3}
\begin{tabular}{@{} l c  l} 
\hline
 Channel & Fraction & Evolutionary History\\
\hline
NSNS 3 & 51.2\% (49.7\%) & RLOF:1$\rightarrow$2, SN:1, CE:2, CE+SN:2 \\
NSNS 4 & 24.7\% (24.0\%) & CE:1, SN:1, CE:2, CE+SN:2\\
NSNS 5 & 6.9\% (8.7\%) & CE:1, CE+SN:1, CE+SN:2 \\
NSNS 6 & 4.5\% (4.4\%) & CE:1, SN:1, CE:2, RLOF:2$\rightarrow$1 CE+SN:2 \\
NSNS 7 & 4.0\% (4.1\%) & DCE, CE+SN:1, CE+SN:2 \\
NSNS others & 8.7\% (9.1\%) & The others\\
\hline
NSBH 7  & 47.7\% (37.7\%) & CE:1, SN:1, CE:2, CE+SN:2  \\
NSBH 8  &13.2\% (13.9\%) & RLOF:1$\rightarrow$2, SN:1, CE:2, SN:2  \\
NSBH 9  & 10.0\%(16.2\%) & CE:1, SN:1, CE:2, SN:2\\
NSBH 10  & 9.1\% (10.8\%) & RLOF:1$\rightarrow$2, SN:1, CE:2, CE+SN:2  \\
NSBH 11 & 7.5\% (5.3\%) & CE:1, SN:1,  RLOF:2$\rightarrow$1, CE+SN:2\\
NSBH others & 12.5\% (16.1\%) & The others\\
\hline
BHBH 8 & 6.2\% (6.0\%) & RLOF:1$\rightarrow$2, DCE+SN:1, SN:2\\
BHBH 12 &80.5\% (78.4\%) &  RLOF:1$\rightarrow$2, SN:1, CE:2, SN:2  \\
BHBH 13 & 8.4\% (9.1\%) & DCE+SN:1, RLOF:2$\rightarrow$1, SN:2 \\
BHBH others & 4.9\% (6.5\%) & The others \\
\hline
\end{tabular}
\end{table*}
\begin{table*}
\caption{The same as Table~4, but for Model I.l.}
\label{channel4}
\begin{tabular}{@{} l c  l} 
\hline
 Channel & Fraction & Evolutionary History\\
\hline
NSNS 3 & 66.9\% (66.4\%) & RLOF:1$\rightarrow$2, SN:1, CE:2, CE+SN:2 \\
NSNS 4 & 19.9\% (19.7\%) & CE:1, SN:1, CE:2, CE+SN:2\\
NSNS 5 & 1.1\% (1.4\%) & CE:1, CE+SN:1, CE+SN:2 \\
NSNS 6 & 3.8\% (3.7\%) & CE:1, SN:1, CE:2, RLOF:2$\rightarrow$1 CE+SN:2 \\
NSNS 7 & 1.5\% (1.5\%) & DCE, CE+SN:1, CE+SN:2 \\
NSNS others & 7.8\% (7.3\%) & The others\\
\hline
NSBH 7  & 51.3\% (43.5\%) & CE:1, SN:1, CE:2, CE+SN:2  \\
NSBH 8  &10.1\% (12.4\%) & RLOF:1$\rightarrow$2, SN:1, CE:2, SN:2  \\
NSBH 9  & 10.9\%(18.0\%) & CE:1, SN:1, CE:2, SN:2\\
NSBH 10  & 6.7\% (6.2\%) & RLOF:1$\rightarrow$2, SN:1, CE:2, CE+SN:2  \\
NSBH 11 & 6.7\% (7.5\%) & CE:1, SN:1,  RLOF:2$\rightarrow$1, CE+SN:2\\
NSBH others & 14.3\% (12.4\%) & The others\\
\hline

BHBH 12 & 80.0\% (76.7\%) &  RLOF:1$\rightarrow$2, SN:1, CE:2, SN:2  \\
BHBH 14  & 11.5\% (12.2\%) & RLOF:1$\rightarrow$2, SN:1, CE:2, CE+SN:2  \\
BHBH others & 18.5\% (11.1\%) & The others \\
\hline
\end{tabular}
\end{table*}

 \begin{figure*}
  \begin{minipage}{.40\linewidth} 
 \includegraphics[width=1.18 \linewidth]{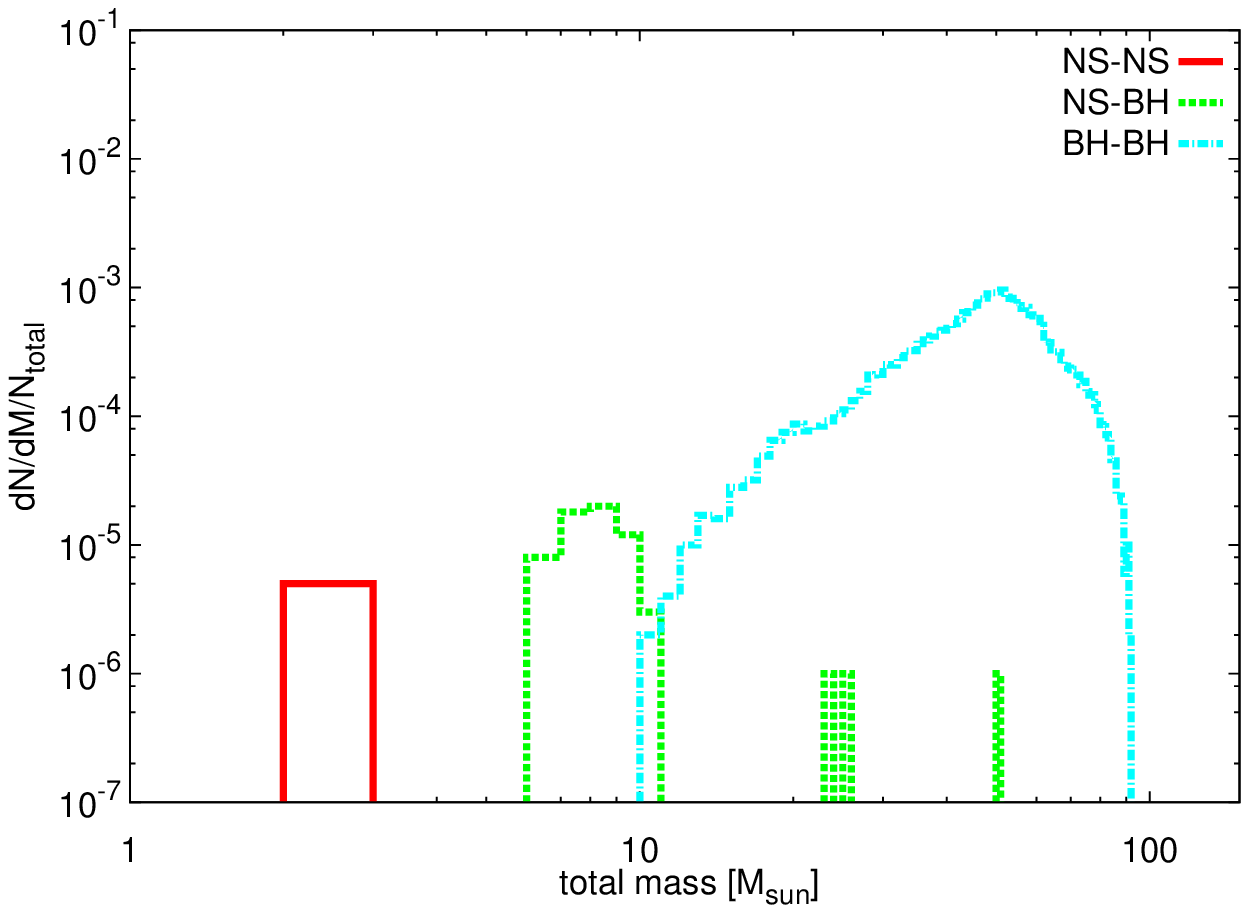} 
\smallskip

~~~~~~~~~~~~~~~~~~~~~~~~~~~~~~~~~~~~~(a) Model III.s
\medskip

  \end{minipage}
  \hspace{2.5pc} 
  \begin{minipage}{.40\linewidth} 
  \includegraphics[width=1.18 \linewidth]{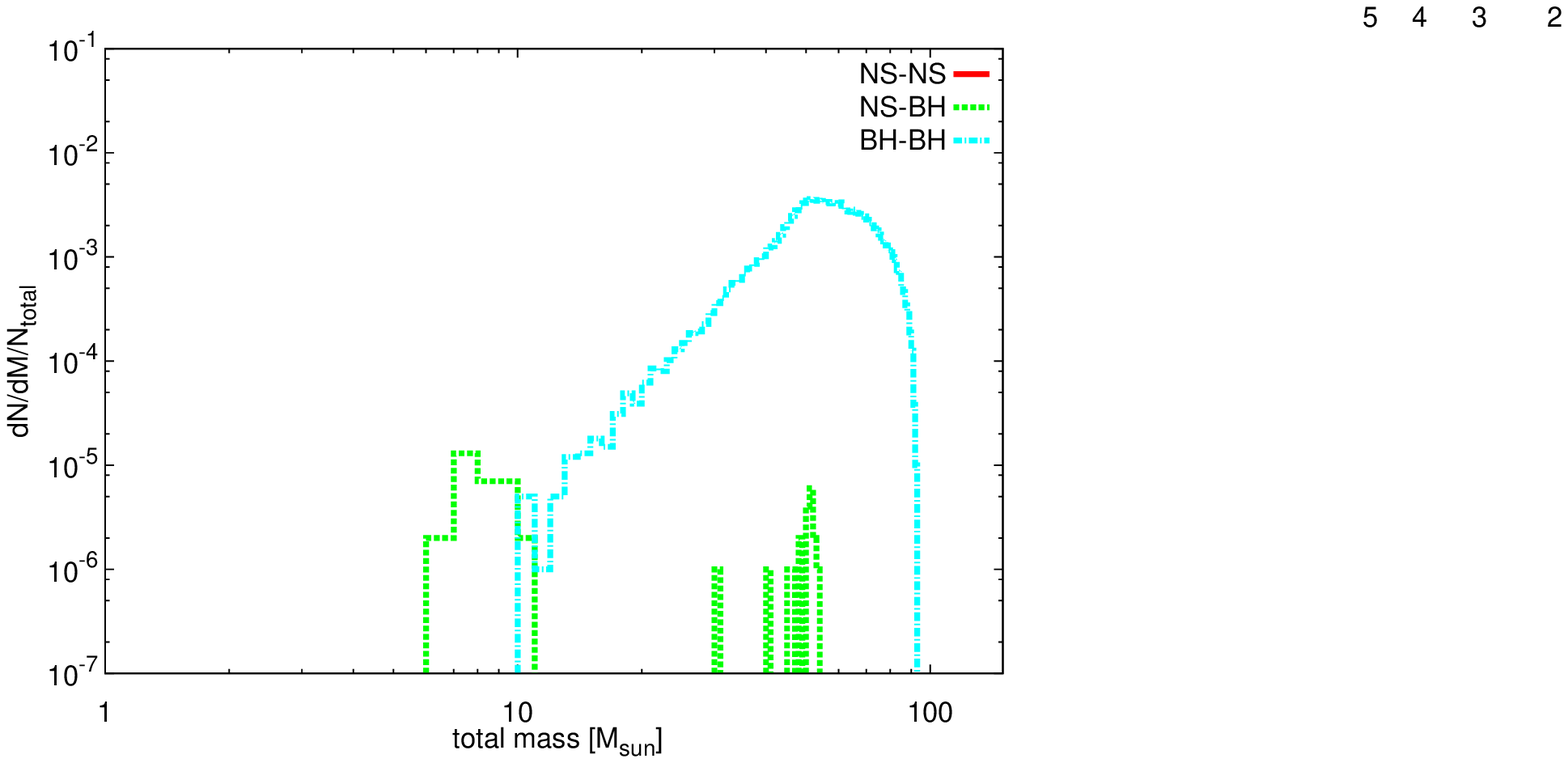} 
\smallskip

~~~~~~~~~~~~~~~~~~~~~~~~~~~~~~~~~~~~~(b) Model III.f
\medskip
  \end{minipage}

  \begin{minipage}{.40\linewidth} 
  \includegraphics[width=1.18 \linewidth]{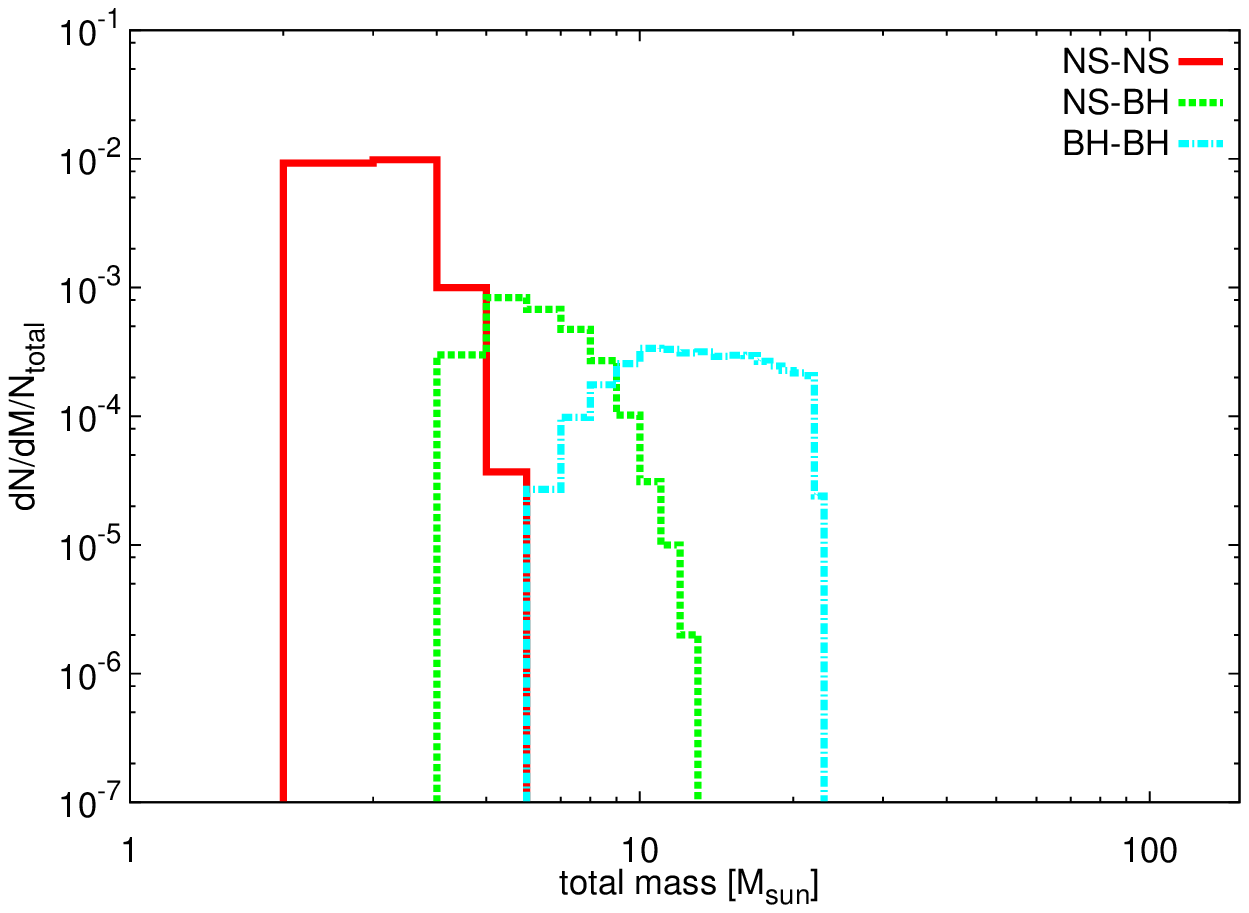} %
\smallskip

~~~~~~~~~~~~~~~~~~~~~~~~~~~~~~~~~~~~~(c) Model I.h

  \end{minipage}
 \hspace{2.5pc}
  \begin{minipage}{.40\linewidth} 
  \includegraphics[width=1.18 \linewidth]{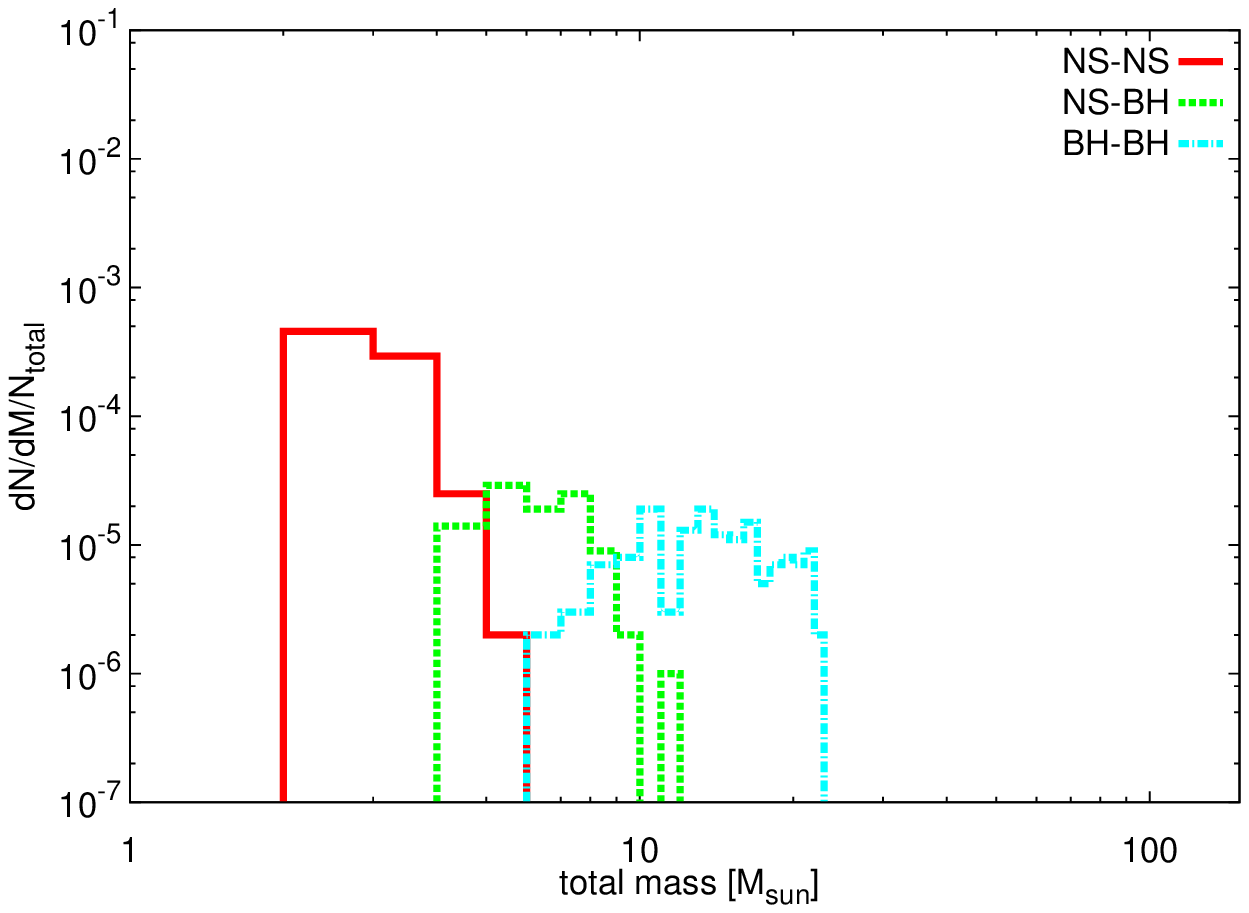}            
\smallskip

~~~~~~~~~~~~~~~~~~~~~~~~~~~~~~~~~~~~~(d) Model I.l

  \end{minipage}
\medskip
\caption{The normalized distribution of the total mass ($M_{\rm tot}=M_1+M_2$) of compact binaries for each model
. Each panel corresponds to Models III.s (a), III.f (b), I.h (c), and I.l (d), respectively. In each figure, the
 red, green, and blue lines correspond to the NS-NSs, NS-BHs, and BH-BHs, respectively.}
\label{mergermass}
\end{figure*}

 \begin{figure*}
  \begin{minipage}{.40\linewidth} 
 \includegraphics[width=1.18 \linewidth]{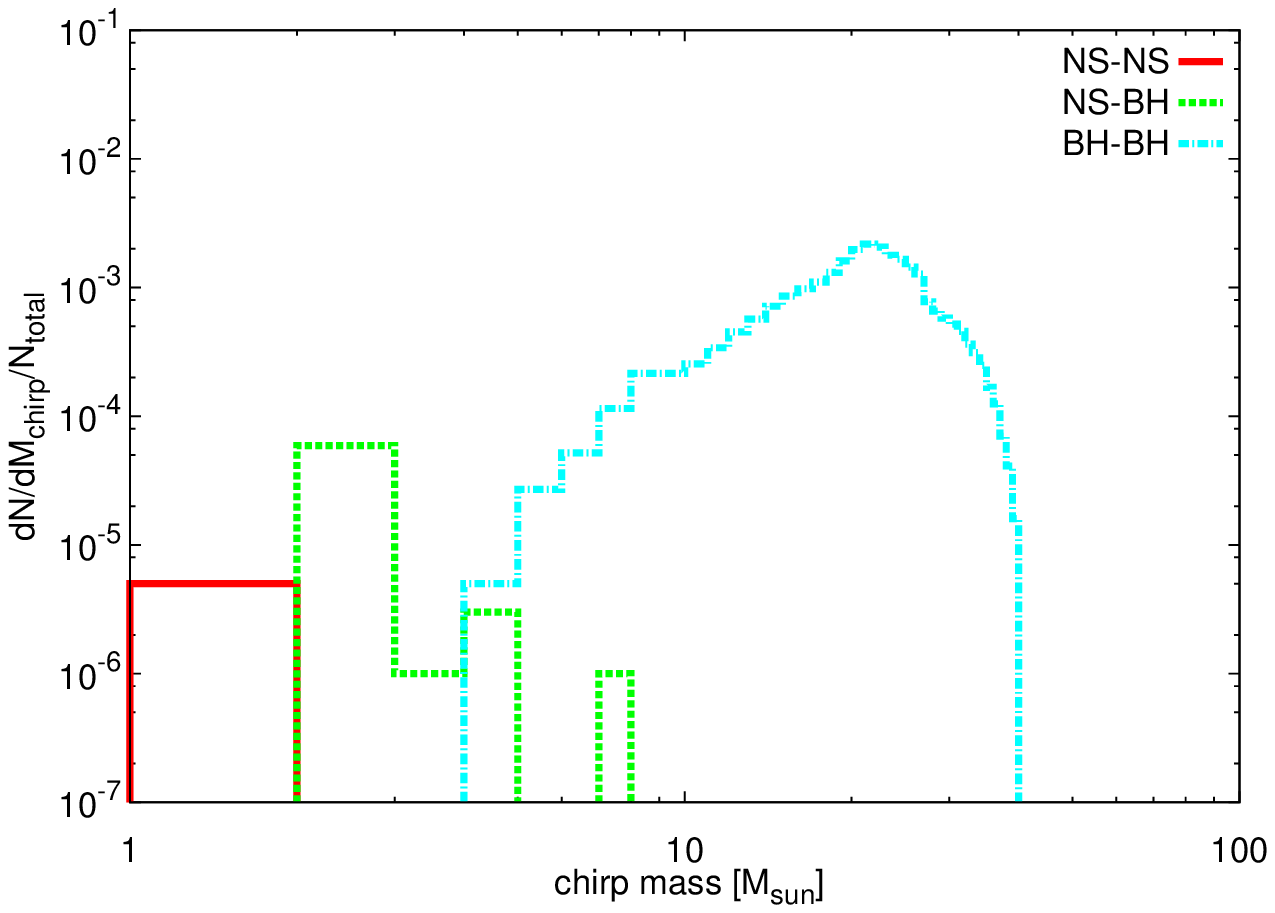}  %
\smallskip

~~~~~~~~~~~~~~~~~~~~~~~~~~~~~~~~~~~~~(a) Model III.s
\medskip

  \end{minipage}
  \hspace{2.5pc} 
  \begin{minipage}{.40\linewidth}
  \includegraphics[width=1.18 \linewidth]{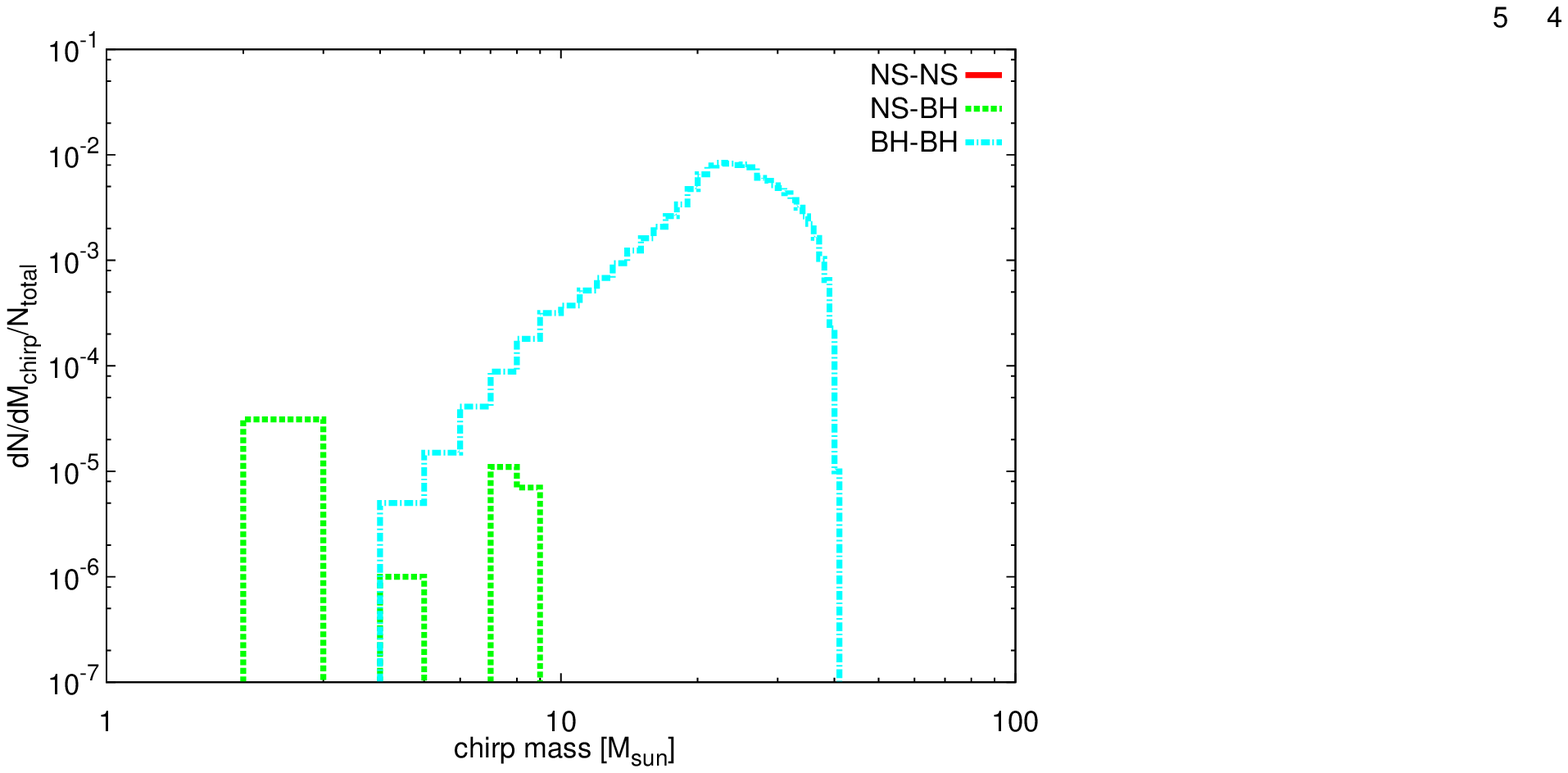} %
\smallskip

~~~~~~~~~~~~~~~~~~~~~~~~~~~~~~~~~~~~~(b) Model III.f
\medskip
  \end{minipage}

  \begin{minipage}{.40\linewidth} 
  \includegraphics[width=1.18 \linewidth]{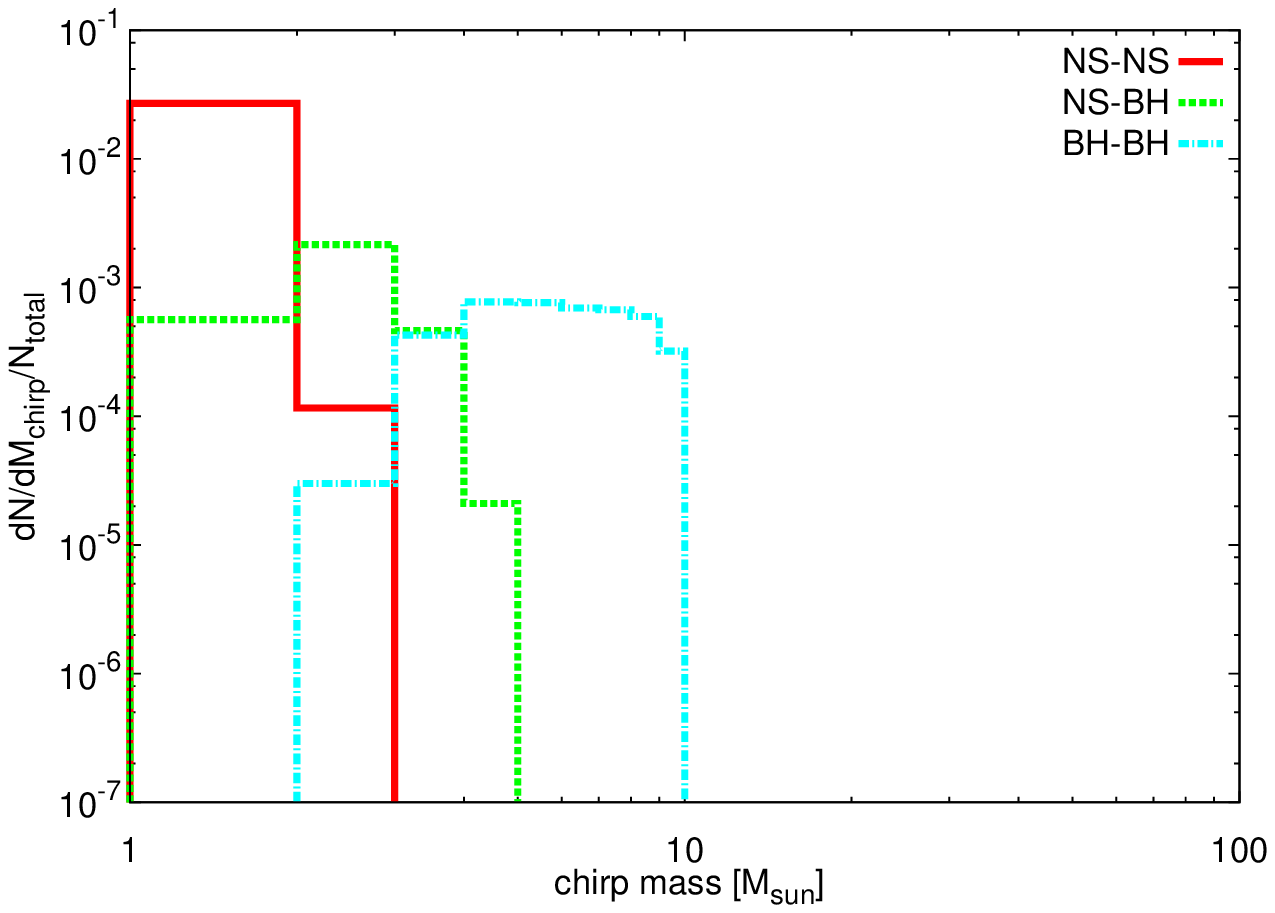} 
\smallskip

~~~~~~~~~~~~~~~~~~~~~~~~~~~~~~~~~~~~~(c) Model I.h

  \end{minipage}
 \hspace{2.5pc} 
  \begin{minipage}{.40\linewidth} 
  \includegraphics[width=1.18 \linewidth]{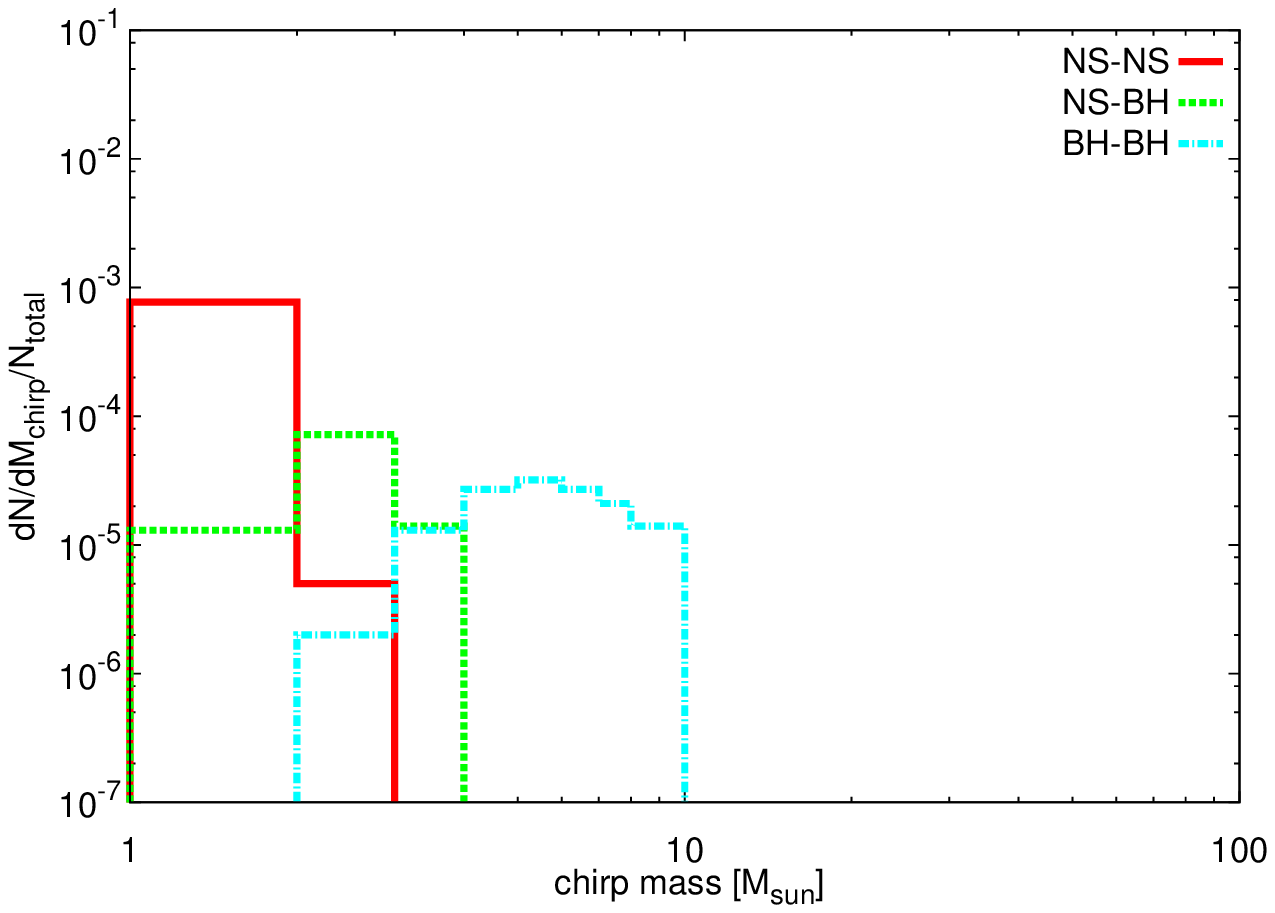}           
\smallskip

~~~~~~~~~~~~~~~~~~~~~~~~~~~~~~~~~~~~~(d) Model I.l

  \end{minipage}
\medskip
\caption{The same as Fig.~\ref{mergermass}, but for distribution of the chirp mass ($M_{\rm chirp}=(M_1M_2)^{3/5
}/(M_1+M_2)^{1/5}$).
}
  \label{mergercmass}
\end{figure*}

\subsection{Model comparisons}
\label{sec:comparison models}
\subsubsection{Pop III with different IMFs}
From the first two rows of Tables~3, we see that the rarity of the  
NS-NSs and NS-BHs which merge within 15Gyr  are similar for Model III.s and Model III.f. The  reason is the same as  we discussed in the 
previous subsection. 
On the other hand, for BH-BHs, these numbers are several times larger
in the flat IMF case~(Model III.f). This is because the number of massive progenitors forming BHs is larger in 
the flat IMF than that in the Salpeter IMF. This feature can also be seen in the distribution of the coalescence time,
the total mass, and the chirp mass~(Figs.~5--7). While the peak mass is the same between
the two IMFs, the fraction of massive stars above the peak becomes larger in the flat IMF.

The formation channels of BH-BHs also reflect the difference of IMFs~(See Table~\ref{channel1} and
\ref{channel2}). Here, we focus on the channel BHBH1, which has nothing to do with the CE phase.
In Model III.s, the channel BHBH1 occupies about 54~\% of all, while in Model III.f, it does about 35~\%.
The reason is that the fraction of massive stars with $\ga 50\ \msun$, which evolve into red supergiants~(RSG;
Fig.~1), is larger in the flat IMF. Since the stars with convective envelope like RSGs have smaller values of
$\zeta_{\rm ad}$ and are easier to satisfy the condition of the onset of the CE phase~($\zeta_{\rm ad} < \zeta_{\rm L}$;
See Sec.~\ref{sec:CE}).

Note that in our calculation, if the stellar mass exceeds 100~$\rm{M}_{\odot}$, the binary evolution
is stopped since the numerical results of \cite{Marigo2001}, and thus, our fitting formulae are given for stars
only up to $100\ \msun$. Therefore, for the binaries with $M_{\rm{total}}>100\ \msun$, our result is an underestimated one.  
We are planning to cover this mass range in  future.

\subsubsection{The differences between Pop III and Pop I}
\subsubsection*{(1) Same initial mass range}
The stellar evolution of Pop III stars is entirely different from that of Pop I stars
as we describe in Sec.~2.1.1.
In the Pop III case, stars more massive than $\ga 50\ \msun$ evolve into RSGs 
and those with $\la 50\ \msun$ evolve into BSGs with radiative envelope~(Fig.~1).
On the other hand, in the Pop I case all stars evolve into RSGs with deep convective envelope.
Therefore, some fractions of Pop III binaries avoid the CE phase.
Here, by comparing Models III.s and I.h, which have the same IMF and mass range, we clarify how the difference in
stellar population affects the formation and coalescence of compact binaries. The clear differences between them
can be seen in the distributions of the coalescence time, total mass, and chirp mass of compact binaries in Figs.~5--7.
In particular, for Pop I, the number of the merging  NS-NSs is the largest, while for Pop III, that of BH-BHs
(Figs.~5--7).

First, we focus on NS-NSs.  From Tables~2 and~3, in Model III.s, much smaller number of NS-NSs 
are formed and merge within 15~Gyr than those in Model I.h. This comes from the fact that Pop III binaries
lose a smaller amount of mass from the system by  the stellar wind
and the mass ejection in the CE phase than Pop I. Therefore Pop III binaries are easier to be disrupted or
separated further away by losing the mass of the system at supernova explosions.

Secondly, the number of NS-BHs formed in Model III.s is almost 
the same as that in Model I.h~(Table~2).
However, in Model III.s, the number of coalescing NS-BHs is 
much smaller than that in Model I.h~(Table~3). 
The reason is the same as the NS-NS case: the major fractions of 
Pop III binaries are separated further away by ejecting some fraction 
of the mass from the system when  the supernova explosion occurs  in the secondary.

Thirdly, apart from the previous two cases, the number of coalescing BH-BHs in Model III.s is much
larger than that in Model I.h~(Table~3). The reason is as follows. Firstly, Pop III binaries which evolve into
BH-BH binaries lose  little mass from the system before the black hole formation and the resultant BH-BHs
typically become more massive than the Pop I cases~(see Figs.~\ref{mergermass} and~\ref{mergercmass}).
Since the coalesce time due to the emission of the gravitational wave  is proportional to
$(M_1 M_2 M_{\rm total})^{-1}$~(Eq.~79), even BH-BHs with larger separations can merge within 15~Gyr 
for the Pop III cases. Secondly, large fraction of Pop III binaries which evolve into BH-BHs avoid the
CE phase, where the separation is decreased and even core-merger occurs before the compact binary
formation. While Pop I binaries with small orbital separations merge before the compact binary formation, 
Pop III binaries do not suffer from such merging even with small orbital separations.  Thus, the number of the coalescing
Pop III BH-BHs is much larger than that of Pop I case.

\subsubsection*{(2) Different initial mass range}
Finally, we briefly mention the results of Model I.l, which are basically the same as Model I.h.
Comparing the 3rd and 4th rows of Table~2, we find that the number of compact binaries formed in Model I.l is 
$\sim 3$ \% of that in Model I.h for each compact binary. This is consistent with the ratio of the number of
massive stars which form compact objects in $10~\rm{M}_{\odot}\le M_1\le 100~\rm{M}_{\odot}$ to that of  $1~\rm{M}_{\odot}\le M_1\le 100~\rm{M}_{\odot}$~($\sim 4$ \%).
Moreover, the ratio of the number of coalescing compact binaries to the total number of compact binaries is 
the same in both models for each compact binary: $\sim 30$ \% for NS-NSs, $\sim 4$ \% for NS-BHs, and $\sim 4$ \% for
BH-BHs~(Tables~2 and~3). It should also be noted that despite the small number of massive stars, the
number of coalescing NS-NSs in Model I.l is larger than or comparable to that in Model III.s.

In these calculations, we do not take into account the angular momentum transfer due to the magnetic braking
even for Models I.h and I.l. This is because we would like to clarify the qualitative difference between the two
populations, by comparing the formation channels of Pop III and Pop I compact binaries under the same conditions.
\\
\\

\section{Pop III compact binary merger rate}\label{sec:merger}

\subsection{Star formation rate of Pop III stars}\label{sec:sfr}

In order to calculate the merger rates and history of compact binaries formed in the early universe, the 
information about Pop III star formation rates (SFRs) is needed. 
We here adopt the SFR calculated by a semi-analytical approach \citep{Souza2011}, 
in which the following three effects are taken into account: 
(1) effect of the radiative feedback on Pop III star formation, 
(2) inhomogeneous reionization of the intergalactic medium~(IGM), and 
(3) chemical evolution of the IGM.

Pop III stars (=the stars without heavy metal) are categorized into two types; 
Pop III.1 and Pop III.2 stars. 
Pop III.1 stars are the {\it very} first stars
\citep{Tegmark1997,  Bromm2002,  Abel2002, Yoshida2006}. 
On the other hand, the Pop III.2 stars are the second generation of stars 
born from pristine gas affected by the some feedbacks from earlier stars, 
e.g., ultraviolet radiations and supernovae (SNe) \citep{Johnson2006,Yoshida2007b}. 
We take into account the contributions from both types of Pop III stars in the Pop III SFR.

\begin{figure*}
\centering
\includegraphics[width=12cm]{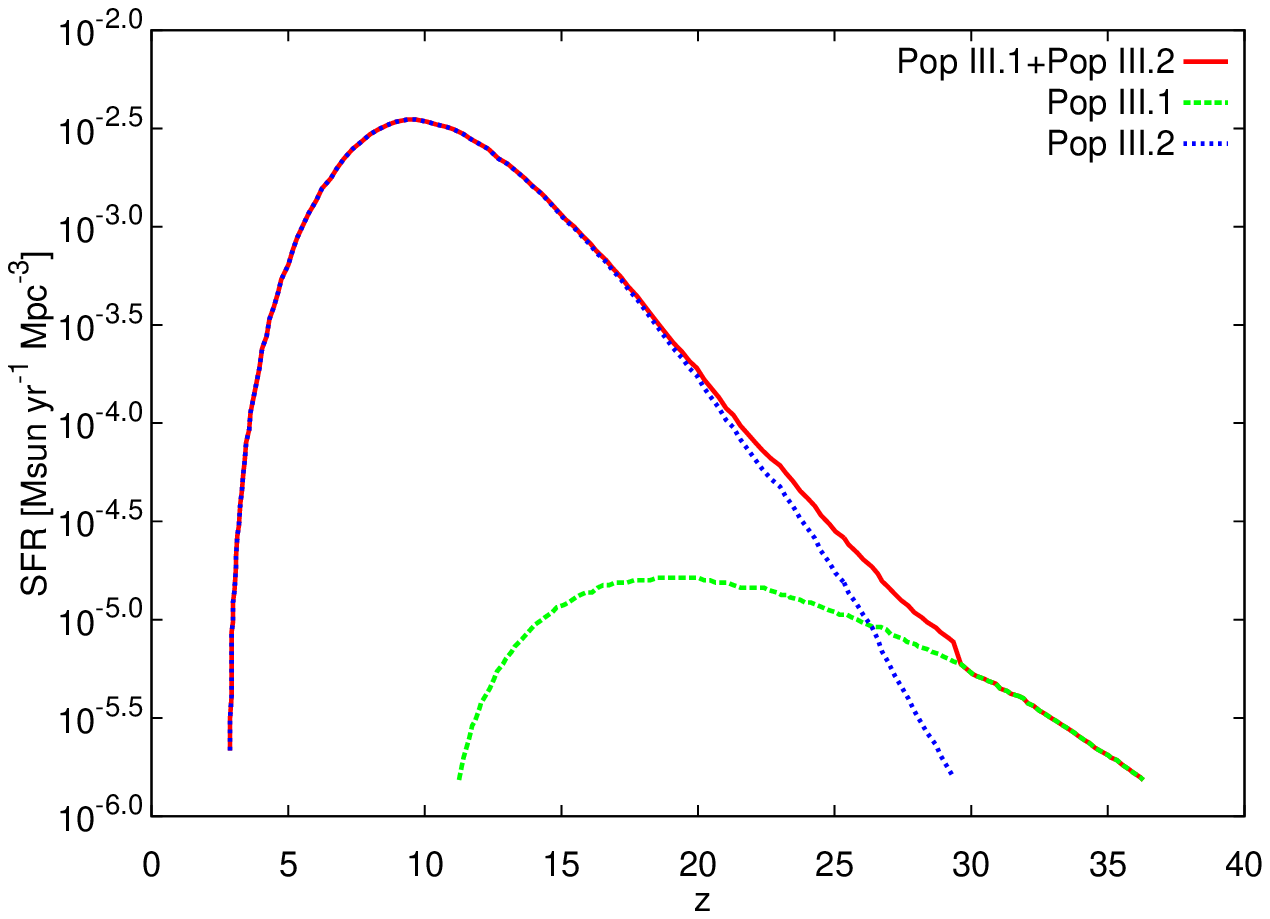}
\caption{The star formation rate density (comoving) calculated by de Souza et al. (2011). The unit of the rate is $\msun$ per comoving volume per proper time. The red line is the the total SFR density of Pop III stars, and the green and blue lines are those of Pop III.1 and Pop III.2 stars, respectively.}
\label{sfr}
\end{figure*}

The metal enrichment in the IGM is also important.
In the early universe, the metal pollution is mainly driven by the Pop III SNe
\citep[e.g.,][]{Madau2001}.
Since the mechanisms of metal pollution are highly uncertain.
\cite{Souza2011} assumes that the metal enrichment proceeds
until the region where the galactic outflows have reached.
In the polluted gas, low mass Pop II stars are expected to be formed
because of the efficient metal and dust cooling
\citep{Omukai2005,  Schneider2006,  Dopcke2013}.
In their model, the Pop III star formation turns off
in metal-enriched regions by the galactic winds.
The star formation rate density (comoving) is shown in figure \ref{sfr}. 
The red line is the the total SFR density of Pop III stars, and the green and blue lines are those of Pop III.1 and Pop III.2 stars, respectively.

\subsection{Pop III compact binary merger rates}\label{sec:merger rates}

\begin{figure*}
  \begin{minipage}{.40\linewidth}
  \includegraphics[width=1.18 \linewidth]{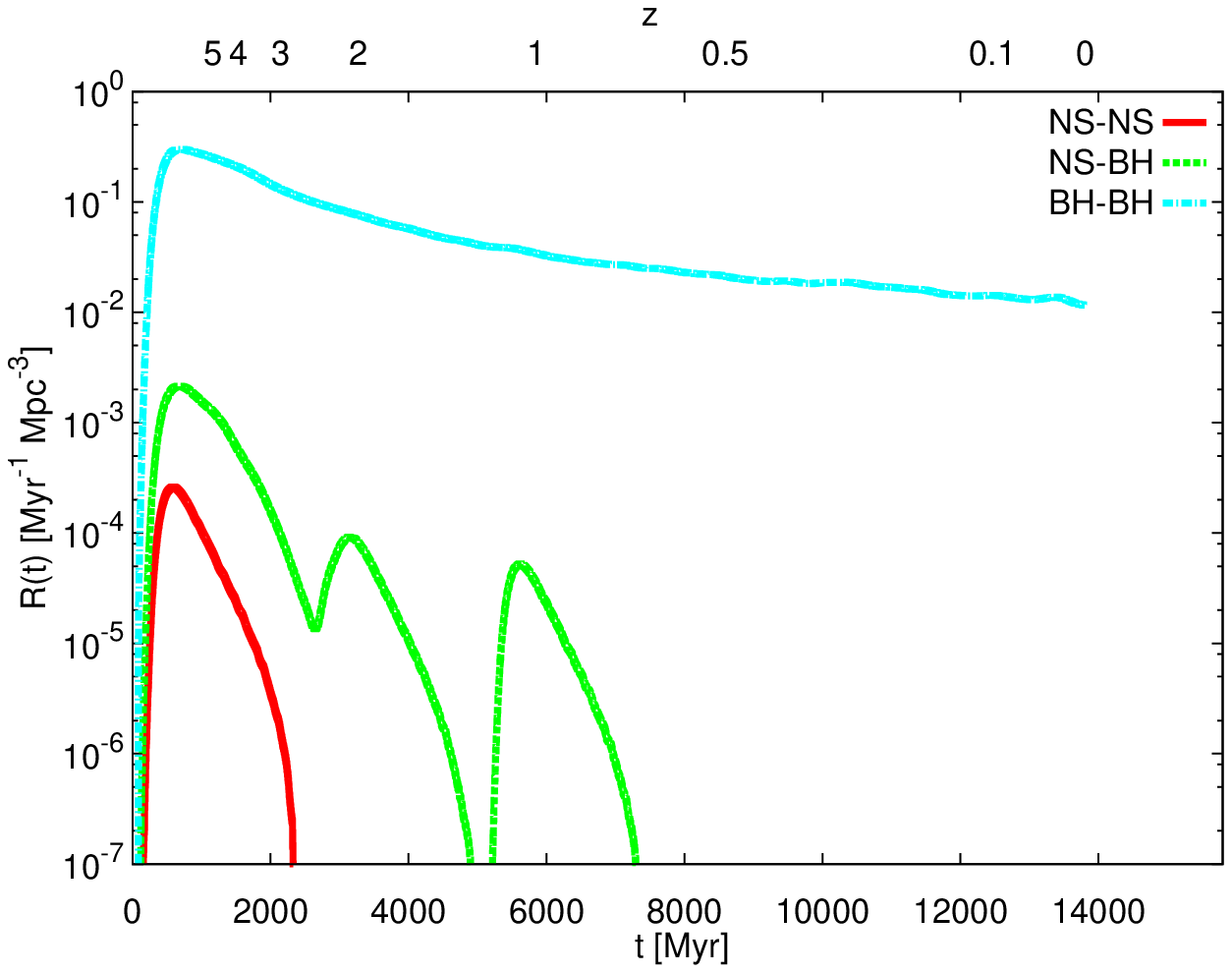} 
\smallskip

~~~~~~~~~~~~~~~~~~~~~~~~~~~~~~~~~(a) Model III.s (conservative)
\medskip
\medskip
\medskip
  \end{minipage}
  \hspace{2.5pc} 
  \begin{minipage}{.40\linewidth} 
   \includegraphics[width=1.18 \linewidth]{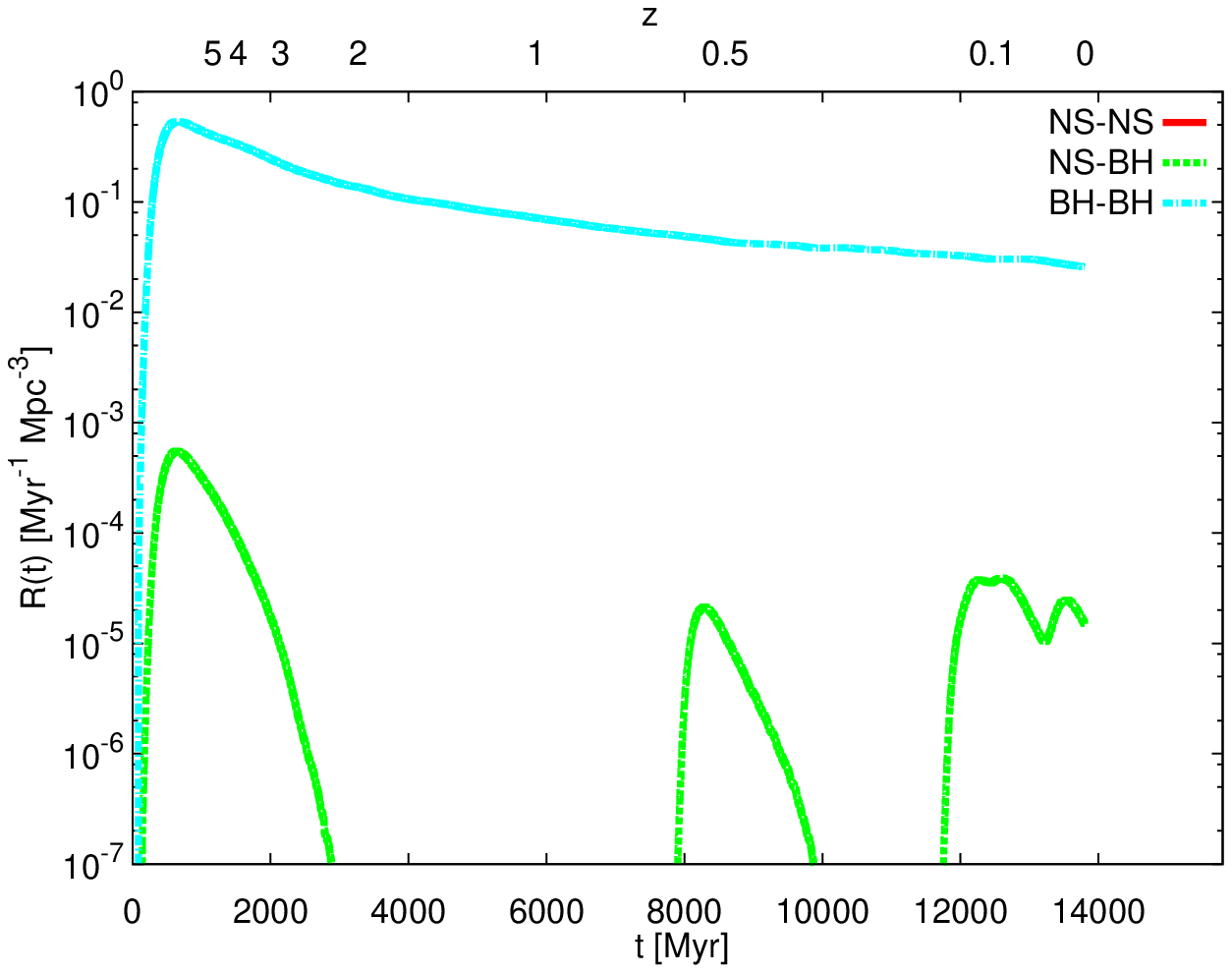}
\smallskip

~~~~~~~~~~~~~~~~~~~~~~~~~~~~~~~~~(b) Model III.f (conservative)
\medskip
\medskip
\medskip
\end{minipage}
  \begin{minipage}{.40\linewidth} 
  \includegraphics[width=1.18 \linewidth]{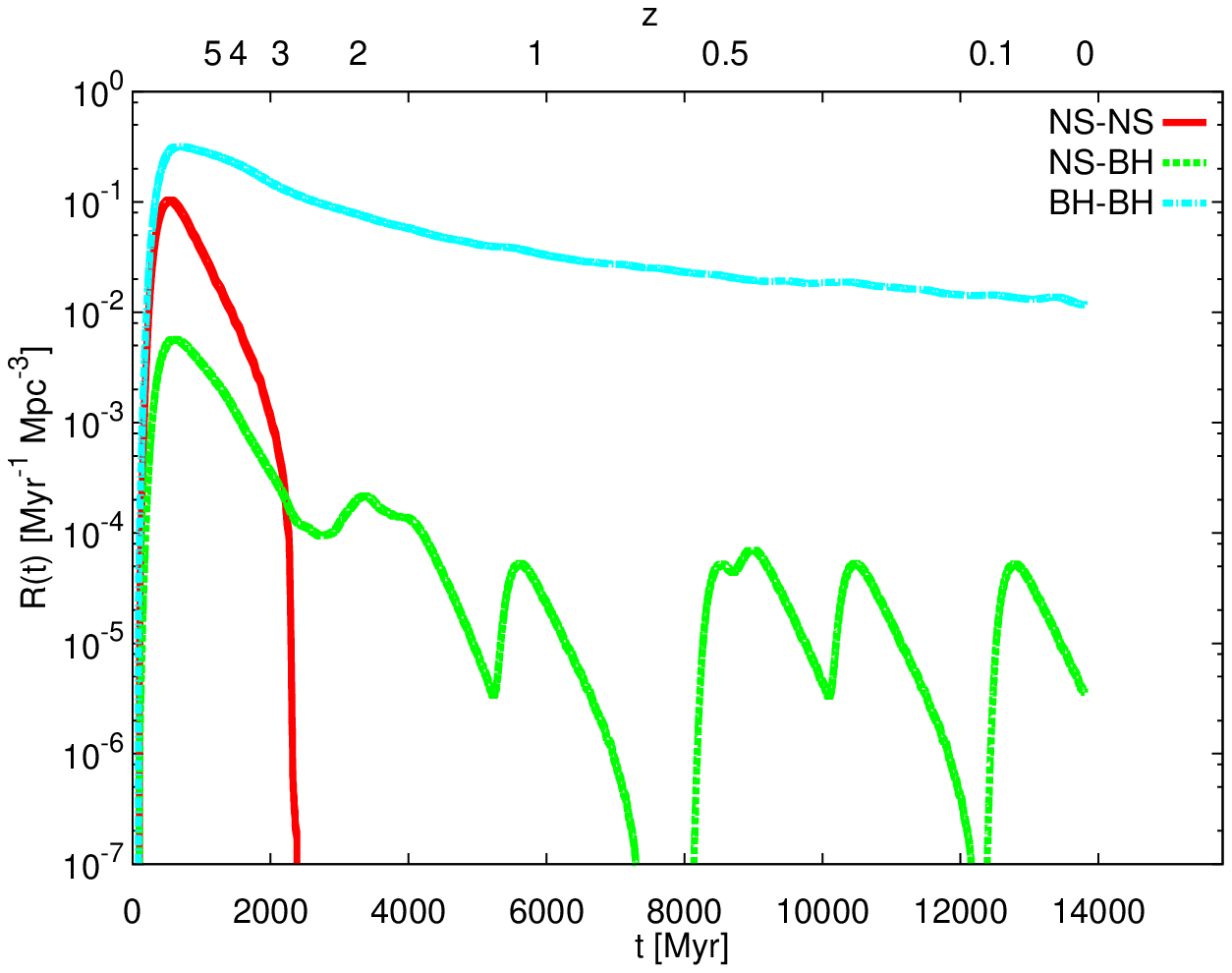} 
\smallskip

~~~~~~~~~~~~~~~~~~~~~~~~~~~~~~~~~(c) Model III.s (optimistic)
  \end{minipage}
  \hspace{2.5pc} 
  \begin{minipage}{.40\linewidth}
   \includegraphics[width=1.18 \linewidth]{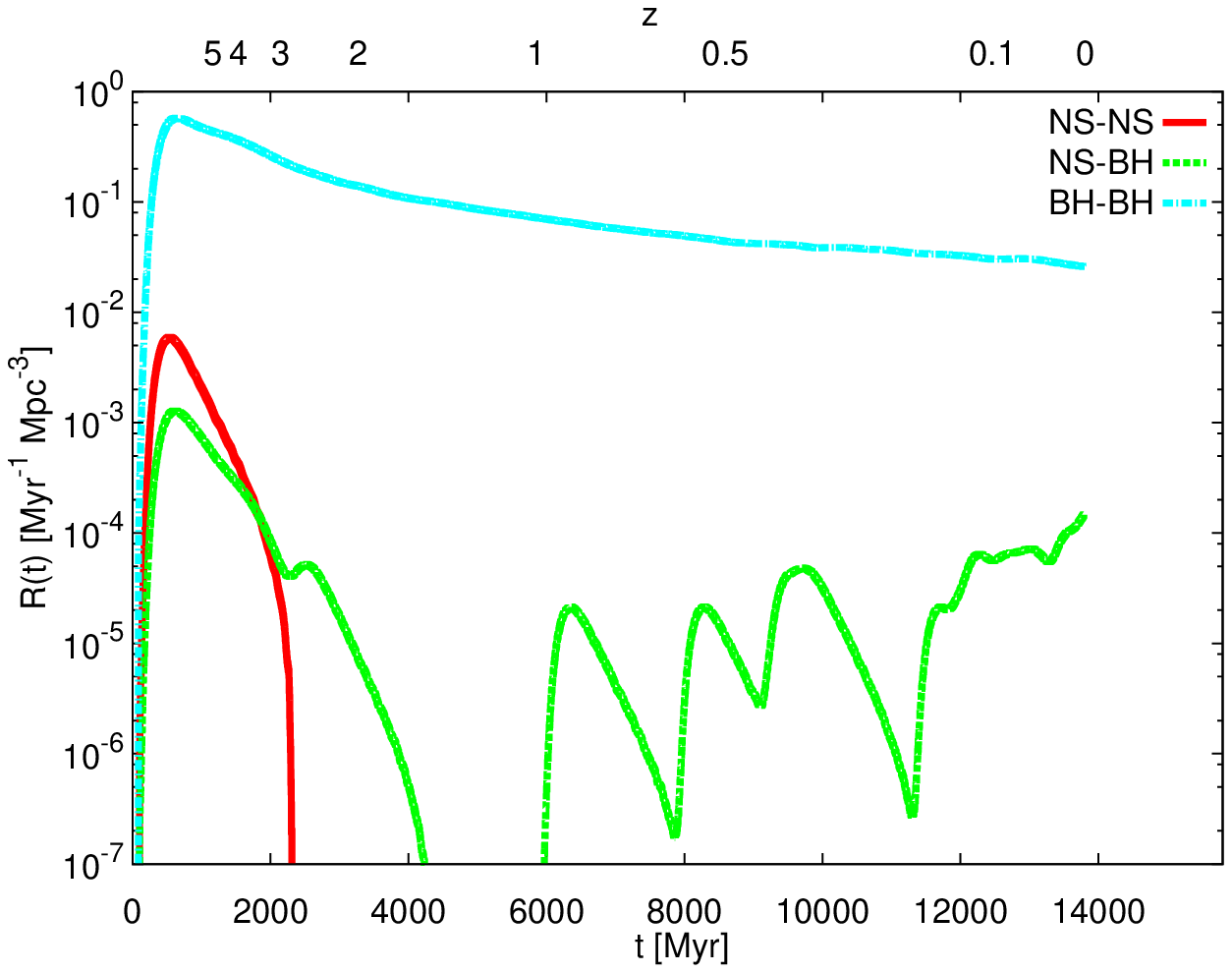}
\smallskip

~~~~~~~~~~~~~~~~~~~~~~~~~~~~~~~~~(d) Model III.f (optimistic)
\end{minipage}
\caption{The evolution of merger rate density of each compact binary. The horizontal axis is the cosmic time.} In this figure, the left panels are Model
III.s (Salpeter IMF) with conservative (a) and optimistic (c) core-merger criteria described in Sec.~2.2.3. The
right panels, (b) and (d) are the same as left ones but for Model III.f (flat IMF). In each panel, the red, 
green, and blue lines correspond to the NS-NSs, NS-BHs, and BH-BHs, respectively.
{As for the reason for multiple peaks of NS-BHs merger rate, the number of coalescing NS-BHs is very small as shown in Table 3 and Fig. 5 so that the coalescence time
distribution of NS-BHs is highly fluctuated. Thus, this fluctuation makes multiple
peaks in Fig. 9. If we increase the number of binary from $10^6$ to $10^7$, we 
can expect that
NS-BH merger rate will be more smooth curve.}
  \label{mergerrate}
\end{figure*}

In this subsection, we show the history of the merger rate density of Pop III compact binaries in the universe.
The merger rate density is calculated using the results in the previous section and the Pop III SFR density
described in Sec.~\ref{sec:sfr}. We define $R_{\rm{i}}(t)~\rm{[Myr^{-1}~Mpc^{-3}]}$ as the merger rate density
at a certain age of the universe $t$ and calculate it from
\begin{equation}
R_{\rm{i}}(t)=\int^{t}_{0} f_{\rm{b}}\frac{{\rm SFR}(t')}{\langle M\rangle}\frac{N_{\rm{i}}(t-t')}{N_{\rm{total}}}dt',
\end{equation}
where the subscript i denotes the type of compact binaries (NS-NS, NS-BH, or BH-BH), $f_{\rm{b}}$ is the initial
binary fraction here taken 1/3, $\langle M\rangle~[\rm{M_{\odot}}]$ is the mean initial stellar mass,
${\rm SFR}(t')~[\rm{M_{\odot}~yr^{-1}~Mpc^{-3}}]$ is the Pop III SFR density at $t'$, $N_{\rm{total}}$ is the total number of simulated stars,
and $N_{\rm{i}}(t-t') dt'$ is the number of compact binaries which are formed during the time interval of
$[t', t'+ dt']$ and merge at the time t.
The resulting $R_i(t)$ is shown in Fig.~\ref{mergerrate} for the four cases with different IMFs
(Salpeter or flat) and core-merger criteria (optimistic or conservative; see Sec.~2.2.3).
Note that the difference in the core-merger criteria does not affect the merger rate density so much,
and the difference in the IMFs varies $R_i(t)$  by a factor of five.

In Model III.s~(the Salpeter IMF), the current merger rate density~($z=0$) of Pop III BH-BHs is estimated
as~(see Fig.~\ref{mergerrate})
\begin{equation}
R_{\rm{BHBH}} \sim 0.012 \left(\frac{{\rm SFR}_{\rm p}}{10^{-2.5}\ \msun {\rm yr}^{-1} {\rm Mpc}^{-3}}\right) {\rm Err}_{\rm sys}\ {\rm Myr}^{-1}{\rm Mpc}^{-3},
\end{equation}
where $\rm SFR_p$ is the peak value of Pop III SFR density in Fig.~8,
and $\rm Err_{sys}$ is the possible systematic errors on the assumption in the Pop III binary population
synthesis. $\rm Err_{sys}=1$ corresponds to adopting distribution functions of semi-major axis $a$, eccentricity
$e$, and the binary parameters for Pop I stars. Note that these might be different for Pop III stars, so that in
general $\rm Err_{sys}\neq 1$.
Adopting that the number density of galaxies is $n_{\rm{galaxy}} \sim 0.01\ \rm{Mpc}^{-3}$~\citep{Cross2001}, the
galactic merger rate of Pop III BH-BHs is estimated as $R_{\rm{BHBH,gal}} \sim 1.2\ {\rm Myr}^{-1}\ 
{\rm galaxy}^{-1}$. For Model III.f~(the flat IMF), the current merger rate density of the Pop III BH-BHs is estimated
as $R_{\rm{BHBH}} \sim 0.025\ {\rm Myr}^{-1}{\rm Mpc}^{-3}$, and the galactic merger rate as $R_{\rm{BHBH,gal}} 
\sim 2.5\ {\rm Myr}^{-1}\ {\rm galaxy}^{-1}$. It is worth to note that these rates are an order of magnitude
smaller than the lower limits of the merger rate of the NS-NSs derived from the observed NS-NSs~\citep{Kalogera2004b}.
We also note that although the number of merging BH-BHs is $\sim 4$ times larger for Model III.s than for Model III.s from Table~3, the merger rate is only a factor 2 larger.
 This comes from the difference in the mass distribution (see Figs. 6 and 7).

\subsection{Expected cumulative distribution as a function of cosmological $z$}\label{sec:cumulative}
\begin{figure*}
\includegraphics[scale=1.5]{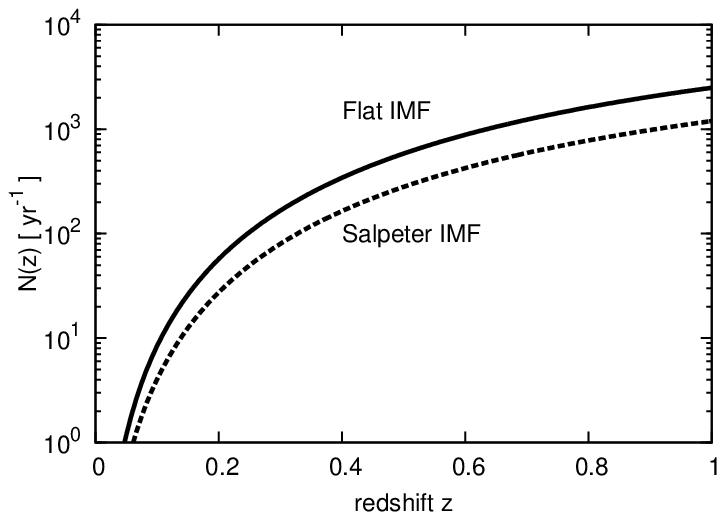}
\caption{The expected cumulative redshift distribution of the merger rate of the Pop III BH-BHs $N(z)$. The solid and 
dashed lines correspond to the flat and Salpeter IMF, respectively.  Here, we assume that the merger rate
density $R_{\rm m}$ of $30\ \msun \rm BH - 30\ \msun \rm BH$ remains constant with redshift. Although the
constant $R_{\rm m}$ is a good approximation for $z < 1$, for $z >1$, $R_{\rm m}$ increases as can be seen in
Fig.~9.}
\label{Fig:N_z}
\end{figure*}

First of all, from the chirp signal of a coalescing compact binary, we obtain the redshifted mass
$M_{1z}=(1+z)M_1$, $ M_{2z}=(1+z)M_2$ and the amplitude of the gravitational waves which is proportional\footnote{{The amplitude of the gravitational waves depends also on the sky position of the binary and the orbital inclination angle.
These quantities will be determined from the amplitude, phase,
and arrival time of the signals with a detector network.
Here we focus only on the quantities depending on the cosmological
redshift z for simplicity.}} to $M_{cz}^{5/6}/d_{\rm L}(z)$, where $M_{cz}$ is the redshifted chirp mass defined by
$(M_{1z}^{3/5}M_{2z}^{3/5})( M_{1z}+ M_{2z})^{-1/5}$ and  the luminosity distance ($d_{\rm L}(z)$)~\citep{Seto2001}. 
The luminosity distance is defined by
\begin{equation}
d_{\rm L}(z)=\frac{c}{H_0}(1+z)\int_0^z\frac{dz}{\sqrt{\Omega_{\rm m}(1+z)^3+\Omega_\Lambda}},
\end{equation}
where $c, H_0, \Omega_{\rm m}$ and $ \Omega_\Lambda$ are the light velocity, the present Hubble parameter, the
matter density parameter and the dark energy parameter, respectively. These values are now well determined
\citep{Planck2013} so that we have the three relations among three unknown variables, $M_1, M_2$ and $z$,
respectively. Then we can determine the values of $M_1, M_2$ and $z$, even if we can not determine the redshift
of the host galaxy or even if we can not determine the angular position of the observed compact binary precisely
by identifying the host galaxy.  The error of these values are order $({\rm S}/{\rm N})^{-1}$.
The comoving distance for a given redshift $z$ is defined by
\begin{equation}
r(z)=\frac{c}{H_0}\int_0^z\frac{dz}{\sqrt{\Omega_{\rm m}(1+z)^3+\Omega_\Lambda}}.
\end{equation}
Now writing the merger rate of a Pop III BH-BH per comoving volume as $R_m(z)$, 
we have the observed cumulative redshift number distribution of the coalescing Pop III BH-BHs $N(z)$ as
\begin{equation}
N(z)=4\pi \int_0^zR_m(z)r(z)^2\frac{1}{1+z}\frac{dr}{dz}dz,
\end{equation}
where $1/(1+z)$ is the effect of the cosmological time dilation. From Fig.~9 (a) and (b), we can regard that
$R_m(z)$ is essentially constant up to $z\sim 1$, so that we expect roughly
\begin{equation}
N(z)\propto \int_0^zr(z)^2\frac{1}{1+z}\frac{dr}{dz}dz.
\label{eq:N_z}
\end{equation}
The above equation shows that the cumulative distribution of Pop III coalescing BH-BHs depends roughly only
on the cosmological parameters $\Omega_{\rm m},\Omega_\Lambda$ and $z$. Figure~\ref{Fig:N_z} shows the $z$
dependence of Eq.~(\ref{eq:N_z}).
{From our simulations, the chirp mass distribution of Pop III BH-BHs is upward to the high mass and has a peak at  
$\sim 30\ \msun$.
The compact objects in IC10 X-1 and NGC300 X-1 are believed to be around 30 $\rm M_{\odot}$ and they can become coalescing massive BH-BHs whose chirp masses are 11-26 $\rm M_{\odot}$(See Bulik, Belczynski \& Prestwich 2011).
Thus, Pop I stars might become coalescing massive BH-BHs.
However, the typical mass of Pop I BH is around $10~\rm M_{\odot}$ and massive BH like IC10 X-1 and NGC300 X-1 would be rare  (See Figure1 in Belczynski et al. 2012) so that the chirp mass distribution of Pop I BH-BHs might be flat or downward to high mass.
Therefore, we might confirm the existence of Pop III black holes by the
determination of the chirp mass distribution.}
From Fig.~9, the merger rate of Pop III BH-BHs per comoving volume up to 
$z \sim 1$ is essentially constant. The expected event rate of the coalescing Pop III BH-BHs by KAGRA, Adv. LIGO, Adv.
Virgo and GEO network is $\sim 140\ (68)\ {\rm yr}^{-1}$ up to $\sim 1500\ {\rm Mpc}\ (z=0.28)$ for the flat
(Salpeter) IMF with the fiducial parameter values of $\rm SFR_p \sim 10^{-2.5}\ \msun {\rm yr}^{-1}{\rm Mpc}^{-3}
$ and $\rm Err_{sys}=1$.
Therefore, by checking if the observed cumulative redshift distribution agrees with
the results of our simulations, we can confirm the existence of Pop III massive stars. 
{Here we note that the density fluctuation of the PopIII BH-BHs at the local Universe does not affect the event rate. Because the spacial distribution for the sources at $>8\rm Mpc~(z=0.002)$ is almost uniform.
Moreover, cosmological simulations studying the hierarchically merging of halos suggest most galaxies could include a large number of the mini-halos where PopIII stars form (e.g., Greif et al. 2008; Bromm \& Yoshida 2011; Rashkov et al. 2012).}

By the third generation of the
detectors such as Einstein Telescope~(ET)\footnote{http://www.et-gw.eu}, 
we expect $\sim 80$ times more events per year.
{If there are more than three third generation detectors, we can determine the redshift up to
$z\sim 3$ so that we might see the evolution of the merger rate and determine the Pop III IMF from the
difference of  detection rate.}
Note that the expected rate of $2.5\ (1.2) \times 10^{-8}\ {\rm events}\ 
{\rm yr}^{-1}\ {\rm Mpc}^{-3}$ for the flat (Salpeter) IMF with the fiducial parameter values is consistent with
the upper limit of $\sim 10^{-7}\ {\rm events}\ {\rm yr}^{-1}\ {\rm Mpc}^{-3}$ by LIGO-Virgo(S6/VSR2/VSR3)~\citep{LIGOS6}. In reality, however, we should consider the mass of each black hole, the inclination angle of the binary orbit,
position of the detector and the detector noise spectrum so that we need to perform Monte Carlo simulations to 
obtain the expected cumulative redshift distribution (Kanda et al. in preparation). 

\section{Discussion}

In this paper, we do not include the magnetic braking since 
Pop III stars are formed from the primordial no metal gas. 
However, there are discussions against this assumption.  
Among them, the enhancement of the magnetic field
during the star formation has been well studied using the numerical simulations 
\citep[e.g.,][]{Maki2004,Machida2008b,Machida2013}.  
According to their results, the turbulent motions driven in the galaxy formation 
might increase the magnetic field up to $< 10^{-6}$ G
\citep{Schleicher2010,Sur2010,Schober2012,Turk2012}, 
which is similar value to that in molecular clouds of the galaxy. 
If this is the case, we should include the effect of magnetic braking, 
which will give rise $\rm Err_{sys}\neq 1$.

\cite{Dominik2012} discussed the metallicity dependence of the compact binary merger by focusing on Pop I stars
with metallicities ${\rm Z_{\odot}}$ and $0.1\ {\rm Z_{\odot}}$. Here, we compare our results for the coalescing
Pop III BH-BHs with those in \cite{Dominik2012} and discuss the implications of our results~(see Table~\ref{dominik}). 
In Table~\ref{dominik}, the second and third columns show the results of \cite{Dominik2012} for
metallicity $\rm Z_\odot$ and $\rm 0.1\ Z_\odot$ stars. Here, Models A and B correspond to the standard case of 
submodels A and B in \cite{Dominik2012}. The last column show our results for Pop III binaries, where we take the
fiducial parameter values: $\rm Err_{sys}= 1$ and $\rm SFR_p=10^{-2.5}\ \msun\ yr^{-1}\ Mpc^{-3}$. As in Table 
\ref{dominik}, \cite{Dominik2012} suggested that for Pop I stars with $Z = {\rm Z_{\odot}}$, the merger rate
of the BH-BHs 
becomes $8.2\ (1.9)\times 10^{-8}\ {\rm events\ yr^{-1}\ Mpc^{-3}}$ in their Model A~(Model B), respectively,
while for Pop I stars with $Z = 0.1\ {\rm Z_{\odot}}$, it becomes $7.33\ (1.36) \times 10^{-7}\ {\rm events\
yr^{-1}\ Mpc^{-3}}$ in their Model A~(Model B), respectively. Here we assume that the galaxy with metallicity
$Z=0.1\ {\rm Z}_\odot$ has the hypothetical number density $\sim 10^{-2}\ \rm Mpc^{-3}$ in order to compare the
results. On the other hand, for our Pop III BH-BHs, the expected merger rate is $2.5\ (1.2)\times 10^{-8}\
{\rm events\ yr^{-1}\ Mpc^{-3}}$ for the flat (Salpeter) IMF, respectively. 
This means that the merger rate of the Pop III BH-BHs 
is smaller than that of Pop I. However, for Pop I BH-BHs with metallicity $Z={\rm Z}_\odot\
(0.1\ {\rm Z}_\odot)$, the typical chirp mass is 6.7~(13.2) $\msun$, while for Pop III ones it is $\sim 30\
\msun$ as we showed in Fig.~7. Since the detection range of merger events increases in proportion to
$M_{\rm chirp}^{5/6}$, the detection rate increases in proportion to $M_{\rm chirp}^{5/2}$. Therefore, the
detectable event rate of the Pop III BH-BHs is 13~(6) times larger for the flat (Salpeter) IMF than that of
Pop I with $Z={\rm Z}_\odot$ in Model A, while it is 0.26~(0.13) times larger than that of Pop I with
$Z=0.1\ {\rm Z}_\odot$, if the galaxy consists of stars with $Z=0.1\ {\rm Z}_\odot$. In Model B, these numbers
become 56~(26) for $Z={\rm Z}_\odot$ case and 1.4~(0.7) for $Z=0.1\ {\rm Z}_\odot$ case, respectively.
Thus, for the fiducial parameters of Pop III binaries, the contribution of Pop III BH-BHs is comparable to or
larger than that of Pop I with $Z={\rm Z}_\odot$ and $0.1\ {\rm Z}_\odot$, since the major part of a galaxy does not necessarily consist of Pop I stars with $Z=0.1\ {\rm Z}_\odot$.

\begin{table*}
\caption[]{The comparison of the merger rate density of the BH-BHs and typical chirp mass between previous studies and our study.
The second and third columns show the results of \cite{Dominik2012} for metallicity $\rm Z_\odot$ and
$\rm 0.1\ Z_\odot$ stars. Here, Models A and B correspond to the standard case of submodels A and B in
\cite{Dominik2012}.
The last column show our results for Pop III binaries. Here, we take the fiducial parameter
values: $\rm Err_{sys}= 1$ and $\rm SFR_p=10^{-2.5}\ \msun\ yr^{-1}\ Mpc^{-3}$.}
\label{dominik}
\begin{tabular}{l c c c} 
\hline
         & Z$_\odot$ & 0.1 Z$_\odot$  & Pop III\\
\hline
Model A [$10^{-8}\ {\rm events\ yr^{-1}\ Mpc^{-3}}$]             & 8.2             & 73.3                 & 2.5 (flat)\\
Model B [$10^{-8}\ {\rm events\ yr^{-1}\ Mpc^{-3}}$]             & 1.9             & 13.6                 & 1.2 (Salpeter)\\
chirp mass [$\msun$]                                                           & 6.7              &13.2                  & 30\\
\hline
\end{tabular}
\end{table*}
 
While \cite{Dominik2012} did not take account of the evolution of the star formation rate, \cite{Dominik2013}
adopted a certain model of the star formation rate~\citep[Eq.~1 of][]{Dominik2013} and the metallicity $Z$
evolution~\citep[Eqs. 3 to 5 of][]{Dominik2013} to compute the cumulative redshift distribution of the coalescing
compact binaries. They also took into account the lower metal stars such as Pop II and even those with $Z<10^{-4}
\ {\rm Z}_\odot$, but not completely metal-free stars, Pop III.1 and Pop III.2 stars. The star formation rate
expressed by Eq.~(1) in \cite{Dominik2013} is completely different from the one shown in Fig.~8 of the present
paper. In the latter case, the star formation rate at $z=0$ is zero, while in the former case, it is the present 
star formation rate of our Galaxy which is not zero.

In Fig.~6 of \cite{Dominik2013},  they show the cumulative merger rate as a function of redshift $z$ for
different four models which corresponds to Fig.~10 of the present paper.  In our Fig.~10, for the second~(third)
generation gravitational wave detectors,  $ z\sim 0.3\ (3)$ is the detection range, respectively. In Fig.~6 of
\cite{Dominik2013}, information on the detectability is not available, since the chirp mass distribution function
is not available. Assuming  it is  similar to that in \cite{Dominik2012}, the merger rate for the second and
third generation gravitational wave detectors is either higher or lower than  Fig.~10 of the present paper taking
into account that the chirp mass of Pop I and Pop II BH-BHs is smaller than that for Pop III.1 and Pop III.2
BH-BHs. If the detection rate of the coalescing Pop I and Pop II BH-BHs is lower than that of Pop III, it might be
possible to confirm the existence of the massive Pop III stars by detecting gravitational waves from their
remnant black hole and identifying the typical chirp mass $\sim 30\ \msun$ and the cumulative redshift distribution which
depends almost only on the cosmological parameters. On the other hand, if the detection rate of the coalescing
Pop I and Pop II BH-BHs is higher, Pop III BH-BHs contribute only some parts of the gravitational wave events of BH-BHs. If 
the detected number is $\sim 10^4$ for the third generation detector like ET, we might select a Pop III BH-BH
from its mass and be able to draw the cumulative redshift distribution function to confirm the existence of
Pop III stars. In any case, it is needless to say that there are many undetermined parameters and distribution
functions for the Pop III population synthesis so that more theoretical study on the evolution and initial
conditions of Pop III binaries including the star formation rate is urgent. 

\section*{Acknowledgements}
We thank Jarrod R. Hurley, Christopher A. Tout and Krzysztof Belczynski, Nobuyuki Kanda and Kei Tanaka for useful comments and Sanemichi Takahashi and Sho Fujibayashi for help of  debugging of numerical codes.
This work is supported in part by the Grant-in-Aid from the Ministry of Education,
Culture, Sports, Science and Technology (MEXT) of Japan, No.25-1284 (T.K.), No.23-838 (K.I.), No.24-1772 (K.H.), No.23540305 (T.N.) and No.24103006 (T.N.).

\section*{Appendix}

\subsection*{A.1 The stellar radius and age after the mass transfer}
In this section, we describe how to calculate the stellar radius and other stellar properties after the mass
transfer. After the mass transfer, the primary mass and the secondary mass change and these changes affect the
stellar evolutions. In some case, the mass losing star becomes the naked-He star losing its H-envelope. 
The evolution of a naked-He star is well represented by that of a He star~\citep{Kippenhahn1990}.
However, for a Pop III star, the numerical calculation of a He main sequence evolution is not presented in the
literature.  So, in this paper, we follow the evolution of a naked-He star using the fitting formulae of
\cite{Hurley2000} with $Z=10^{-4}$.

\subsubsection*{(1)Main sequence}
We consider main sequence like stars, that is, the main sequence and naked He main sequence.
For example, if the main sequence star which has the convective core gains mass due to the mass transfer from the
companion giant star, it will get mass and mix the H in the convective core so that the star will
appear younger. Thus, the radius after the mass transfer of $\delta M$ is calculated as
$R_{\rm{fit}}(M+\delta M, t'(M,M+\delta M,t))$ where  $t'(M,M+\delta M,t)$ is the correction of the stellar
age due to the mass gain. In order to treat this rejuvenation, we use the approximation by Hurley et al. (2002).
Hurley's approximation treat the rejuvenation as three cases below (For details,  see Hurley et al. 2002, Hurley
et al. 2000, Tout et al. 1997).

Firstly, we consider that the main sequence star which has the convective core gains or loses the mass.
Assuming that the convective core mass and the fraction of the burnt hydrogen fuel at the core are proportional to the stellar mass and   the fraction of the main sequence lifetime $\tau_{\rm{H}}=t/t_{\rm{H}}$, respectively,
we have that the burnt H-mass at the convective core is proportional to $M\tau_{\rm{H}}$.
Since the burnt H-mass at the convective core does not change before and after the mass transfer,
we have 
\begin{equation}
M\frac{t}{t_{\rm{H}}(M)}=(M+\delta M)\frac{t'}{t_{\rm{H}}(M+\delta M)}.
\end{equation}
Therefore, the effective age $t'$ is
\begin{equation}
t'=\frac{M}{M+\delta M}\frac{t_{\rm{H}}(M+\delta M)}{t_{\rm{H}}(M)}t,
\end{equation}
and the stellar radius after the mass transfer is calculated as $R_{\rm{fit}}(M+\delta M, t'(M,M+\delta M,t))$.

Secondly, we consider that the naked He main sequence star gets the mass from the naked He giant star.
In the same way as the main sequence, we approximate
\begin{equation}
t'=\frac{M}{M+\delta M}\frac{t_{\rm{HeMS}}(M+\delta M)}{t_{\rm{HeMS}}(M)}t,
\end{equation} 
where
\begin{equation}
\left(\frac{t_{\rm{HeMS}}}{\rm Myr}\right)=\frac{0.4129+18.81(M/\msun)^4+1.853(M/\msun)^6}{(M/\msun)^{6.5}},
\end{equation}
 is the naked He main sequence star lifetime (Hurley et al. 2000).
The stellar radius after the mass transfer is calculated as $R_{\rm{fit}}(M+\delta M, t'(M,M+\delta M,t))$.

Thirdly, if the naked He main sequence star gets the H-rich envelope due to the mass transfer,
the star becomes the He-burning star.
Under the same assumption as the case of the main sequence,  the core mass of the new He-burning star is the same as the naked He main sequence star mass. 
Then the  age of the He-burning star is approximated as
\begin{equation}
t'=t_{\rm{H}}(M+\delta M)+\frac{t_{\rm{He}}(M+\delta M)}{t_{\rm{HeMS}}(M)}t.
\end{equation}
In this case, the stellar radius after the mass transfer is calculated by the same method as the giant star below.

\subsubsection*{(2) The giant stars}
For a giant star, we assume that the stellar age is decided by the core mass which is not affected by the mass transfer at the envelope.
Therefore, we do not change the age of the giant star at the mass transfer.
However the stellar radius changes due to the change of the mass ratio of the core to the envelope.
In order to calculate the  radius of the giant which is in the thermal equilibrium state after the star loses the envelope due to the mass transfer, 
we use the Hurley's formula (Hurley et al. 2000).

If the star loses the envelope mass, we calculate 
\begin{equation}\label{eq:muenv}
\mu = \left( \frac{M - M_{\rm{c}}}{M} \right) {\rm{min}} \left[ 5.0 , {\rm{max}} \left( 1.2 , \left( \frac{L}{L_0} \right)^{-1/2} \right) \right], 
\end{equation}
where $L_0 = 7.0 \times 10^4~\rm L_{\odot}$ for giant stars which have the H-envelope and 
\begin{equation}
\mu = 5 \left( \frac{M - M_{\rm{c}}}{M} \right), 
\end{equation}
for naked He giant stars. 
Then if $\mu < 1.0$ we calculate the radius as
\begin{equation} 
R_{\rm{th}}  =  R_{\rm{c}}(M_{\rm{c}}) \left( \frac{R_{\rm{fit}}(M+\delta M,t+\delta t)}{R_{\rm{c}}(M_{\rm{c}})} \right)^{r},
\end{equation}
where $R_{\rm{fit}}(M, t)$ is calculated using fitting formulae (Section \ref{sec:fitting formula}), 
\begin{equation}
r  = \frac{\left( 1 + c^3 \right) \left( \mu / c \right)^3 
{\mu}^{\rm fac}}
{1 + \left( \mu / c \right)^3},
\end{equation}
with
\begin{eqnarray}
c & = & 0.006, \\
\rm fac & = & {\rm{min}} \left[0.1/\ln \left( \frac{R}{R_{\rm{c}}} \right), \frac{-14}{\log\mu}\right].
\end{eqnarray}

In our calculation, the ratio of the stellar envelope mass to the total mass is typically about 1/2, i.e., $\mu \simeq 1/2$ and ${\rm fac}=0.1/(\ln(R/R_{\rm{c}})\simeq0.1/(\ln(100))\simeq 0.02 $.
Thus, the exponent of perturbation is typically $r \simeq \mu^{\rm fac} \simeq 1$.
Therefore, $R_{\rm{th}}\simeq R_{\rm{fit}}(M+\delta M,t+\delta t)$. 

\subsection*{A.2 The comparison with Hurley's code}
\begin{table*}
\caption{The models}
\label{model}
\begin{center}
\begin{tabular}[H!]{c c c c c c c} 
\hline
code & model A  & model B & model C & model D & model E&model F\\ 
\hline
Tidal evolution & ON & OFF & ON & ON & ON & ON  \\
$\alpha$ & 3.0 & 3.0 & 1.0 & 3.0 & 3.0 & 3.0 \\
e & 0 & 0 & 0 & 0 & 0 & Eq.\ref{eq:initial ecc} \\
$M_2$ & Eq.\ref{eq:mass ratio} & Eq.\ref{eq:mass ratio} & Eq.\ref{eq:mass ratio} & Eq.\ref{eq:IMF} & Eq.\ref{eq:mass ratio}  & Eq.\ref{eq:mass ratio}\\
Z & 0.02 & 0.02 & 0.02& 0.02 & 0.0001& 0.02 \\
\hline
\end{tabular}
\end{center}
\end{table*}

\begin{table*}
\caption{The comparison with Hurley's  code}
\label{comparison}
\begin{center}
\begin{tabular}[H!]{c c c c c c c} 
\hline
code & model A  & model B & model C & model D & model E&model F\\ 
\hline
Hurley et al. (2002) & $1.131\times10^{-1}$&$1.229\times10^{-1}$  &$7.572\times10^{-2}$ &$1.334\times10^{-2}$ & $2.290\times10^{-1}$ & $8.631\times10^{-2}$  \\
Hurley's open code &$1.683\times10^{-1}$& 1.799$\times10^{-1}$  &$1.197\times10^{-1}$ &$4.389\times10^{-2}$&$2.851\times10^{-1}$&$1.399\times10^{-1}$ \\
Our code & $1.189\times10^{-1}$ &$1.270\times10^{-1}$ &$8.491\times10^{-2}$&$1.031\times10^{-2}$ &$2.293\times10^{-1}$& $9.065\times10^{-2}$ \\
\hline
\end{tabular}
\end{center}
\end{table*}

Hurley's binary evolution open  code has a number of differences from BSE code in Hurley et al. (2002) (Private communication with Hurley and Tout).
The main differences of codes are the following.
First, in Hurley's binary evolution open  code, CE parameter $\lambda$ is not constant.
Hurley fitted $\lambda$ as a function of mass and luminosity and rewrote the code.
Second, they changed the treatments of compact remnant  mass from the formula of Hurley et al. (2002) to
Belczynski et al. (2002).
Furthermore, there are a lot of differences smaller than above two differences.
So we show the difference in the number of compact binary  between Hurley et al. (2002), Hurley's binary evolution open  code and our binary evolution model (see Section\ref{sec:binary}). 
We use the number of Pop I white dwarf-white dwarf (WD-WD) binaries in order to compare the binary evolution codes.
We use six models.
In these models, there are differences in  five quantities  such as the tidal evolution effect, the CE parameter $\alpha$, the initial eccentricity, the initial secondary mass and the metallicity (See Table \ref{model}, \cite{Hurley2002}). 
We show the comparison of WD-WD formation rates in  Table \ref{comparison}. We can see that results of \citet{Hurley2002} agree rather well with ours.

\subsection*{A.3 The convergence check}
In order to check whether the results depend on the total number of the binaries,
we calculate how many compact binaries can merge within 15~Gyr for Model III.s
setting $N_{\rm total}=10^5,~10^6$, and$~10^7$. 
The result is shown in Table \ref{convergence}

We assume that the result for $N_{\rm total}=10^7$ is the correct one 
and check the convergence of the results for $N_{\rm total}=10^6$ and $10^5$ 
using the Poisson distribution. 
First, we consider the result of NS-NSs.
In the case of $10^6$ binaries, the expected mean of the number of the coalescing NS-NSs is 
$n_{\rm{NSNS}}=27/10=2.7$. 
The number of the NS-NSs for $N_{\rm total}=10^6$ is 5, which is within 2$\sigma$ error, i.e., 
$n_{\rm{NSNS}}\pm 2\sqrt{n_{\rm{NSNS}}}$=$2.7\pm 3.2$.
Next, for NS-BHs and BH-BHs, the expected mean of the numbers are $n_{\rm{NSBH}}=56.2$
and $n_{\rm{BHBH}}=25434.6$, respectively.
Our results (64 and 25536) are within 1$\sigma$ error, i.e.,
$n_{\rm{NSBH}}\pm\sqrt{n_{\rm{NSBH}}}$=$56.2\pm 7.5$
and 
$n_{\rm{BHBH}}\pm\sqrt{n_{\rm{BHBH}}}$=$25434.6\pm 159.5$
Similar argument can be done for the results of $10^5$ binaries.
In conclusion the convergence of the results is confirmed within 1$\sigma$ statistical error except for small 
number of events like NS-NSs.
\begin{table*}
\caption{The convergence check of Monte Carlo simulations for Model III.s.  Each column means the number of 
binaries, the number of the coalescing NS-NSs, NS-BHs, and BH-BHs, respectively. }
\label{convergence}
\begin{center}
\begin{tabular}{c c c c } 
\hline
Total number & NSNS  & NSBH & BHBH \\ 
\hline
$10^5$ & 0 & 11 & 2593\\
$10^6$ & 5 & 64  & 25536 \\
$10^7$ & 27  &562& 254346 \\
\hline
\end{tabular}
\end{center}
\end{table*}  
\bibliographystyle{mn2e}

\begin{thebibliography}{}

\bibitem[\protect\citeauthoryear{Aasi et al.}{2013}]{LIGOS6} Aasi, J. et al.  \ 2013,  PRD, 87, 022002
\bibitem[\protect\citeauthoryear{Abel, Bryan \& Norman}{2002}]{Abel2002} Abel, T., Bryan, G.~L., 
\& Norman, M.~L.\ 2002, Science, 295, 93 

\bibitem[\protect\citeauthoryear{Abt}{1983}]{Abt1983} Abt, H.~A.\ 1983, ARA\&A, 21, 343

\bibitem[\protect\citeauthoryear{Accadia et al.}{2011}]{Accadia2011} Accadia, T., Acernese, 
F., Antonucci, F., et al.\ 2011, Classical and Quantum Gravity, 28, 114002

\bibitem[\protect\citeauthoryear{Amaro-Seoane et al.}{2013}]{Seoane2013} Amaro-Seoane, P., 
Aoudia, S., Babak, S., et al.\ 2013, GW Notes, Vol.~6, p.~4-110, 6, 4  

\bibitem[\protect\citeauthoryear{Baraffe, Heger \& Woosley}{2001}]{Baraffe2001} Baraffe, I., Heger, A., 
\& Woosley, S.~E.\ 2001, ApJ, 550, 890 

\bibitem[\protect\citeauthoryear{Belczynski, Kalogera \& Bulik}{2002}]{Belczynski2002} Belczynski, K., 
Kalogera, V., \& Bulik, T.\ 2002, ApJ, 572, 407 

\bibitem[\protect\citeauthoryear{Belczynski, Bulik \& Rudak}{2004}]{Belczynski2004} Belczynski, K., 
Bulik, T., \& Rudak, B.\ 2004, ApJL, 608, L45 

\bibitem[\protect\citeauthoryear{Belczynski et al.}{2007}]{Belczynski2007} Belczynski, K., 
Taam, R.~E., Kalogera, V., Rasio, F.~A., \& Bulik, T.\ 2007, ApJ, 662, 504 

\bibitem[\protect\citeauthoryear{Belczynski et al.}{2008}]{Belczynski2008} Belczynski, K., 
Kalogera, V., Rasio, F.~A., et al.\ 2008, ApJS, 174, 223 

\bibitem[\protect\citeauthoryear{Belczynski 
\& Dominik}{2012}]{Belczynski2012} Belczynski, K., \& Dominik, M.\ 2012, arXiv:1208.0358 

\bibitem[Belczynski et al.(2012)]{2012ApJ...757...91B} Belczynski, K., 
Wiktorowicz, G., Fryer, C.~L., Holz, D.~E., 
\& Kalogera, V.\ 2012, ApJ, 757, 91 

\bibitem[\protect\citeauthoryear{Bromm, Coppi \& Larson}{2002}]{Bromm2002} Bromm, V., Coppi, P.~S.,\& Larson, R.~B.\ 2002, ApJ, 564, 23



\bibitem[Bromm \& Yoshida(2011)]
{2011ARA&A..49..373B} Bromm, V., \& Yoshida, N.\ 2011, ARA\&A, 49, 373

\bibitem[Bulik et al.(2011)]{2011ApJ...730..140B} Bulik, T., Belczynski, 
K., \& Prestwich, A.\ 2011, ApJ, 730, 140

\bibitem[\protect\citeauthoryear{Bulik, Belczy{\'n}ski \& Rudak}{2004}]{Bulik2004} Bulik, T., Belczy{\'n}ski, K., \& Rudak, B.\ 2004, A\&A, 415, 407 

\bibitem[\protect\citeauthoryear{Cameron}{1967}]{Cameron1967} Cameron, A.~G.~W.\ 1967, Nature, 
215, 464 

\bibitem[\protect\citeauthoryear{Cappellaro et al.}{1997}]{Cappellaro1997} Cappellaro, E., Turatto, M., Tsvetkov, D.~Y., et al.\ 1997, A \& A, 322, 431

\bibitem[\protect\citeauthoryear{Cappellaro et al.}{1999}]{Cappellaro1999} Cappellaro, E., Evans, R., \& Turatto, M.\ 1999, A \& A, 351, 459


\bibitem[\protect\citeauthoryear{Clark et al.}{2011}]{Clark2011} Clark, P.~C., Glover, 
S.~C.~O., Klessen, R.~S., \& Bromm, V.\ 2011, ApJ, 727, 110 

\bibitem[\protect\citeauthoryear{Coward et al.}{2012}]{coward12}
Coward, D.M. , Howell, E.J., Piran, T. et al.\  2012, MNRAS, 425, 2668

\bibitem[\protect\citeauthoryear{Cross et al.}{2001}]{Cross2001} Cross, N., Driver, S.~P., 
Couch, W., et al.\ 2001, MNRAS, 324, 825

\bibitem[\protect\citeauthoryear{de Souza, Yoshida \& Ioka}{2011}]{Souza2011} de Souza, R.~S., Yoshida, N., \& Ioka, K.\ 2011, A\&A, 533, A32 


\bibitem[\protect\citeauthoryear{Doi \& Susa}{2011}]{Doi2011} Doi, K., \& Susa, H.\ 2011, ApJ, 741,
93

\bibitem[\protect\citeauthoryear{Dominik et al.}{2012}]{Dominik2012} Dominik, M., Belczynski, K., Fryer, C., et al.\ 2012, ApJ, 759, 52 

\bibitem[\protect\citeauthoryear{Dominik et al.}{2013}]{Dominik2013} Dominik, M., Belczynski, K., Fryer, C., et al.\ 2013, ApJ, 779, 72

\bibitem[\protect\citeauthoryear{Dopcke et al.}{2013}]{Dopcke2013} Dopcke, G., Glover, S.~C.~O., Clark, P.~C., \& Klessen, R.~S.\ 2013, ApJ, 766, 103 

\bibitem[\protect\citeauthoryear{Duquennoy, Mayor \& Halbwachs}{1991}]{Duquennoy1991} Duquennoy, A., Mayor, M., \& Halbwachs, J.-L.\ 1991, A\&AS, 88, 281 

\bibitem[\protect\citeauthoryear{Eggleton}{1983}]{Eggleton1983} Eggleton P.P.,1983, ApJ, 268, 368


\bibitem[\protect\citeauthoryear{Fryer}{1999}]{Fryer1999} Fryer, C.~L.\ 1999, ApJ, 522, 
413 

\bibitem[\protect\citeauthoryear{Fryer et al.}{2012}]{Fryer2012} Fryer, C.~L., Belczynski, 
K., Wiktorowicz, G., et al.\ 2012, ApJ, 749, 91 

\bibitem[Greif et al.(2008)]{2008MNRAS.387.1021G} Greif, T.~H., Johnson,
J.~L., Klessen, R.~S., \& Bromm, V.\ 2008, MNRAS, 387, 1021

\bibitem[\protect\citeauthoryear{Greif et al.}{2012}]{Greif2012} Greif, T.~H., Bromm, V.,Clark, P.~C., et al.\ 2012, MNRAS, 424, 399

\bibitem[\protect\citeauthoryear{Grindlay,  Portegies Zwart \& McMillan}{2006}]{Grindlay2006} Grindlay, J.,  Portegies Zwart, S. \&  McMillan, S. \ 2006, Nature Phys., 2, 116

\bibitem[\protect\citeauthoryear{Heggie}{1975}]{Heggie1975} Heggie, D.~C.\ 1975, MNRAS, 
173, 729 

\bibitem[\protect\citeauthoryear{Hjellming \& Webbink}{1987}]{Hjellming1987} Hjellming, M.~S., \& Webbink, R.~F.\ 1987, ApJ, 318, 794

\bibitem[\protect\citeauthoryear{Hjellming}{1989}]{Hjellming1989} Hjellming, M.~S.\ 1989, 
Ph.D.~Thesis, 

\bibitem[\protect\citeauthoryear{Hirano et al.}{2013}]{Hirano2013} Hirano, S., Hosokawa, 
T., Yoshida, N., et al.\ 2013, arXiv:1308.4456

\bibitem[\protect\citeauthoryear{Hosokawa et al.}{2011}]{Hosokawa2011} Hosokawa, T., Omukai,
K., Yoshida, N., \& Yorke, H.~W.\ 2011, Science, 334, 1250 

\bibitem[\protect\citeauthoryear{Hosokawa et al.}{2012}]{Hosokawa2012} Hosokawa, T., Yoshida, 
N., Omukai, K., \& Yorke, H.~W.\ 2012, ApJL, 760, L37

\bibitem[\protect\citeauthoryear{Hurley, Pols \& Tout}{2000}]{Hurley2000} Hurley, J.~R., Pols, O.~R., \& Tout, C.~A.\ 2000, MNRAS, 315, 543 

\bibitem[\protect\citeauthoryear{Hurley, Tout \& Pols}{2002}]{Hurley2002} Hurley, J.~R., Tout, 
C.~A., \& Pols, O.~R.\ 2002, MNRAS, 329, 897 

\bibitem[\protect\citeauthoryear{Hut}{1981}]{Hut1981} Hut, P.\ 1981, A\&A, 99, 126 

\bibitem[\protect\citeauthoryear{Inayoshi, Hosokawa \& Omukai}{2013}]{Inayoshi2013} Inayoshi, K., Hosokawa, T., \& Omukai, K.\ 2013, MNRAS, 431, 3036 

\bibitem[\protect\citeauthoryear{Ivanova, Podsiadlowski \& Spruit}{2002}]{Ivanova2002} Ivanova, N., 
Podsiadlowski, P., \& Spruit, H.\ 2002, MNRAS, 334, 819 

\bibitem[\protect\citeauthoryear{Ivanova et al. }{2008}]{Ivanova2008} Ivanova, N., Heinke, C. O., Rasio, F. A.  et al. \ 2008, MNRAS, 386, 553

\bibitem[\protect\citeauthoryear{Johnson \& Bromm}{2006}]{Johnson2006} Johnson, J.~L., \& Bromm, V.\ 2006, MNRAS, 366, 247 

\bibitem[\protect\citeauthoryear{Kalogera et al.}{2004a}]{Kalogera2004a} Kalogera, V., Kim, C., Lorimer, D. R. et al. \ 2004, ApJ, 601, L179 

\bibitem[\protect\citeauthoryear{Kalogera et al.}{2004b}]{Kalogera2004b} Kalogera, V., Kim, C., Lorimer, D. R. et al. \ 2004, ApJ, 614, L137

\bibitem[\protect\citeauthoryear{Kawamura et al.}{2011}]{Kawamura2011} Kawamura, S., Ando, 
M., Seto, N., et al.\ 2011, Classical and Quantum Gravity, 28, 094011

\bibitem[\protect\citeauthoryear{Kippenhahn \& Weigert}{1990}]{Kippenhahn1990} Kippenhahn, R., \& Weigert, A.\ 1990, Stellar Structure and Evolution,~ Springer-Verlag Berlin Heidelberg New York.~Also Astronomy and Astrophysics Library,  

\bibitem[\protect\citeauthoryear{Kobulnicky \& Fryer}{2007}]{Kobulnicky2007} Kobulnicky, H.~A., \& Fryer, C.~L.\ 2007, ApJ, 670, 747 

\bibitem[\protect\citeauthoryear{Kobulnicky et al.}{2012}]{Kobulnicky2012} Kobulnicky, H.~A., Smullen, R.~A., Kiminki, D.~C., et al.\ 2012, ApJ, 756, 50 


\bibitem[\protect\citeauthoryear{Kowalska, Bulik \& Belczynski}{2012}]{Kowalska2012} Kowalska, I., Bulik, T., \& Belczynski, K.\ 2012, A\&A, 541, A120 

\bibitem[\protect\citeauthoryear{Kulczycki et 
al.}{2006}]{Kulczycki2006} Kulczycki, K., Bulik, T., Belczy{\'n}ski, K., \& Rudak, B.\ 2006, A\&A, 459, 1001 


\bibitem[\protect\citeauthoryear{Kulsrud et al.}{Kulsrud et al.}{1997}]{Kulsrud1997} Kulsrud, R.~M., Cen,
R., Ostriker, J.~P., \& Ryu, D.\ 1997, ApJ, 480, 481

\bibitem[\protect\citeauthoryear{Langer, Puget \& Aghanim}{Langer et al.}{2003}]{Langer2003} Langer, M., Puget,
J.-L., \& Aghanim, N.\ 2003, PRD, 67, 043505

\bibitem[\protect\citeauthoryear{Li et al.}{2011}]{Li2011} Li, W., Chornock, R., Leaman, J., et al.\ 2011, MNRAS, 412, 1473


\bibitem[\protect\citeauthoryear{Loveridge, van der Sluys \& Kalogera}{2011}]{Loveridge2011} Loveridge, A.~J., van 
der Sluys, M.~V., \& Kalogera, V.\ 2011, ApJ, 743, 49  

\bibitem[\protect\citeauthoryear{Machida et al.}{2008}]{Machida2008a} Machida, M.~N., Omukai, K., Matsumoto, T., \& Inutsuka, S.-i.\ 2008, ApJ, 677, 813

\bibitem[\protect\citeauthoryear{Machida, Matsumoto \& Inutsuka}{Machida et al.}{2008}]{Machida2008b} Machida, M.~N.,
Matsumoto, T., \& Inutsuka, S.-i.\ 2008, ApJ, 685, 690

\bibitem[\protect\citeauthoryear{Machida
\& Doi}{2013}]{Machida2013} Machida, M.~N., \& Doi, K.\ 2013,
MNRAS, 435, 3283

\bibitem[\protect\citeauthoryear{Madau, Ferrara \& Rees}{2001}]{Madau2001} Madau, P., Ferrara, A., \& Rees, M.~J.\ 2001, ApJ, 555, 92

\bibitem[\protect\citeauthoryear{Maki \& Susa}{2004}]{Maki2004} Maki, H., \& Susa, H.\ 2004, ApJ, 609,
467

\bibitem[\protect\citeauthoryear{Marigo et al.}{2001}]{Marigo2001} Marigo, P., Girardi, L., Chiosi, C., \& Wood, P.~R.\ 2001, A\&A, 371, 152 

\bibitem[\protect\citeauthoryear{Matsumoto \& Nishimura}{1998}]{Matsumoto1998} Matsumoto, M. \& Nishimura, T.,\ 1998, TOMACS, 8, 1, 3

\bibitem[\protect\citeauthoryear{McKee \& Tan}{2008}]{McKee2008} McKee, C.~F., \& Tan,J.~C.\ 2008, ApJ, 681, 771



\bibitem[\protect\citeauthoryear{Omukai \& Nishi}{1998}]{Omukai1998} Omukai, K., \& Nishi,R.\ 1998, ApJ, 508, 141

\bibitem[\protect\citeauthoryear{Omukai et al.}{2005}]{Omukai2005} Omukai, K., Tsuribe, T., Schneider, R., \& Ferrara, A.\ 2005, ApJ, 626, 627 

\bibitem[\protect\citeauthoryear{Paczy{\'n}ski 
\& Sienkiewicz}{1972}]{Paczynski1972} Paczy{\'n}ski, B., \& Sienkiewicz, R.\ 1972, Acta Astronomica, 22, 73 

\bibitem[\protect\citeauthoryear{Paczynski}{1976}]{Paczynski1976} Paczynski, B.\ 1976, 
Structure and Evolution of Close Binary Systems, 73, 75 

\bibitem[\protect\citeauthoryear{Peters \& Mathews}{1963}]{Peters1963} Peters, P.~C., \& Mathews, J.\ 1963, Physical Review, 131, 435 

\bibitem[\protect\citeauthoryear{Peters}{1964}]{Peters1964} Peters, P.~C.\ 1964, Physical Review, 136, 1224 

\bibitem[\protect\citeauthoryear{Planck Collaboration}{2013}]{Planck2013} Planck                                                                                     
Collaboration, Ade, P.~A.~R., Aghanim, N., et al.\ 2013, arXiv:1303.5076

\bibitem[\protect\citeauthoryear{Podsiadlowski et al.}{2010}]{Podsiadlowski2010} Podsiadlowski, P., Ivanova, N., Justham, S., \& Rappaport, S.\ 2010, MNRAS, 406, 840 

\bibitem[\protect\citeauthoryear{Prince et al.}{1991}]{Prince1991}  Prince, T. A., Anderson, S. B.,  Kulkarni, S. R.\& Wolszczan, A. \ 1991, ApJ, 374, L41

\bibitem[\protect\citeauthoryear{Pudritz
\& Silk}{1989}]{Pudritz1989} Pudritz, R.~E., \& Silk, J.\ 1989,
ApJ, 342, 650

\bibitem[\protect\citeauthoryear{Punturo et al.}{2010}]{Punturo2010} Punturo, M., Abernathy, M., Acernese, F., et al.\ 2010, Classical and Quantum Gravity, 27, 194002 

\bibitem[Rashkov et al.(2012)]{2012ApJ...745..142R} Rashkov, V., Madau, P.,
Kuhlen, M., \& Diemand, J.\ 2012, ApJ, 745, 142

\bibitem[\protect\citeauthoryear{Rasio et al.}{1996}]{Rasio1996} Rasio, F.~A., Tout, 
C.~A., Lubow, S.~H., \& Livio, M.\ 1996, ApJ, 470, 1187 

\bibitem[Sathyaprakash
\& Schutz(2009)]{Sathyaprakash2009} Sathyaprakash, B.~S., \& Schutz,
B.~F.\ 2009, Living Reviews in Relativity, 12, 2


\bibitem[\protect\citeauthoryear{Salpeter}{1955}]{Salpeter1955} Salpeter, E.~E.\ 1955, ApJ, 121, 161

\bibitem[\protect\citeauthoryear{Schaerer}{2002}]{Schaerer2002} Schaerer, D.\ 2002, A\&A, 382, 28 

\bibitem[\protect\citeauthoryear{Schleicher et
al.}{2010}]{Schleicher2010} Schleicher, D.~R.~G., Banerjee, R., Sur,
S., et al.\ 2010, A \& A, 522, A115


\bibitem[\protect\citeauthoryear{Schneider et al.}{2006}]{Schneider2006} Schneider, R., Omukai, K., Inoue, A.~K., \& Ferrara, A.\ 2006, MNRAS, 369, 1437 


\bibitem[\protect\citeauthoryear{Schober et al.}{2012}]{Schober2012} Schober, J.,
Schleicher, D., Federrath, C., et al.\ 2012, ApJ, 754, 99

\bibitem[\protect\citeauthoryear{Seto, Kawamura \& Nakamura}{2001}]{Seto2001} Seto, N., Kawamura, S., \& Nakamura, T.\ 2001, Physical Review Letters, 87, 221103 

\bibitem[\protect\citeauthoryear{Stacy, Greif \& Bromm}{Stacy et al.}{2010}]{Stacy2010} Stacy, A., Greif, T.~H.,
\& Bromm, V.\ 2010, MNRAS, 403, 45

\bibitem[\protect\citeauthoryear{Stacy, Greif \& Bromm}{Stacy et al.}{2012}]{Stacy2012} Stacy, A., Greif, T.~H., \& Bromm, V.\ 2012, MNRAS, 422, 290

\bibitem[\protect\citeauthoryear{Stothers}{1966}]{Stothers1966} Stothers, R.\ 1966, ApJ, 144, 959 

\bibitem[\protect\citeauthoryear{Sur et al.}{2010}]{Sur2010} Sur, S., Schleicher,
D.~R.~G., Banerjee, R., Federrath, C., \& Klessen, R.~S.\ 2010, ApJL, 721,
L134


\bibitem[\protect\citeauthoryear{Tegmark et al.}{1997}]{Tegmark1997} Tegmark, M., Silk, J., Rees, M.~J., et al.\ 1997, ApJ, 474, 1

\bibitem[\protect\citeauthoryear{Timmes, Woosley \& Weaver}{1996}]{Timmes1996} Timmes, F.~X., Woosley, S.~E., \& Weaver, T.~A.\ 1996, ApJ, 457, 834 

\bibitem[\protect\citeauthoryear{Tout et al.}{1997}]{Tout1997} Tout, C.~A., Aarseth, S.~J., Pols, O.~R., \& Eggleton, P.~P.\ 1997, MNRAS, 291, 732

\bibitem[\protect\citeauthoryear{Turk et al.}{2012}]{Turk2012} Turk, M.~J., Oishi, J.~S.,
Abel, T., \& Bryan, G.~L.\ 2012, ApJ, 745, 154

\bibitem[\protect\citeauthoryear{Verbunt \& Phinney}{1995}]{Verbunt1995} Verbunt, F., \& Phinney, E.~S.\ 1995, A\&A, 296, 709  

\bibitem[\protect\citeauthoryear{Webbink}{1984}]{Webbink1984} Webbink, R.~F.\ 1984, ApJ, 277, 355 

\bibitem[\protect\citeauthoryear{Webbink}{1985}]{Webbink1985} Webbink, R.~F.\ 1985, Interacting Binary Stars, 39 

\bibitem[\protect\citeauthoryear{Webbink}{1988}]{Webbink1988} Webbink, R.~F.\ 1988, IAU Colloq.~103: The Symbiotic Phenomenon, 311 

\bibitem[\protect\citeauthoryear{Widrow}{2002}]{Widrow2002} Widrow, L.~M.\ 2002, Reviews of
Modern Physics, 74, 775


\bibitem[\protect\citeauthoryear{Woosley}{1986}]{Woosley1986} Woosley, S.~E.\ 1986, Saas-Fee 
Advanced Course 16: Nucleosynthesis and Chemical Evolution, 1 

\bibitem[\protect\citeauthoryear{Xu \& Li}{2010}]{Xu2010} Xu, X.-J., \& Li, X.-D.\ 2010, ApJ, 716, 114 

\bibitem[\protect\citeauthoryear{Yoshida et al.}{2006}]{Yoshida2006} Yoshida, N., Omukai, K., Hernquist, L., \& Abel, T.\ 2006, ApJ, 652, 6 


\bibitem[\protect\citeauthoryear{Yoshida, Omukai \& Hernquist}{2007}]{Yoshida2007b} Yoshida, N., Omukai, K., \& Hernquist, L.\ 2007, ApJL, 667, L117 

\bibitem[\protect\citeauthoryear{Yoshida, Omukai \& Hernquist}{2008}]{Yoshida2008} Yoshida, N., Omukai,K., \& Hernquist, L.\ 2008, Science, 321, 669

\bibitem[\protect\citeauthoryear{Zahn}{1975}]{Zahn1975} Zahn, J.-P.\ 1975, A\&A, 41, 329 

\bibitem[\protect\citeauthoryear{Zahn}{1977}]{Zahn1977} Zahn, J.-P.\ 1977, A\&A, 57, 383 


\end{thebibliography}

\end{document}